\documentclass[journal,]{IEEEtran}

\usepackage{enumerate}
\usepackage{cite}
\usepackage{tikz}
\usepackage{xcolor}
\graphicspath{{figures/}}
\usepackage{algorithm }
\usepackage{algorithmic}

\usepackage{makecell}
\usepackage{bm}
\usepackage{amsmath,amsfonts,amssymb,amsthm}
\usepackage{subfigure}
\usepackage{threeparttable}
\usepackage{booktabs}
\usepackage{bm}
\usepackage{mathrsfs}

\usepackage{url}
\usepackage{hyperref}
\makeatletter
\def\UrlAlphabet{%
      \do\a\do\b\do\c\do\d\do\e\do\f\do\g\do\h\do\i\do\j%
      \do\k\do\l\do\m\do\n\do\o\do\p\do\q\do\r\do\s\do\t%
      \do\u\do\v\do\w\do\x\do\y\do\z\do\A\do\B\do\C\do\D%
      \do\E\do\F\do\G\do\H\do\I\do\J\do\K\do\L\do\M\do\N%
      \do\O\do\P\do\Q\do\R\do\S\do\T\do\U\do\V\do\W\do\X%
      \do\Y\do\Z}
\def\UrlDigits{\do\1\do\2\do\3\do\4\do\5\do\6\do\7\do\8\do\9\do\0}
\g@addto@macro{\UrlBreaks}{\UrlOrds}
\g@addto@macro{\UrlBreaks}{\UrlAlphabet}
\g@addto@macro{\UrlBreaks}{\UrlDigits}
\makeatother

\usepackage{footnote}
\makesavenoteenv{tabular}

\begin{document}

\title{ Location-Enabled IoT (LE-IoT): A Survey of Positioning Techniques, Error Sources, and Mitigation  \thanks{
}}
\author{You Li,~\IEEEmembership{Member,~IEEE},
  Yuan Zhuang,~\IEEEmembership{Member,~IEEE},
  Xin Hu,~\IEEEmembership{Senior Member,~IEEE},
  Zhouzheng Gao,\\
  Jia Hu,
  Long Chen,~\IEEEmembership{Senior Member,~IEEE},  
  Zhe He,~\IEEEmembership{Member,~IEEE}, 
  Ling Pei,~\IEEEmembership{Member,~IEEE}, \\
  Kejie Chen,
  Maosong Wang, 
  Xiaoji Niu,  
  Ruizhi Chen, 
  John Thompson,~\IEEEmembership{Fellow,~IEEE}, \\
  Fadhel Ghannouchi,~\IEEEmembership{Fellow,~IEEE}, 
  and Naser El-Sheimy
   \thanks{Y. Li, Z. He, F. Ghannouchi, and N. El-Sheimy are with Department of Geomatics Engineering, University of Calgary (liyou331@gmail.com; zhehe@ieee.org; fghannou@ucalgary.ca; elsheimy@ucalgary.ca). 
Y. Zhuang and R. Chen are with the State Key Laboratory of Surveying, Mapping and Remote Sensing, Wuhan University (zhy.0908@gmail.com; ruizhi.chen@whu.edu.cn). 
X. Hu is with the School of Electronic Engineering, Beijing University of Posts and Telecommunications (huxin2016@bupt.edu.cn). 
Z. Gao is with the Department of Land Sciences, China University of Geosciences (Beijing) (zhouzhenggao@126.com).  
J. Hu is with the Department of Computer Science, the University of Exeter (j.hu@exeter.ac.uk).  
L. Chen is with the School of Data and Computer Science, Sun Yat-sen University (chenl46@mail.sysu.edu.cn).  
L. Pei is with the Shanghai Key laboratory of Navigation and Location Based Services, School of Electronic Information and Electrical Engineering, Shanghai Jiao Tong University (ling.pei@sjtu.edu.cn). 
K. Chen is with the the Department of Earth and Space Sciences, Southern University of Science and Technology (chenkj@sustech.edu.cn). 
M. Wang is with the College of Intelligence Science and Technology, National University of Defense Technology (wangmaosong12@hotmail.com). 
X. Niu is with the GNSS Research Center, Wuhan University (xjniu@whu.edu.cn).
J. Thompson is with the School of Engineering, University of Edinburgh  (john.thompson@ed.ac.uk). 
Corresponding author: Y. Zhuang. 
   This paper is partly supported by the National Natural Science Foundation of China (NSFC) Grants (No. 41804027, 61771135, and 61873163), the Natural Sciences and Engineering Research Council of Canada (NSERC) Discovery Grants, NSERC CREATE Grants, NSERC Strategic Partnership Grants, the Canada Research Chair (CRC) Grants, and the Alberta Innovates Technology Future (AITF) Grants}
}

\markboth{In Preparation for IEEE COMMUNICATIONS SURVEYS \& TUTORIALS}%
         {Shell \MakeLowercase{\textit{et al.}}: Bare Demo of IEEEtran.cls for Journals}

         \maketitle

         \begin{abstract}
The Internet of Things (IoT) has started to empower the future of many industrial and mass-market applications. Localization techniques are becoming key to add location context to IoT data without human perception and intervention. Meanwhile, the newly-emerged Low-Power Wide-Area Network (LPWAN) technologies have advantages such as long range, low power consumption, low cost, massive connections, and the capability for communication in both indoor and outdoor areas. These features make LPWAN signals strong candidates for mass-market localization applications. However, there are various error sources that have limited the localization performance by using such IoT signals. This paper reviews the IoT localization system through the following sequence: IoT localization system review - localization data sources - localization algorithms - localization error sources and mitigation - localization performance evaluation. Compared to the related surveys, this paper has a more comprehensive and state-of-the-art review on IoT localization methods, an original review on IoT localization error sources and mitigation, an original review on IoT localization performance evaluation, and a more comprehensive review of IoT localization applications, opportunities, and challenges. Thus, this survey provides comprehensive guidance for peers who are interested in enabling localization ability in the existing IoT systems, using IoT systems for localization, or integrating IoT signals with the existing localization sensors.          \end{abstract}
		 

         \IEEEpeerreviewmaketitle

\section{Introduction}  
\label{sec-intro}
         \IEEEPARstart{T}{he} Internet of Things (IoT) is shaping the future of many industrial and mass-market applications \cite{Zanella-IoT-2014}. As a core technology to acquire spatial IoT data, localization techniques are both an important application scenario and a distinguished feature for the next-generation IoT \cite{Zafari-2019}. In particular, Location-Enabled IoT (LE-IoT) is becoming key to add location context to IoT data without human perception and intervention.

         This section answers three questions: (1) why is it necessary to review localization techniques for IoT systems; (2) what are the advantages and challenges for IoT-signal-based localization; and (3) what are the differences between this survey and the previous ones. Table \ref{tab-notations} illustrates the notations and symbols that will be used in this survey. 
        
          \begin{figure*}
           \centering
           \includegraphics[width=0.86 \textwidth]{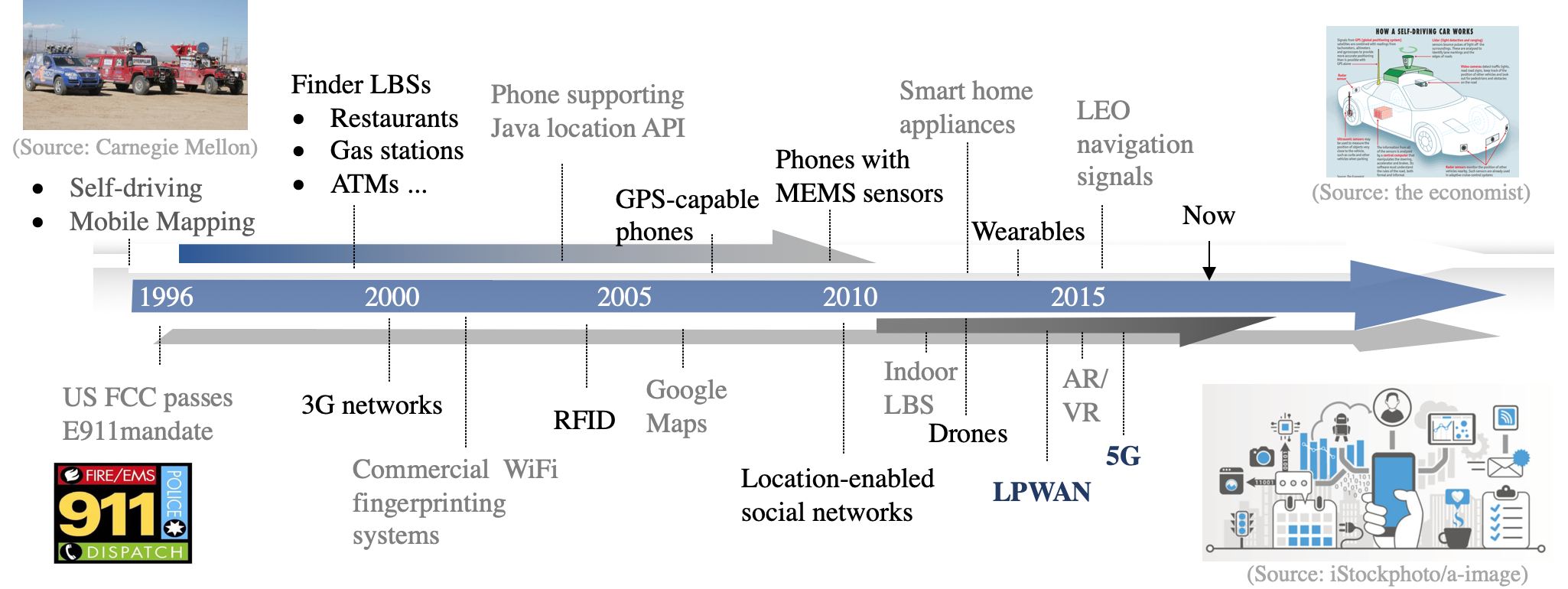}
           \caption{Timeline of location-based applications. Devices used in these applications are ``things" in IoT}
           \label{fig:iot-history}
         \end{figure*} 
         
\subsection{Localization Technologies and Applications}
\label{sec-exist-loc-tech-and-app}
As a military-to-civilian application, localization has been intensively researched and successfully commercialized in outdoor areas. Figure \ref{fig:iot-history} demonstrates the timeline of location-based services (LBS). The devices used in these applications are actually ``things" in the IoT. The main technologies for these use cases are Global Navigation Satellite Systems (GNSS) and Inertial Navigation Systems (INS) \cite{GaoZ-Sensors-2019}. In contrast, robust localization in indoor and urban areas is still an open challenge \cite{Shit-2018}. The development of indoor localization technologies has two directions: professional and mass-market applications. Professional applications (e.g., underground construction and machine industry) are commonly implemented in small areas, require high (e.g., decimeter or centimeter level) location accuracy; thus, they need specific network infrastructure, devices, and manpower. Furthermore, many professional applications also require intellisense techniques for environment sensing. 

In contrast, mass-market applications commonly do not need high localization accuracy. Meter-level accuracy is already within the human sensing range. However, mass-market applications are usually implemented in wide areas with varying environments; meanwhile, these applications are not affordable for specific devices or manpower. In particular, for mass-market IoT applications, power consumption is new key factor that should be considered. In general, mass-market IoT localization is more challenging due to the following factors:
         
\begin{itemize}
\item Many IoT nodes (including user end-devices) cannot afford GNSS receivers due to their high power consumption and cost. 
\item The complexity of environments. For example, the occurrence of Non-Line-of-Sight (NLoS) \cite{Ramadan-2018}, multipath \cite{Meissner-2014}, wide-area \cite{Erceg-MV-1999} and multi-floor \cite{LiY-Access-2018} effects, and interference by moving objects and human bodies \cite{Trogh-2016}. 
\item The necessity of using low-cost sensors which have significant sensor errors. The sensor errors also change over time and are susceptible to environmental factors (e.g., temperature \cite{Niu-thermal-2013}). 
\item The variety of node motions, such as changes of speed and orientation \cite{LiY-IF}. Also, it may be difficult to constrain node motion with predefined paths \cite{Saeedi-2013}. 
\end{itemize}          
         
There are various types of localization technologies. Their advantages and disadvantages are

\begin{itemize}
\item GNSS. Because of its capability to provide global weather-independent positioning solutions, GNSS has been commercialized successfully. However, its performance can be degraded by signal outages, degradations, and multipath in indoor and urban environments \cite{GaoZ-Sensors-2019}.
\item Wireless localization. Localization using wireless signals can provide long-term location accuracy. However, its performance is highly dependent on signal availability and geometry \cite{ChengY2013}; meanwhile, its accuracy can be degraded by signal fluctuations and interferences due to NLoS conditions \cite{Ramadan-2018}, reflections \cite{Torres-Solis2010}, and multipath \cite{Meissner-2014}, and the outage \cite{mi} and time variance \cite{LiY-IF} of radio maps.
\item Environmental signals (e.g., magnetic, air pressure, light, and sound intensity). Database matching (DB-M) is one main technique for localization using environmental signals. The challenges include the dependency on features in environmental signals \cite{LiY-sens-j-mag}, the low signal dimension \cite{LiY-IF}, and the time variance and outage \cite{Solin-2018} of environmental signal feature maps. 
\item Dead-reckoning (DR). Motion sensor (e.g., inertial sensor, magnetometer, and odometer) based DR can provide autonomous outdoor/indoor localization solutions \cite{YuN2018}. Nevertheless, it is challenging to obtain long-term accurate DR solutions with low-cost sensors because of the existence of sensor errors \cite{Niu-thermal-2013}, the misalignment angles between vehicle (e.g., human body and land vehicles) and device \cite{PeiL-access}, and the requirement for position and heading initialization. 
\item Vision localization. Vision sensors (e.g., cameras and Light Detection and Ranging (LiDAR)) can provide high location accuracy when loop closures have been correctly detected \cite{LeeT2019}. Meanwhile, some previous issues, such as a large computational load, are being eliminated by modern processors and wireless transmission technologies. However, the performance of vision localization systems is highly dependent on whether the measured features are distinct in space and stable over time. It is difficult to maintain accuracy in environments with indistinct or reduplicative features (e.g., areas with glass or solid-color walls) \cite{Mur-Artal-2015}.
\end{itemize} 

In general, the existing technologies have their own advantages and limitations \cite{LiY-thesis}. Thus, it is difficult to generate low-cost but high-performance localization solutions through the use of a stand-alone technology. Due to the complementary characteristics of various technologies, multi-sensor integration has become a trend to achieve reliable, continuous, and accurate outdoor/indoor seamless localization. 

The Low-Power Wide-Area Network (LPWAN) and 5G technologies have been used for communication in pilot sites. However, their localization capability has not been fully developed. Many of the existing IoT systems still rely on location solutions from the existing localization technologies such as GNSS \cite{Wang-2019} and Wireless Fidelity (WiFi) \cite{Janssen-2017}. There are two reasons for this phenomenon. First, the deployment density (i.e., the number within a given area) of current IoT Base Stations (BS) are not high enough for accurate localization. Second, there are several error sources and challenges in IoT localization uses. For the second factor, this survey provides detailed investigation and guidance. 

\begin{table*}
           \centering
\begin{tabular}{p{1.6cm} p{5.0cm} p{0.4cm} | p{0.4cm} p{1.6cm} p{5.0cm}}
\hline
\textbf{Abbreviation} & \textbf{Definition}  &   &  &  \textbf{Abbreviation} & \textbf{Definition} \\ \hline
2D/3D  &  Two/Three-Dimensional &   &  &  LoRaWAN & Long Range Wide-Area Network \\ 
3GPP &  3rd Generation Partnership Project &   &  &  LoS &Line-of-Sight \\ 
5G & 5th Generation cellular network  &   &  &   LPWAN & Low-Power Wide-Area Network \\ 
A3C &  Asynchronous Advantage Actor-Critic &   &   & LSTM&Long Short-Term Memory  \\ 
AHRS & Attitude and Heading Reference System  &   &  & LTE-M   &Long Term Evolution for Machines \\ 
AI & Artificial Intelligence  &   &  &   MAC&Media Access Control \\ 
ANN & Artificial Neural Network   &   &   & MIMO & Multiple-Input and Multiple-Output \\  
AoA & Angle of Arrival  &   &  &  ML & Machine Learning \\ 
AR &  Augmented Reality &   &  &  MLP &  Multi-Layer Perceptron\\ 
BLE & Bluetooth Low Energy  &   &   &  NB-IoT & Narrowband Internet of Things \\ 
BS & Base Station  &   &  &   NHC &Non-Holonomic Constraint \\ 
CNN &  Convolution Neural Network &   &  &  NLoS & Non-Line-of-Sight\\ 
CRLB &  Cramér-Rao Lower Bound &   &   & PF& Particle Filter \\   
CSI & Channel State Information  &   &  &   PHY & Physical layer \\ 
D2D & Device-to-Device  &   &  &  PLM & Path-Loss Model \\ 
DB-M & Database Matching  &   &   & PLM-P & Path-Loss Model Parameter  \\ 
DOP &  Dilution of Precision &   &  & PoA  & Phase-of-Arrival\\ 
DQN & Deep Q-Networks  &   &  &  RBF &Radial Basis Function \\ 
DR &  Dead-Reckoning &   &   & RFID& Radio-Frequency IDentification \\  
DRL & Deep Reinforcement Learning  &   &  &   RNN&Recurrent Neural Network \\ 
DRSS & Differential RSS  &   &  &  RP & Reference Points \\ 
DTM &  Digital Terrain Model &   &   & RPMA &Random Phase Multiple Access  \\ 
EKF & Extended Kalman Filter   &   &  &   RSS & Received Signal Strength \\ 
eMBB & Enhanced Mobile BroadBand  &   &  &   RTT & Round-Trip Time\\ 
ETSI & European Telecommunications Standards Institute  &   &   & SLAM &  Simultaneous Localization And Mapping\\   
GNSS &  Global Navigation Satellite Systems &   &  &  SNR & Signal-to-Noise Ratio\\ 
GP & Gaussian Processes  &   &  &  SVM &Support Vector Machine \\ 
GSM & Global System for Mobile Communications  &   &   & TDoA& Time Difference of Arrival \\ 
HMM & Hidden Markov Model  &   &  &  ToA & Time of Arrival \\ 
IEEE & Institute of Electrical and Electronics Engineers  &   &  &   UKF &Unscented Kalman Filter  \\ 
IETF & Internet Engineering Task Force  &   &   & UNB & Ultra NarrowBand \\  
INS & Inertial Navigation Systems   &   &  &   UNREAL & UNsupervised REinforcement and Auxiliary Learning \\ 
IoT &  Internet of Things &   &  &  URLLC  & Ultra-Reliable and Low Latency Communication \\ 
IPv6 & Internet Protocol version 6  &   &   & UWB & Ultra-Wide Band \\ 
ITU-R & International Telecommunication Union Radiocommunication Sector  &   &  &   VLP & Visible Light Positioning \\ 
KF & Kalman Filter  &   &  &  VR & Virtual Reality \\ 
LBS & Location-Based Services  &   &  & Weightless-SIG & Weightless Special Interest Group \\
LE-IoT & Location-Enabled IoT  &   &   &  WiFi & Wireless Fidelity  \\   
LE-LPWAN & Location-Enabled Low-Power Wide-Area Network   &   &  & ZARU & Zero Angular Rate Updates  \\ 
LEO & Low Earth Orbits  &   &  &  ZUPT & Zero velocity UPdaTes  \\ 
LF & Localization Feature  &   &   &  & \\ 
LiDAR & Light Detection And Ranging  &   &  &   &  \\
LoRa & Long Range  &   &  &   & \\ 

 \hline
           \end{tabular}
           \caption{ List of abbreviations   }
           \label{tab-notations}
         \end{table*}
         
\subsection{Advantages and Challenges of IoT Signals for Localization}
\label{sec-iot-pro-con}
\begin{figure}
           \centering
           \includegraphics[width=0.45 \textwidth]{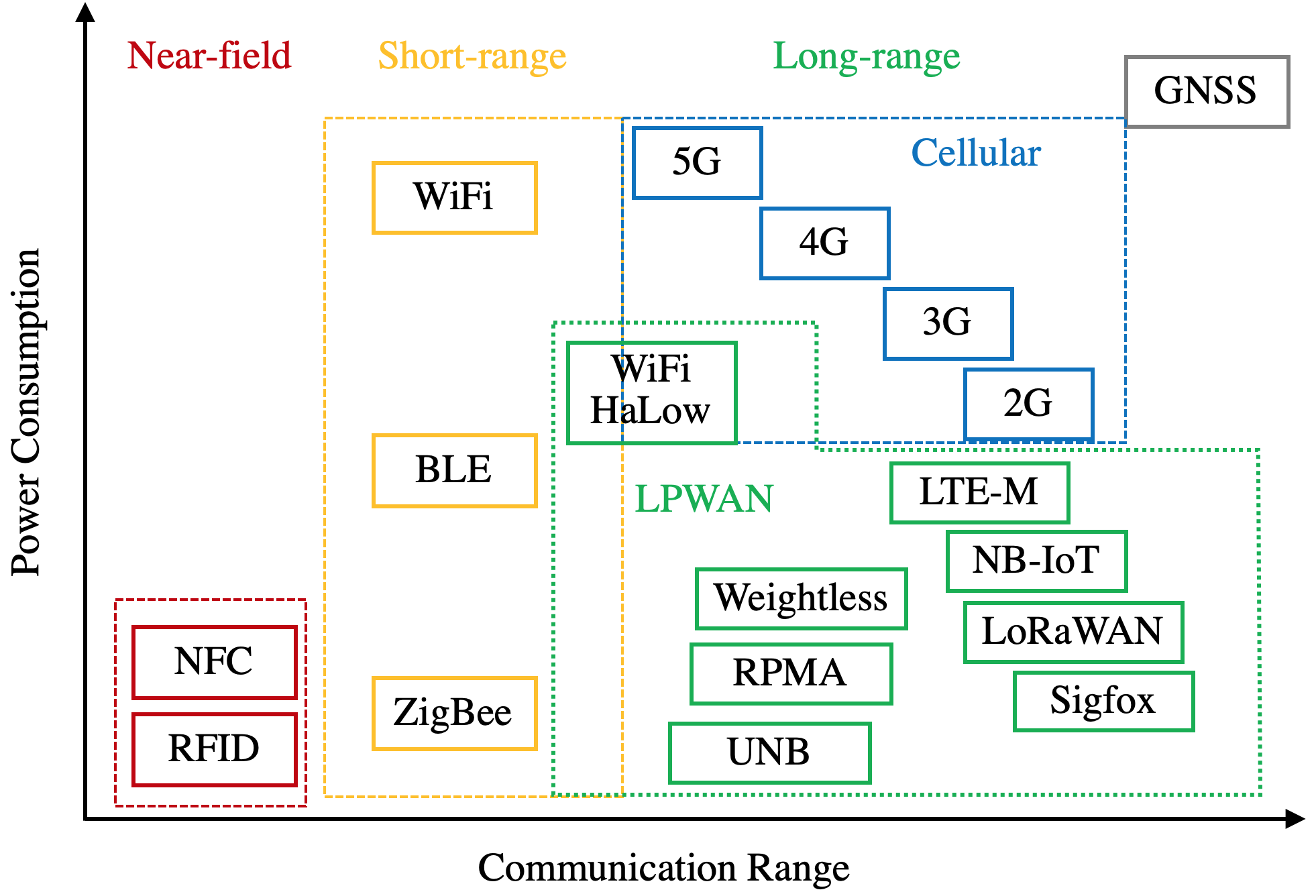}
           \caption{Coverage ranges and power consumption of IoT signals (modified on \cite{Thielemans-Bezunartea-2017}\cite{GuC-2018}\cite{ChoiW-2018})}
           \label{fig:wireless-signals}
         \end{figure}

The latest communication infrastructure is beginning to support the research on IoT signal based localization because: (1) IoT signals have been supported by mainstream IoT devices and are expected to be supported by more intelligent consumer devices. (2) IoT systems can already provide various localization-signal measurements such as Received Signal Strength (RSS), Time Difference of Arrival (TDoA), and Channel State Information (CSI). (3) The popularization of IoT/5G small BSs and the possibility to enable the communication capability of smart home appliances (e.g., lamps, routers, speakers, and outlets) are increasing the density of localization BSs. This survey focuses on LPWAN signals but also covers other IoT technologies such as cellular networks (e.g., 5G) and local wireless networks (e.g., WiFi, Bluetooth Low Energy (BLE), Zigbee, and Radio-Frequency IDentification (RFID)). Figure \ref{fig:wireless-signals} demonstrates the coverage ranges and power consumption of the main IoT signals. As a type of newly-emerged IoT signals, LPWAN has the following advantages:

\begin{itemize}
\item Communication capability. LPWAN nodes do not require extra costly communication modules.
\item Long range. Theoretically, 5 to 40 kilometers in rural areas and 1 to 5 kilometers in urban areas can be achieved \cite{Mekki-2019}.
\item Low power consumption. LPWAN nodes are expected to have over 10 years of battery lifetime \cite{Podevijn-2018}. With the same battery, LPWAN can support transmissions that are two orders more than GNSS \cite{Semtech-design-2019}. 
\item Low cost. The cost of a LPWAN radio chipset are being reduced to within 2 dollars, while the operation cost of each node can reach 1 dollar per year \cite{Raza-2017}. 
\item Massive connections. It is expected to support millions of nodes per BS (or gateway) per square kilometers \cite{Raza-2017}. 
\item The capability to work both outdoors and indoors \cite{LiY-LoRa-2018}.
\end{itemize}

The research paper \cite{Raza-2017} has reviewed the physical structures, techniques, and parameters for LPWAN, as well as the specific techniques for meeting the requirements such as long range, low power consumption, and low cost. 

On the other hand, the emergence of LPWAN has brought new challenges for localization techniques. Such challenges include the existence of wide-area scenarios, the high requirement for power consumption, the necessity of using low-data-rate and low-cost nodes, the high density of nodes, and the existence of complex node motions. Section \ref{sec-err-mitigation} reviews the IoT localization error sources and their mitigation in detail.

\subsection{Related Surveys and Tutorials}
\label{sec-rel-sur-tut}
The survey paper \cite{Mekki-2019} describes the development and technical differences among three main LPWAN technologies: Sigfox, NarrowBand IoT (NB-IoT), and Long-Range Wide-Area Network (LoRaWAN). Meanwhile, it describes several key IoT parameters, such as quality of service, scalability, latency, network coverage, battery life, payload length, cost, and deployment model. Also, it discusses the potential IoT applications, such as electric metering, manufacturing automation, smart building, and smart farming.

The review paper \cite{Raza-2017} investigates the LPWAN physical principles and techniques, such as band selection, modulation, narrowband, and spread spectrum techniques to meet long range requirements; topology, duty cycling, lightweight medium access control, and offloading complexity techniques to meet low-power-consumption requirements; hardware complexity reduction, minimum infrastructure, and license-free bands to meet low-cost requirements; and diversification, densification, and channel and data rate adaptive selection techniques for scalability. Moreover, the IoT standardizations from main standards organizations, such as the 3rd Generation Partnership Project (3GPP), the LoRa Alliance, the Institute of Electrical and Electronics Engineers (IEEE), the European Telecommunications Standards Institute (ETSI), the Internet Engineering Task Force (IETF), the Weightless Special Interest Group (Weightless-SIG), and the Dash7 Alliance, are described.

The paper \cite{Zafari-2019} reviews the existing indoor localization technologies and methods. Specifically, localization signals, such as RSS, Time of Arrival (ToA), TDoA, Angle of Arrival (AoA), CSI, and fingerprints, are described, followed by the sensors that provide these measurements. The sensors include WiFi, BLE, ZigBee, RFID, Ultra-Wide Band (UWB), Visible Light Positioning (VLP), ultrasound, and acoustic ones. Furthermore, the paper illustrates several key indoor localization evaluation indicators, including availability, cost, accuracy, scalability, coverage range, latency, and energy efficiency. These evaluation indicators are also suitable for IoT localization. For LPWAN, the paper describes the principles, design parameters, and possibility for localization. 

The review paper \cite{Haxhibeqiri-2018} surveys the hardware design and applications of LoRaWAN. Specifically, tools and methodologies such as system-level simulators, testbed deployments, physical layer (PHY) performance evaluation (e.g., coverage and interference impact), and Media Access Control (MAC) layer performance evaluation (e.g., network models, power usage, and security) are investigated. Meanwhile, the paper provides methods for enhancing LPWAN communication. The methods include network scalability assessment and improvement, scheduling and synchronization, new MAC design, Internet Protocol version 6 (IPv6) networks, multihop networks, and multi-modal networks. Furthermore, the strengths, weaknesses, opportunities, and threats analysis for LoRaWAN has been provided.

The paper \cite{Peral-Rosado-2018} reviews the development of cellular communication technologies (i.e., 1G to 5G). It also reviews the cellular localization methods, such as proximity, scene analysis, trilateration, and hybrid localization. In particular, it predicts the impact of several 5G features on localization. These 5G features include mmWave massive Multiple-Input and Multiple-Output (MIMO), multipath-assisted localization, and Device-to-Device (D2D) communication. The localization application aspects, such as indoor positioning, heterogeneity, synchronization, interference, power consumption, device-centric and network-centric, network planning, and commercial exploitation, are covered as well.

\begin{table*}
\centering
\begin{tabular}{p{5.5cm} p{0.6cm} p{0.6cm} p{0.6cm} p{0.6cm} p{0.6cm} p{0.6cm} p{0.6cm} p{0.6cm} p{0.6cm} p{0.6cm} p{0.6cm}}
\hline
\textbf{Features} & \textbf{\cite{Mekki-2019}}  & \textbf{\cite{Raza-2017}} & \textbf{~~\cite{Zafari-2019}} & \textbf{\cite{Haxhibeqiri-2018}} & \textbf{\cite{Peral-Rosado-2018}}  & \textbf{\cite{Whitepaper-2018}} & \textbf{\cite{Ikpehai-2018}} & \textbf{\cite{Petajajarvi-2017}}  & \textbf{\cite{WenF-2019}}  & \textbf{~~\cite{Shit-2018}} & \textbf{~~This}  \\ \hline
\textbf{System Principle and Architecture} &~~ S* &~~ S	&~~ M	&~~ M	&~~ M	&~~ S	&~~ S	&~~ S	&~~ M	&~~ M	&~~ M   \\ 
\textbf{Network Structure}	&~~ M	&~~ S	&~~ M	&~~ S	&~~ S	&~~ S	&~~ S	&~~ S	&~~ M	&~~ W	&~~ W   \\ 
\textbf{Hardware Technique}	&~~ M	&~~ S	&~~ W	&~~ S	&~~ S	&~~ M	&~~ M	&~~ M	&~~ S	&~~ W	&~~ W   \\ 
\textbf{MAC Layer}	&~~ S	&~~ S	&~~ W	&~~ S	&~~ S	&~~ S	&~~ S	&~~ S	&~~ M	&~~ W	&~~ W   \\ 
\textbf{PHY}	&~~ S	&~~ S	&~~ W	&~~ S	&~~ S	&~~ S	&~~ S	&~~ S	&~~ M	&~~ W	&~~ W   \\ 
\textbf{Standardization}	&~~ S	&~~ S	&~~ W	&~~ M	&~~ M	&~~ S	&~~ S	&~~ S	&~~ M	&~~ W	&~~ W   \\ 
\textbf{Existing System}	&~~ M	&~~ S	&~~ S	&~~ W	&~~ M	&~~ M	&~~ S	&~~ S	&~~ W	&~~ W	&~~ S   \\ 
\textbf{Localization Signal Source} 	&~~ W	&~~ W	&~~ S	&~~ W	&~~ M	&~~ M	&~~ W	&~~ M	&~~ M	&~~ M	&~~ S   \\ 
\textbf{Localization Algorithm}	&~~ W	&~~ W	&~~ M	&~~ W	&~~ M	&~~ M	&~~ W	&~~ W	&~~ M	&~~ S	&~~ S   \\ 
\textbf{Localization Error Source and Mitigation}	&~~ W	&~~ W	&~~ W	&~~ W	&~~ W	&~~ M	&~~ W	&~~ W	&~~ W	&~~ W	&~~ S   \\ 
\textbf{Localization Performance Evaluation} 	&~~ W	&~~ W	&~~ W	&~~ W	&~~ M	&~~ M	&~~ W	&~~ W	&~~ W	&~~ W	&~~ S   \\ 
\textbf{Localization Application}	&~~ W	&~~ W	&~~ S	&~~ M	&~~ M	&~~ S	&~~ W	&~~ M	&~~ M	&~~ S	&~~ S   \\ 
\textbf{New Localization Opportunity}	&~~ W	&~~ W	&~~ M	&~~ M	&~~ M	&~~ S	&~~ W	&~~ M	&~~ S	&~~ M	&~~ S   \\ \hline
\end{tabular}
\begin{tablenotes}
        		\item[*] ~~~* S-Strong; M-Medium; W-Weak
     		\end{tablenotes}
\caption{Comparison of Previous Works and This Survey}
\label{tab:existing-surveys}
\end{table*}

The whitepaper \cite{Whitepaper-2018} has systematically introduced the 5G and IoT standards, new features, applications, and limitations. Meanwhile, it points out several localization challenges, such as heterogeneity, multipath propagation, Line-of-Sight (LoS) availability, time synchronization, hardware complexity in large antenna array systems, power consumption and computational burden, and MAC latency and bandwidth usage. Also, the theoretical accuracy limitation for 5G ToA and AoA localization methods are derived. 

The review paper \cite{Ikpehai-2018} surveys the characteristics of various LPWAN technologies, including Sigfox, LoRaWAN, NB-IoT, Long Term Evolution for Machines (LTE-M), Random Phase Multiple Access (RPMA), and WavIoT. The investigated characteristics include network model and methodology, link budget and its impact on the implication, signal propagation, and network performance analysis (e.g., coverage, sensitivity analysis and network optimization, transmission delay, and energy consumption).

The paper \cite{Petajajarvi-2017} focuses on LPWAN PHY features (e.g., link layer and network architecture) and the analysis of LoRaWAN performance, such as the Doppler effect, node data rate, scalability, and network capacity. Localization scenario factors such as angular/linear velocity and outdoor coverage are considered in experiments.

The survey paper \cite{WenF-2019} reviews the 5G system principles, channel models and improvements, channel-parameter estimation, and localization. The channel-parameter estimation approaches involve subspace methods, compressed sensing, and distributed sources. The localization methods include LoS/NLoS localization, non-cooperative/cooperative localization, and indirect/direct localization. Furthermore, the paper analyzes 5G-localization opportunities and challenges, including efficient channel parameter estimation, accurate mmWave propagation modeling, cooperative localization, and the use of Artificial Intelligence (AI) techniques.

The survey paper \cite{Shit-2018} reviews the IoT localization and path-planning approaches. The localization techniques include multilateration, multiangulation, centroid, energy attenuation, region overlapping, bionics, verification, landmark design, clustering, and historical-information based methods. Furthermore, it summarizes the motion models, such as those for random walk, random waypoint, group mobility, self-organizing, and probability distribution. Meanwhile, it introduces the estimation approaches, such as multidimensional scaling, least squares, semi-definite programming, maximum likelihood estimation, Bayesian estimation, and Monte Carlo estimation. This paper provides the most systematical survey on IoT localization and estimation approaches. 

In general, the existing surveys (e.g., \cite{Mekki-2019}\cite{Raza-2017}\cite{Whitepaper-2018}) and numerous on-line resources have introduced the LPWAN and 5G principles, developments, technologies, and applications. Most of these resources focus on their communication capability. For localization purposes, the papers \cite{Zafari-2019}\cite{Shit-2018}\cite{Whitepaper-2018}\cite{WenF-2019} already have a systematical review on localization sensors and approaches. However, none of these papers has reviewed localization error sources and mitigation, which are key to design, use, and improve an LE-IoT system. This survey fills this gap. Table \ref{tab:existing-surveys} compares this paper with the related surveys.

\subsection{Main Contributions and Structure}
\label{sec-main-cont-str}
This paper reviews the IoT localization system in the following order: IoT localization system review - localization data sources - localization algorithms - localization error sources and mitigation - localization performance evaluation. Thus, it provides a comprehensive guidance for peers who are interested in enabling localization ability in the existing IoT systems, using IoT systems for localization, or integrating IoT signals with the existing localization sensors. In particular, this paper is the first survey on IoT localization error-source analysis, error mitigation, and performance evaluation. Compared to the related surveys, this paper has

\begin{itemize}
\item A more comprehensive and state-of-the-art review on IoT localization methods.
\item The first review on IoT localization error sources.
\item The first review on IoT localization error mitigation.
\item The first review on IoT localization performance analysis and evaluation.
\item A more comprehensive review of IoT localization applications, opportunities, and challenges.
\end{itemize}

Table \ref{tab-structure} illustrates the paper structure and the questions that are answered in each section. This survey is organized as follows

\begin{table*}
\centering
\begin{tabular}{p{2.4cm} p{7.0cm} p{6.5cm}}
\hline
\textbf{Section} & \textbf{Subsection}  & \textbf{Questions Answered} \\ \hline
\ref{sec-intro}. Introduction  &  \begin{itemize}
            \item \ref{sec-exist-loc-tech-and-app}. Localization Technologies and Applications 
            \item \ref{sec-iot-pro-con}. Advantages and Challenges of IoT Signals for Localization
            \item \ref{sec-rel-sur-tut}. Related Surveys and Tutorials
            \item \ref{sec-main-cont-str}. Main Contributions and Structure
			\end{itemize}
			 &   \begin{itemize}
            \item Why is it necessary to review localization techniques for IoT systems
            \item What are the advantages and challenges for IoT-signal-based localization
            \item What are the differences between this survey and the previous ones
			\end{itemize}  
\\ 
\ref{sec-overview}. Overview  &  \begin{itemize}
            \item \ref{sec-exist-lpwan-tech}. LPWAN Technologies 
            \item \ref{sec-iot-loc-app}. IoT Localization Applications
            \item \ref{sec-iot-loc-sys-stru}. IoT Localization System Architecture
            \item \ref{sec-sig-meas}. IoT Localization Signal Measurements
			\end{itemize}
			 &   \begin{itemize}
            \item How to select a LPWAN technology from the perspective of localization system designers  
            \item What can a LE-IoT system be used for 
            \item Which types of localization signals can be used
			\end{itemize}  
\\ 
\ref{sec-loc-method}. IoT Localization Methods  &  \begin{itemize}
            \item \ref{sec-db-m}. DB-M Localization Methods
            \item \ref{sec-geometric-loc}. Geometrical Localization Methods
			\end{itemize}
			 &   \begin{itemize}
            \item What are the state-of-the-art localization approaches 
            \item What are the advantages and challenges for each type of localization method
			\end{itemize}  
\\ 
\ref{sec-err-mitigation}. IoT Localization Error Sources and Mitigation  &  \begin{itemize}
            \item \ref{sec-err-end-device}. End-Device-Related Errors 
            \item \ref{sec-environm-error}. Environment-Related Errors
            \item \ref{sec-bs-error}. Base-Station-Related Errors
            \item \ref{sec-data-error}. Data-Related Errors
			\end{itemize}
			 &   \begin{itemize}
            \item Which types of localization error sources should be considered when designing a LE-IoT system  
            \item How to mitigate the effect of each type of localization error source
			\end{itemize}  
\\ 
\ref{sec-loc-per-eval}. IoT Localization-Performance Evaluation Methods  &  \begin{itemize}
            \item \ref{sec-thero-analysis}. Theoretical Analysis 
            \item \ref{sec-simu-analysis}. Simulation Analysis
            \item \ref{sec-in-the-lab}. In-the-lab Testing
            \item \ref{sec-field-test}. Field Testing
            \item \ref{sec-sig-grafting}. Signal Grafting
			\end{itemize}
			 &   \begin{itemize}
            \item What are the existing localization-performance evaluation methods
            \item What are the advantages and limitations of these localization-performance evaluation methods
			\end{itemize}  
\\ 
\ref{sec-loc-opport}. Localization Opportunities From LPWAN and 5G  &  \begin{itemize}
            \item \ref{sec-coop-loc}. Cooperative Localization
            \item \ref{sec-ml-ai}. Machine Learning / Artificial Intelligence
            \item \ref{sec-ms-integration}. Multi-Sensor Integration
            \item \ref{sec-fog-edge}. Fog/Edge Computing
            \item \ref{sec-blockchain}. Blockchain
            \item \ref{sec-airborne-land}. Airborne-Land Integrated Localization
            \item \ref{sec-multipath-assist}. Multipath-Assisted Localization 
			\end{itemize}
			 &   \begin{itemize}
            \item What are the new opportunities for localization due to the emergence of LPWAN and 5G signals
			\end{itemize}  
\\ 
\\ \hline
\end{tabular}
\caption{ Structure of This Survey }
\label{tab-structure}
\end{table*}

Section \ref{sec-overview} overviews the existing IoT technologies, followed by IoT localization applications, system architecture, and signal measurements. 

Section \ref{sec-loc-method} demonstrates the state-of-the-art IoT localization methods, including DB-M and geometrical localization. Specifically, DB-M approaches include deterministic DB-M methods such as nearest neighbors, stochastic DB-M methods such as Gaussian-distribution and histogram-based ones, and Machine-Learning (ML)-based DB-M methods such as Artificial Neural Network (ANN), random forests, Gaussian Processes (GP), and Deep Reinforcement Learning (DRL). Meanwhile, geometrical methods involve multilateration, hyperbolic positioning, multiangulation, multiangulateration, min-max, centroid, and proximity. 

Afterwards, Section \ref{sec-err-mitigation} systematically reviews IoT localization error sources and mitigation. The location errors are divided into four parts: (1) end-device-related errors (e.g., device diversity, motion/attitude diversity, low response rate/low sampling rate/data loss/data latency, and channel diversity), (2) environment-related errors (e.g., multipath, NLoS, wide-area effects, multi-floor effects, human-body effects, weather effects, and signal variations), (3) BS-related errors (e.g., number of BSs, BS geometry, BS location uncertainty, BS Path-Loss Model Parameter (PLM-P) uncertainty, and BS time synchronization errors), and (4) data-related errors (e.g., database timeliness/training cost, Reference Point (RP) location uncertainty, database outage, data and computational loads, and localization integrity). 

Then, Section \ref{sec-loc-per-eval} illustrates the localization performance evaluation methods, including theoretical analysis, simulation analysis, in-the-lab testing, field testing, and signal grafting. 

Finally, Section \ref{sec-loc-opport} shows the new localization opportunities, such as cooperative localization, AI, multi-sensor integration, motion constraints, fog/edge computing, blockchain, airborne-land integration, and multipath-assisted localization.

\begin{table*}
\centering
\begin{tabular}{p{1.7cm} p{1.6cm} p{1.6cm} p{1.6cm} p{1.6cm} p{1.6cm} p{1.6cm} p{1.6cm} p{1.8cm}}
\hline
\textbf{Features} & \textbf{NB-IoT}  & \textbf{LoRa} & \textbf{Sigfox} & \textbf{LTE-M} & \textbf{Weightless-N}  & \textbf{RPMA} & \textbf{HaLow} & \textbf{5G}  \\ \hline
\textbf{Standardization}	& 3GPP	& LoRa-Alliance 	& Sigfox and ETSI	& 3GPP	& Weightless SIG	& Ingenu	& IEEE	&  3GPP	   \\ 
\textbf{Licensed}	& Yes	& No	& No	& Yes	& No	& No	& No	& Yes	  \\ 
\textbf{Frequency} 	& In band LTE 	& 868/915/433 MHz ISM 	& 862-928 MHz ISM	& In band LTE	& Sub-1Ghz ISM    & 2.4 GHz ISM	& 900 MHz	&  Low/Mid/ mmWave bands  	  \\ 
\textbf{Bandwidth}	& 200 kHz	& 250/125 kHz	& 100 Hz	& 1.4 MHz	& 12.5 kHz	& 80 MHz 	& 1/2/4/8/16MHz	& 100/400 MHz	  \\ 
\textbf{Range} 	& 1/10 km in urban/rural	& 5/20 km in urban/rural	& 10/40 km in urban/rural	& 0.7/7 km in urban/rural	& 3 km in rural	& 5 km in rural	& 1 km	& Few hundreds of meters \\ 
\textbf{Peak data rate} 	& 250 kbps	& 3-50 kbps	& 600 bps 	& 1 mbps	& 30-100kbps	& 8bps-8 kbps	& 347 mbps	& Gbps level	  \\ 
\textbf{Messages per day} 	&  Limited	& Unlimited	& 140/4	& Unlimited	& Unlimited	& Undisclosed	& Unlimited	& Unlimited	  \\ 
\textbf{Peak payload length} 	& 1600 bytes	& 243 bytes	& 12/8 bytes	& 100 byte level	& 20 bytes	& 10 kb	& Very large	& Very large	  \\ 
\textbf{Cellular network}	& Yes	& No	& No	& Yes	& No	& No	& No	& Yes	  \\
\textbf{Network type}	& Nationwide	& Nationwide/ Private 	& Nationwide	& Nationwide	& Nationwide/ Private	& Nationwide/ Private	& Private	& Nationwide	 \\ 
\textbf{Authentication \& Encryption} 	& LTE encryption 	& AES 128b	& Not supported	& LTE encryption	& AES 128b	&  AES 256b	&  IEEE 802.11 high-level	& LTE encryption	 \\ 
\textbf{Capacity of Nodes}	& 50 k/BS	& 200 k/ gateway	& 1 m/gateway	& 50 k/BS	& Undisclosed	& 100 k/BS level	& 8 k/BS	& 1m/km$^2$	 \\ 
\textbf{Power Consumption}	& Low	& Low	& Low	& Medium, band dependent	& Low	& Low	& Medium	& Medium	 \\ 
\textbf{Time latency}	& 5 sec	& 1-10 sec	& 1-30 sec	&  100 ms	& 5-10 s	& over 25 sec	& ms level	& ms level	 \\ 
\textbf{BS Cost}	& \$15k/BS	& \$100/gateway, \$3k/BS 	& \$4k/BS	& \$15k/BS	& \$3k/BS	& Undisclosed	& \$100 level/gateway	& \$10k/BS	 \\ 
\textbf{Node Cost}	& \$5-15 	& \$3-10	& \$3-10	& \$10-20	& \$3-10	& \$5-10	& \$10-15	& Higher	 \\ 
\textbf{Topology}	& Star	& Star of stars	& Star	& Star, mesh	& Star of stars		& Star	& Star	& Star, mesh	 \\  \hline
\end{tabular}
\caption{Comparation of LPWAN and 5G\cite{Mekki-2019}\cite{Raza-2017}\cite{Haxhibeqiri-2018}\cite{Peral-Rosado-2018}\cite{Whitepaper-2018}\cite{Ikpehai-2018}\cite{Petajajarvi-2017}\cite{Almeida-Mendes-2019}\cite{Actility-2019}\cite{Rewers-2019} }
\label{tab:comparison-existiong-lpwan}
\end{table*}

\section{Overview}
\label{sec-overview}
This section first compares the existing LPWAN systems from the perspective of localization-system users, followed by the application scenarios of IoT localization. Afterwards, the IoT localization system architecture and the types of localization signal measurements are described. This section answers the following questions: (1) how to choose a LPWAN system for localization purposes; (2) what are the potential application scenarios for LE-IoT systems; and (3) what are the possible measurements that can be used for localization. Table \ref{tab:comparison-existiong-lpwan} compares the features of the main LPWAN techniques.

\subsection{LPWAN Technologies}
\label{sec-exist-lpwan-tech}
\begin{figure}
           \centering
           \includegraphics[width=0.43 \textwidth]{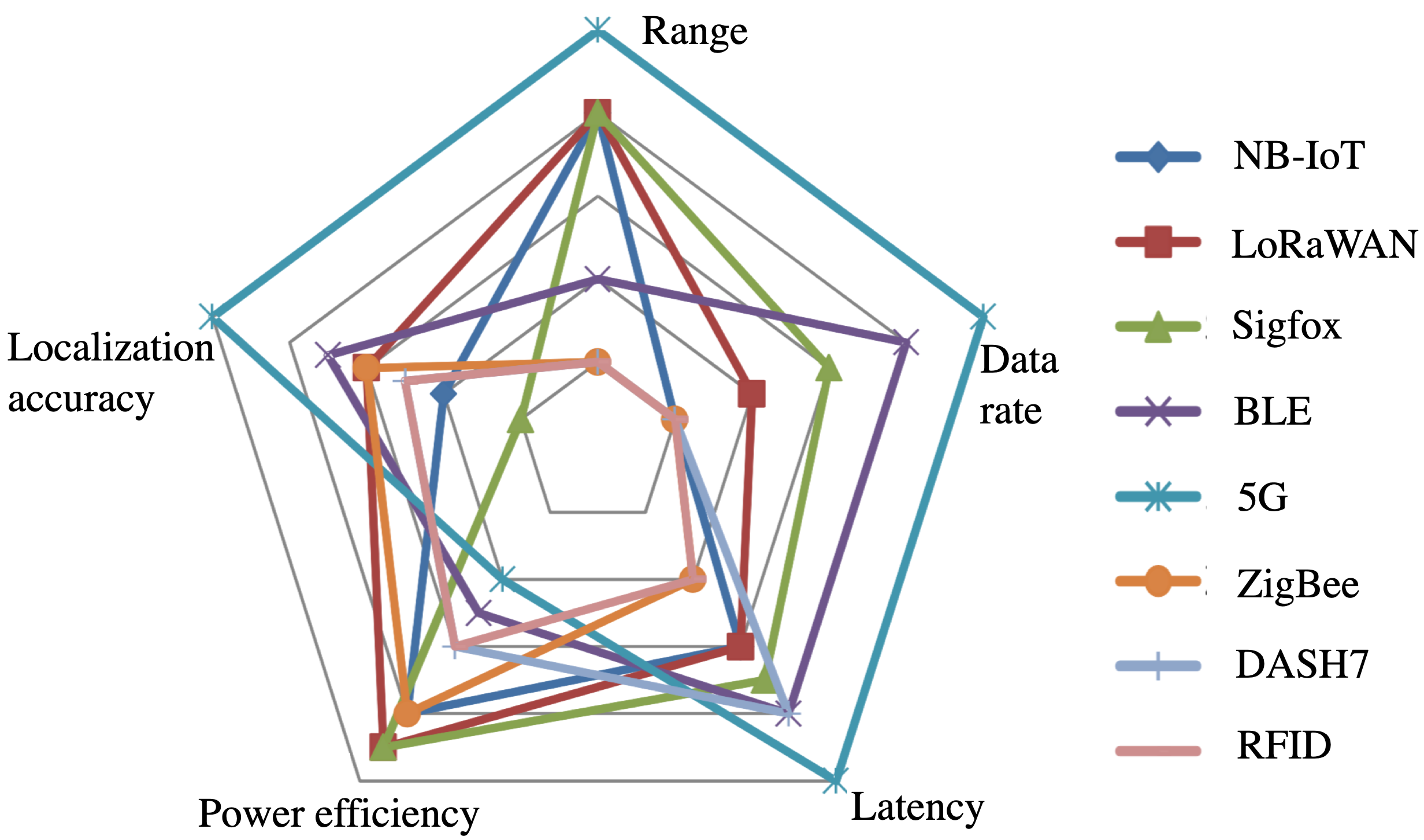}
           \caption{Comparative aspects of main IoT technologies \cite{Whitepaper-2018}}
           \label{fig:lpwan-compare}
         \end{figure}

	IoT, which was started as embedded internet or pervasive computing in 1970s and was termed in 1999, has a history of decades. The early-state IoT systems mainly used local communication technologies (e.g., RFID, WiFi, BLE, and Zigbee) and GNSS localization until LPWAN systems became available in around 2013 \cite{Mekki-2019}. Afterwards, over ten LPWAN technologies, licensed or license-free, were presented. Among them, LoRaWAN \cite{LoRa-Alliance-2019} and Sigfox \cite{Sigfox-2019}, which use license-free bands, and NB-IoT \cite{3GPP-R13-2019} and LTE-M \cite{3GPP-R13-2019}, which use licensed bands, are most widely used. Meanwhile, there are other LPWAN technologies, such as WiFi HaLow \cite{HaLow-2019}, Weightless \cite{Weightless-2019}, Ingenu RPMA \cite{rpma-2019}, Telensa \cite{Telensa-2019}, and Qowisio \cite{Qowisio-2019}. The papers \cite{Mekki-2019}\cite{Raza-2017}\cite{Haxhibeqiri-2018}\cite{Ikpehai-2018} have detailed descriptions on these LPWAN technologies. Figure \ref{fig:lpwan-compare} illustrates the comparative aspects of several IoT technologies. The following subsections describe the characteristics and applications of the main LPWAN systems.
	
\subsubsection{LoRaWAN}
\label{sec-lora}
LoRa was developed by the startup company Cycleo and was acquired by Semtech in 2012. Afterwards, the LoRa Alliance was founded in 2015. Until July 2019, the LoRa Alliance already has over 500 members and has deployed 76 LoRa public networks in 142 countries \cite{LoRa-Alliance-2019}. Thus, LoRa is one of the LPWAN technologies that have attracted extensive attention.

LoRaWAN has the following advantages: (1) its deployment is flexible. The user can deploy either a public or private network without the license from telecommunication operators. Thus, it is suitable for independent areas, such as communities, campuses, farms, and industrial parks, especially those in indoor or underground areas where the telecommunication signals are degraded. In these areas, localization is usually needed. (2) LoRaWAN has a well-developed ecosystem. A complete chain of LoRa chipsets - sensors and modules - BSs and gateways - network services - application services has already been set up. The standardization from the LoRa Alliance assures the interoperability among LoRaWAN in different countries. (3) In contrast to most of LPWAN systems which use a star topology, LoRaWAN uses a star of stars topology. That is, gateways are introduced to bridge nodes and BSs. The use of gateways, which have a lower cost and are more flexible to deploy, may enhance network coverage and localization performance in urban and indoor areas. (4) LoRaWAN support various classes (i.e., A, B, and C) for applications with various power and latency requirements. 

There are two types of challenges for LoRaWAN. The first type is the challenges for all LPWAN systems that use license-free bands: (1) it does not support upgrade using existing telecommunication BSs. That is, it always needs specific network deployment. (2) it is vulnerable to attacks due to its license-free bands and open standards. Although LoRaWAN has a relatively strong security standard, attackers may use LoRa nodes to jam the signal channel. (3) Spectral interference may occur with the increase of LoRaWAN operators. The second type of challenges only exists in LoRaWAN: (4) LoRa chipsets have been patented by Semtech. The excessive concentration of chip patents is not conducive to industrial growth.

\subsubsection{Sigfox}
\label{sec-sigfox}
Sigfox was developed by the startup Sigfox and then has experienced rapid development in the recent years. Until July 2019, Sigfox has already deployed networks in over 60 countries and regions \cite{Sigfox-2019}. It has generally completed the coverage of western Europe and is promoting to Asia and America. Although Sigfox and LoRaWAN use license-free bands, they have different operation modes. A main difference is that the Sigfox company itself acts as the global network operator.

The main advantages of Sigfox include: (1) it has a low node hardware cost and power consumption. To reduce cost, Sigfox uses the Ultra NarrowBand (UNB) technology and limits the data rate (100 bit/s/node level), message length (12 bytes) and the number of messages (140 message/day/node). Such a low data rate can reduce the node cost because even low-cost Binary Phase Shift Keying (BPSK) modules can meet the requirement \cite{Mekki-2019}. (2) Due to the use of UNB and short messages, a Sigfox network can be deployed by using a smaller number of long-range BSs. (3) It is straightforward to deploy a Sigfox network. The Sigfox company, which acts as the global operator, helps the users to deploy networks. Meanwhile, Sigfox is a global network, which does not need roaming between countries. (4) In contrast to LoRaWAN, Sigfox allows users to select chipsets from various chip manufacturers. 

The challenges for Sigfox include: (1) having the Sigfox company as a global operator has limited user permission and application flexibility. The users need to register and pay to the Sigfox company for services. Furthermore, the data has to be stored on the Sigfox server. (2) Both the UNB technology and the narrow downlink (i.e., the link from BS to node) have strongly limited the application scenarios. It is not cost-effective to use a Sigfox network for low-data-rate applications and use another LPWAN network for other applications. (3) From the localization perspective, having high-density BSs is generally beneficial. However, increasing the density of BSs contradict Sigfox's advantage of a lower BS density.

\subsubsection{NB-IoT}
\label{sec-nbiot}
Both NB-IoT and LTE-M started later than LoRaWAN and Sigfox. The 3GPP released the R13 NB-IoT standardization in 2016. However, the NB-IoT community is growing quickly. Until March 2019, it already has 140 operators in 69 countries \cite{GSA-2019}.

NB-IoT is mainly promoted by telecommunication operators. The main advantages of NB-IoT include: (1) it is highly valued by telecommunication operators. First, it supports upgrade on existing telecommunication BSs. Second, it can increase the number of users and bring extra service fees to the operators. (2) NB-IoT is based on licensed bands, which has an operator-level security and quality assurance. (3) The promotion from the government. For example, the China Ministry of Industry and Information Technology has released a policy to promote the development of NB-IoT in June 2017.
On the other hand, NB-IoT has met the following challenges: (1) it is difficult for independent companies, which are not cooperating with telecommunication operators, to participate. (2) The use of licensed bands increases the costs of both BSs and nodes. (3) NB-IoT started later than LoRa and Sigfox and thus may need time to achieve the same industry and market maturity.

\subsubsection{LTE-M}
\label{sec-ltem}
LTE-M, which includes enhanced Machine Type Communication (eMTC), was released in the 3GPP R13 standardization. This is similar to NB-IoT. Meanwhile, LTE-M is also designed for low-bandwidth cellular communications for the internet devices that transmit small amounts of data and have lower costs but higher battery life. Until March 2019, it has 60 operators in 35 countries \cite{GSA-2019}. Generally, LTE-M is relatively more supported by telecommunication operators in North America, while NB-IoT is more popular in China and Europe. 

Compared to NB-IoT, LoRa, and Sigfox, LTE-M has its own advantages: (1) LTE-M has much higher (i.e., up to 1 Mbps) data speeds. (2) It supports voice communication. (3) It has better mobility for devices in movement. Thus, LTE-M is suitable for applications such as vehicle networks, transportation, and security cameras.

The challenges for LTE-M include: (1) its nodes have a higher complexity and cost. LTE-M has a higher bandwidth (1.4 MHz) than NB-IoT (200 KHz); thus, both front end and digital processing are more complex for LTE-M. (2) LTE-M has a smaller signal coverage compared to other LPWAN technologies.

\subsubsection{Summary and Insight on LPWAN Technologies}
\label{sec-comm-lpwan}

\begin{figure}
           \centering
           \includegraphics[width=0.50 \textwidth]{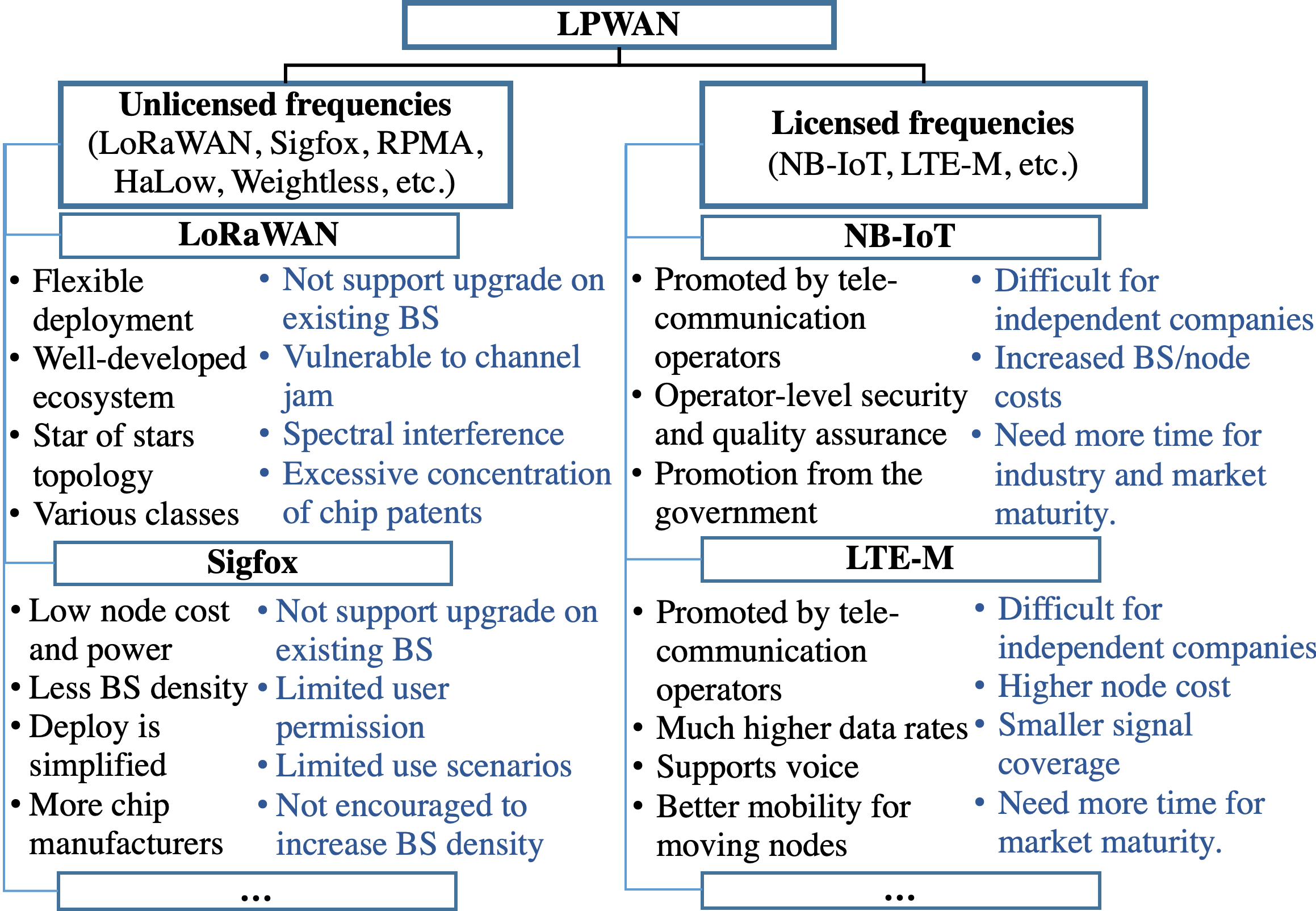}
           \caption{Advantages and challenges for mainstream LPWAN technologies}
           \label{fig:lpwan-pros-cons}
         \end{figure}

Figure \ref{fig:lpwan-pros-cons} illustrates the advantages and challenges of NB-IoT, LTE-M, LoRaWAN, and Sigfox. Although they have attracted most interests, there are various other LPWAN systems that have their own advantages. Therefore, it is expected that the users can select LPWAN technologies according to their requirements and integrate multiple LPWAN signals for better communication and localization performances. 

\subsubsection{Relation Between LPWAN and 5G}
\label{sec-rel-lpwan-5g}
In addition to LPWAN, the development of 5G (i.e., the fifth-generation cellular network) technology has brought opportunities for both communication and localization. Thus, it is worthwhile to introduce the relation between LPWAN and 5G. 

5G has been well-known for its high speed, massive connection, high reliability, and low latency in communication. Compared to 4G, 5G has innovatively designed various standards and solutions for different application scenarios. The International Telecommunication Union Radiocommunication Sector (ITU-R) has defined three 5G application categories: Ultra-Reliable and Low Latency Communication (URLLC), Enhanced Mobile BroadBand (eMBB), and massive Machine-Type Communication (mMTC) \cite{Restrepo-2019}. Specifically, URLLC has the advantages of high reliability (e.g., 99.999 \% reliable under 500 km/h high-speed motion) and low-latency (e.g., millisecond level); thus, it is suitable for applications such as vehicle networks, industrial control, and telemedicine. By contrast, eMBB has an extremely high data rate (e.g., Gbps level, with a peak of 10 Gbps) and strong mobility; thus, it is suitable for video, Augmented Reality (AR), Virtual Reality (VR), and remote officing applications. In contrast, mMTC is designed for application scenarios that have massive nodes which have a low cost, low power consumption, and low data rate, and are not sensitive to latency. Examples of these applications include Intelligent agriculture, logistics, home, city, and environment monitoring.

Therefore, LPWAN can aid 5G as follows: (1) LPWAN provides an important application direction for 5G, especially mMTC, applications. (2) Besides NB-IoT and LTE-M, which are within the 5G standardization group, there are various LPWAN networks and systems. They can provide complementary supports to 5G techniques and applications. 

On the other hand, 5G can help the development of LPWAN: (1) 5G is expected to build the fundamental infrastructure for communication services. In particular, the coverage range for 5G BSs may shrunk from kilometers to hundreds of meters or even under 100 m \cite{Andrews-2014}. The existence of such small BSs can help LPWAN communication and localization. (2) 5G features (e.g., mmWave MIMO, large-scale antenna, beamforming, and D2D communication) may enhance the LPWAN performance and experience. (3) 5G has a stronger connection with the new-generation information technologies (e.g., big data, cloud/edge computing, and AI) and thus can extend the application space of LPWAN.

\subsection{IoT Localization Applications }
\label{sec-iot-loc-app}
According the above analysis, IoT systems are especially suitable for applications that have massive connection, low data rate, low power consumption, and are not sensitive to latency. Many of such applications have a strong requirement for localization. Examples of these applications include

\begin{itemize}
\item Emergency service: determining people location is a feature of increasing importance for emergency systems such as the Enhanced 911 (E-911) in North America \cite{FCC-2015} and the E-112 in Europe \cite{Chochliouros-2015}. The current E-911 system uses cellular signals and has a typical localization accuracy of 80 \% for an error of 50 m. To enhance location accuracy in urban and indoor areas, other localization signals, including IoT signals, are needed.
\item Smart community: through the deployed IoT BSs and nodes around a community, it is possible to localize and track the residents as well as obtain their surrounding facilities (e.g., security alarms, fire alarms, lamps, air conditioners, and surveillance cameras).
\item Shopping mall: LE-IoT can be used for wide-area product positioning and management. Meanwhile, the data pertinent to people and products can be used for big-data analysis and service optimization. LE-IoT can provide the localization and management of public infrastructure such as vending machines, point-of-sale terminals, and advertising light boxes.
\item Intelligent transportation: LE-IoT nodes or chips in vehicles (e.g., cars or bikes) can be used for positioning and information tracking. The vehicle and related infrastructure (e.g., charging piles and parking spaces) locations can be used for traffic monitoring and parking guidance. 
\item Smart logistics: LE-IoT can provide city-level wide-area product tracking and management.
\item Environmental monitoring: LE-IoT can be used for localizing environmental hazards such as debris flows, sewer abnormities, and hazardous wastes.
\item Smart animal husbandry: LE-IoT can be used to track livestock locations and motions and thus provide services such as diet monitoring and meat traceability.
\item Animal tracking: LE-IoT can be used for wildlife tracking, pet monitoring, and animal movement data analysis. 
\item Smart agriculture: LE-IoT nodes around farms can be used to localize fertilization devices and monitoring environmental factors (e.g., temperature and humidity).
\item Smart home: LE-IoT nodes in smart home appliances (e.g., smart speakers, lamps, and outlets) and those on the human body can provide personalized services such as automatic temperature adjustment and light control. 
\item Health care: LE-IoT can localize patients and medical devices and then provide services such as remote monitoring, fall detection, and motion analysis.
\end{itemize}

\begin{figure}
           \centering
           \includegraphics[width=0.45 \textwidth]{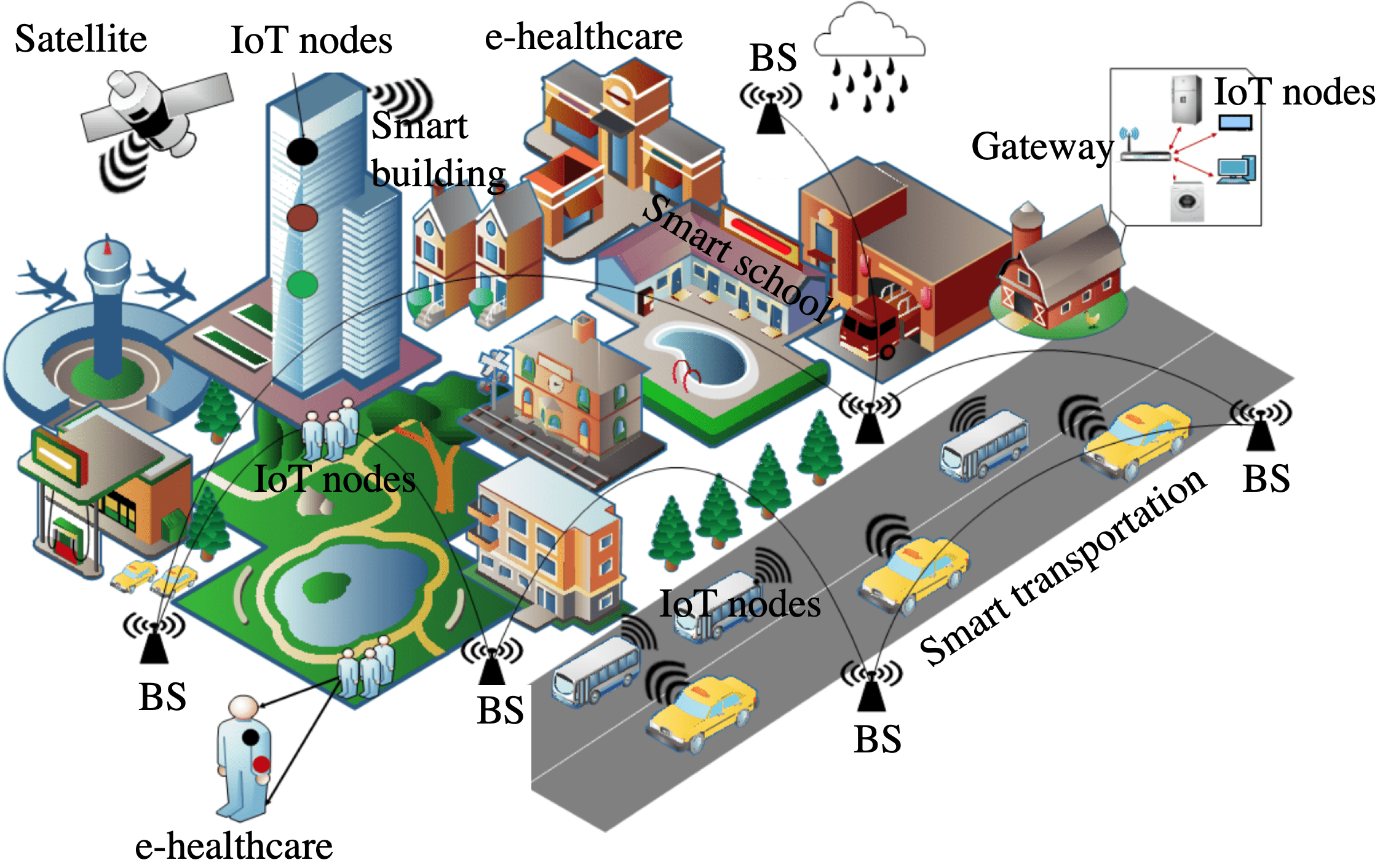}
           \caption{Example of LE-IoT applications \cite{Mehmood-Ahmad-2017}}
           \label{fig:iot-application}
         \end{figure}

Figure \ref{fig:iot-application} demonstrates some of the LE-IoT applications. Many of the current LE-IoT nodes use existing localization sensors (e.g., GNSS, inertial sensors, WiFi, and RFID) and an extra communication module (e.g., LTE). Thus, these nodes have relatively high cost and high power consumption, which have limited their applications. With the advent of LPWAN, some costly communication modules can be replaced by LPWAN based modules. For example, the configuration of LPWAN plus GNSS have been used for applications such as bus tracking \cite{GuanP-2018}, highway tracking \cite{Silva-2018}, and patient-motion monitoring \cite{Nugraha-2018}. Also, it is feasible to use LPWAN to send GNSS raw measurements to a server for processing \cite{Wang-2019}, or using WiFi instead of GNSS for localization \cite{Janssen-2017}. In these applications, the node hardware cost has been significantly reduced. Furthermore, if the LPWAN localization capability can be explored, it will be possible to remove all or parts of other localization sensors and thus further reduce node cost and power consumption.

\subsection{IoT Localization System Architecture}
\label{sec-iot-loc-sys-stru}

		 \begin{figure}
           \centering
           \includegraphics[width=0.48 \textwidth]{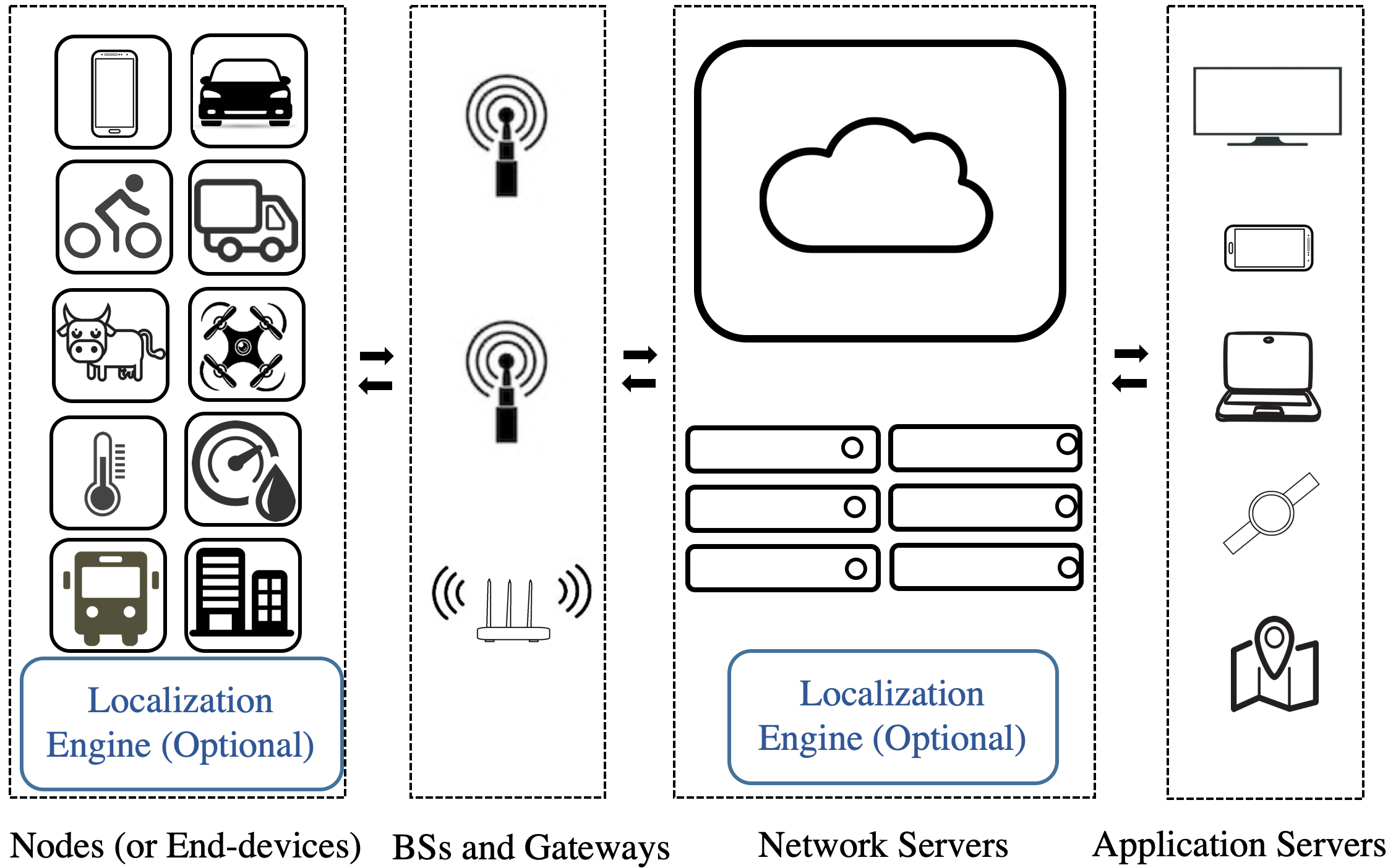}
           \caption{LE-IoT system architecture}
           \label{fig:le-iot-structure}
         \end{figure} 

A LE-IoT system is comprised of four components: nodes (including end-devices), BSs (including gateways or anchors), network servers, and application servers \cite{Raza-2017}. Figure \ref{fig:le-iot-structure} shows an LE-IoT system architecture. The LE-IoT system has an extra localization engine compared to an ordinary IoT system. The localization module may be located at either nodes or network servers, depending on user requirements on factors such as node computational load, communication load, and data security. The main functionalities of its components are as follows.

\subsubsection{Nodes}
A node contains a transponder, which transmits signals, and optionally a micro-controller with on-board memory. Meanwhile, the node optionally has application sensors such as a GNSS receiver for precise positioning, inertial sensors for motion tracking, and environmental sensors for monitoring temperature, humidity, smoke, gas, light, magnetic, and sound. The sensors may be connected to or integrated within the transponder chip. Also, the nodes may be fixed for static monitoring, mounted on dynamic objects (e.g., products, vehicles, and animals) as tags, or put on human body as user devices. In some IoT applications, the nodes only broadcast signals frequently, instead of processing data, to save energy. There are also applications in which motion-tracking or localization data processing is implemented on nodes to reduce communication load. Meanwhile, some applications may use ToA localization, which requires precise timing on nodes. This requirement can only be met in relatively high-end IoT applications. 

\subsubsection{BSs}
The main communication function of BSs is to route the data between nodes and network servers. The BSs may connect to network servers via the standard user datagram protocol / internet protocol and transmit the data from node sensors to network servers and vice versa. BSs usually have fixed and known locations as well as globally unique IDentities (IDs, e.g., MAC addresses). For LE-IoT, BSs also need to measure localization signals, such as node ID, BS ID, the data reception time, channel, RSS, payload, and Signal-to-Noise Ratio (SNR). Furthermore, BS time synchronization may be required for TDoA or ToA based localization \cite{Leugner-2016}. Meanwhile, multi-array antennae and phase detection may be needed for AoA localization \cite{Badawy2014}. 

\subsubsection{Network servers}
A network server is responsible for decoding data from BSs, recording data into databases, optionally implementing localization computation, and transmitting processed data to application servers. Network servers can be used for both sensor-to-application and application-to-sensor communication. For TDoA based localization, it is important that the packets from different BSs arrive at a network server. Furthermore, for localization applications, there are extra localization-signal databases on network servers. Meanwhile, motion-tracking and localization data-processing engines are located at network servers for many LPWAN applications. Network servers may be either cloud or edge servers.

\subsubsection{Application servers}
Their main functions are to obtain data from network servers, parse it, and process it for further applications.

\subsection{IoT Localization Signal Measurements}
\label{sec-sig-meas}
Compared to IoT, LE-IoT systems measure localization signals and process them to estimate motion states such as location, velocity, attitude, and motion modes. This subsection illustrates the commonly used localization signals.

\subsubsection{RSS}
\label{sec-rss}
RSS is measured when a BS or node receives the data packet from the other side. The advantages of using RSS include: (1) RSS can be straightforwardly collected without extra hardware on either nodes or BSs. (2) RSS can be flexibly used for various localization algorithms, such as proximity, region-determination, multilateration, and DB-M. On the other hand, the challenges for using RSS include: (1) it is difficult to determine the PLM-P accurately in wide-area \cite{Sallouha-2017}, urban \cite{Bor-2016}, and indoor \cite{Seidel-1992} scenarios, where many IoT applications take place. (3) The RSS-ranging resolution degrades over node-BS distance. Specifically, an RSS change of one dBm may lead to distance differences of meters in small areas but hundreds of meters in wide areas. Meanwhile, RSS measurements vary significantly when the node-BS distance changes within a certain range but less significantly when the node-BS distance becomes far \cite{LiY-LoRa-2018}. (4) RSS variations and interference due to environmental factors are issues inherent to wireless signals.

\subsubsection{ToA}
\label{sec-tof}
ToA is obtained by measuring the time interval between signal transmission and reception. The advantages of using ToA include: (1) theoretically, ToA measurements can be linearly converted to node-BS distances without any known PLM-P. (2) Take UWB \cite{Dardari-2009} and ultrasonic \cite{marvelmind} as examples, ToA ranging can achieve high accuracy (e.g., decimeter or even centimeter level) in light-of-sight (LoS) environments. (3) ToA localization has a well-researched theoretical-derivation and accuracy-assessment mechanism \cite{Shen-2012}. The challenges for ToA localization include: (1) ToA measurements require precise timing on both nodes and BSs, or precise time synchronization between them. A ten-nanosecond-level timing accuracy is required to achieve meter-level ranging. Such timing accuracy is not affordable for many IoT nodes. Thus, TDoA operating across multiple BSs is commonly used in IoT localization to eliminate the requirement for precise timing on nodes. (2) A high accuracy is commonly expected when ToA is used. In this case, the degradations from environmental factors (e.g., NLoS and multipath) are relatively more significant.

\subsubsection{TDoA}
\label{sec-tdpa}
TDoA is measured by computing the signal arrival time differences among multiple BSs. TDoA localization has the following advantages: (1) it does not need precise timing on nodes or precise time synchronization between nodes and BSs. Instead, it only requires precise BS time synchronization, which is affordable for many IoT systems such as LoRa, Sigfox, and NB-IoT \cite{Leugner-2016}. (2) The impact of node diversity can be mitigated through the use of differential measurements between BSs. (3) TDoA localization methods, such as hyperbolic localization, have a well-researched theoretical-derivation and accuracy-assessment mechanism \cite{Kaune-2011}. On the other hand, the challenges for TDoA localization include: (1) the requirement of precise time synchronization increases the BS cost. (2) The use of differential measurements enhances the impact of noise in localization signals. 

\subsubsection{AoA}
\label{sec-aoa}
 AoA systems provide the node position by measuring BS-node angles \cite{BaiL2008}. The advantages of AoA positioning include: (1) typical AoA localization systems (e.g., the HAIP system \cite{Quuppa2018}) can provide high-accuracy (e.g., decimeter or centimeter level) locations. (2) AoA requires less BSs than ToA and TDoA. It is feasible to use two BS-node angle measurements, or one BS-node angle and one BS-node distance, for two-dimensional (2D) localization. By fixing the AoA BS on the ceiling with known height, it is even possible to provide accurate localization with one BS \cite{Quuppa2018}. (3) AoA localization approaches, such as multiangulation, have a well-researched theoretical-derivation and accuracy-assessment mechanism \cite{WangY-2015}. The challenges for AoA localization include: (1) AoA systems need specific hardware such as multi-array antennae and phase detection \cite{Badawy2014}. The high node cost has limited the use of AoA in low-cost IoT applications. (2) Although there are low-cost RSS-based AoA systems \cite{LiY-TIM-2019}, the accuracy of both angular-measuring and positioning degrade significantly when the BS-node distance increases. Thus, a high-density BS network is still needed for wide-area applications.

\subsubsection{Round-Trip Time (RTT)}
\label{sec-rtt}
 RTT can be collected by measuring the round-trip signal propagation time to estimate the distance between nodes and BSs \cite{Ciurana-2007}. The use of RTT has advantages such as: (1) compared to ToA, RTT needs less accurate clock synchronization between BSs and nodes \cite{Zafari-2019}. (2) RTT can be collected from the MAC layer, instead of the PHY \cite{Gunther-2005}. (3) It is straightforward to use ToA-localization methods for RTT localization. The challenges for RTT localization include: (1) Modification on nodes is not affordable for many low-cost IoT applications. (2) The response delay between signal reception and transmission, which is difficult to eliminate, directly leads to ranging errors \cite{Zafari-2019}. (3) RTT-estimation accuracy is degraded by the same error sources as ToA.

\subsubsection{CSI}
\label{sec-csi}
It is becoming possible to collect CSI between IoT nodes and BSs \cite{Halperin-2011}. The advantages of using CSI include (1) CSI localization can achieve a high accuracy (e.g., decimeter level or higher) \cite{WuK-2013}. (2) CSI measurements have more features than RSS \cite{Chen-2013}. (3) CSI is more robust to multipath and indoor noise \cite{Zafari-2019}. (4) Many existing localization approaches, such as DB-M and multilateration, can be used for CSI localization. The challenges for CSI localization include: (1) CSI may not be available on off-the-shelf Network Interface Controllers (NICs). (2) The CSI measurements may suffer from deviations because of factors such as limitations in channel parameter estimation \cite{Chen-2013}. (3) It is challenging to assure the CSI-based ToA measurement accuracy due to the limited IoT signal bandwidth \cite{Chen-2013}. To mitigate this issue, techniques such as frequency hopping may be needed \cite{Vasisht2016}. 

\subsubsection{Phase-of-Arrival (PoA)}
\label{sec-poa}
PoA is obtained by measuring the phase or phase difference of carrier signals between nodes and BSs. PoA measurements can be converted to BS-node distances \cite{GuC-2018}. The advantages for PoA localization include: (1) PoA measurements can achieve high (e.g., centimeter-level or higher) ranging accuracy \cite{Scherhaufl-2013}. (2) The existing ToA and TDoA algorithms can be directly used for PoA localization. The challenges for PoA localization include: (1) Extra node and BS hardware are needed to measure PoA \cite{GuC-2018}. Meanwhile, accurate PoA-ranging requires a relative high data rate, which is not suitable for many IoT applications. (2) A high accuracy is commonly expected when PoA is used. In this case, the degradations from environmental factors (e.g., NLoS and multipath) are relatively more significant \cite{GuC-2018}. (3) PoA measurements may suffer from the integer-ambiguity issue \cite{GengJ-2010} and cycle slips \cite{JiangW-2017}.

\subsubsection{Summary and Insight on IoT Localization-Signal Measurements}
\label{sec-comm-loc-sig-meas}

\begin{itemize}
\item Similar to other engineering problems, the selection of IoT localization signals is a tradeoff between performance and cost. Some measurements (e.g., ToA, AoA, RTT, and PoA) can be used to achieve high localization accuracy but require extra hardware or modifications on nodes, which are not affordable for many low-cost IoT applications. In contrast, measurements such as RSS can be collected without any change on hardware; however, their localization accuracy is lower, especially for wide-area applications. Another types of measurements (e.g., TDoA and CSI) may be realized by adding extra hardware or modifications on the BS side, which is affordable for some IoT applications. Besides performance and cost, other factors should be considered when selecting localization signals. Example of these factors include environment size, outdoors or indoors, node motion modes, and the number of nodes. 

\item Because each type of localization measurement has advantages and limitations, it is common to combine various types of measurements (e.g., TDoA/RSS \cite{Kazikli-2019}	 and AoA/TDoA \cite{YinJ-2019}) for a higher localization performance. Furthermore, data from other sensors (e.g., inertial sensors, magnetometers, barometers, and maps) can be introduced to enhance localization solutions by mitigating the impact of error sources that are inherent to wireless signals.

\item Moreover, all the localization signals suffer from both deterministic and stochastic measurement errors. The impact of deterministic errors (e.g., sensor biases, scale factor errors, and thermal drifts) may be mitigated through calibration \cite{Figuera-2011} or on-line estimation. In contrast, it is difficult to compensate for stochastic measurement errors. These errors can be modeled as stochastic processes \cite{Maybeck-1982}. The statistical parameters for stochastic models may be estimated through methods such as correlation, power spectral density analysis, Allan variance \cite{LuoC-2018}, and multisignal wavelet variance \cite{Radi-2019}.
\end{itemize}

		 \begin{figure}
           \centering
           \includegraphics[width=0.48 \textwidth]{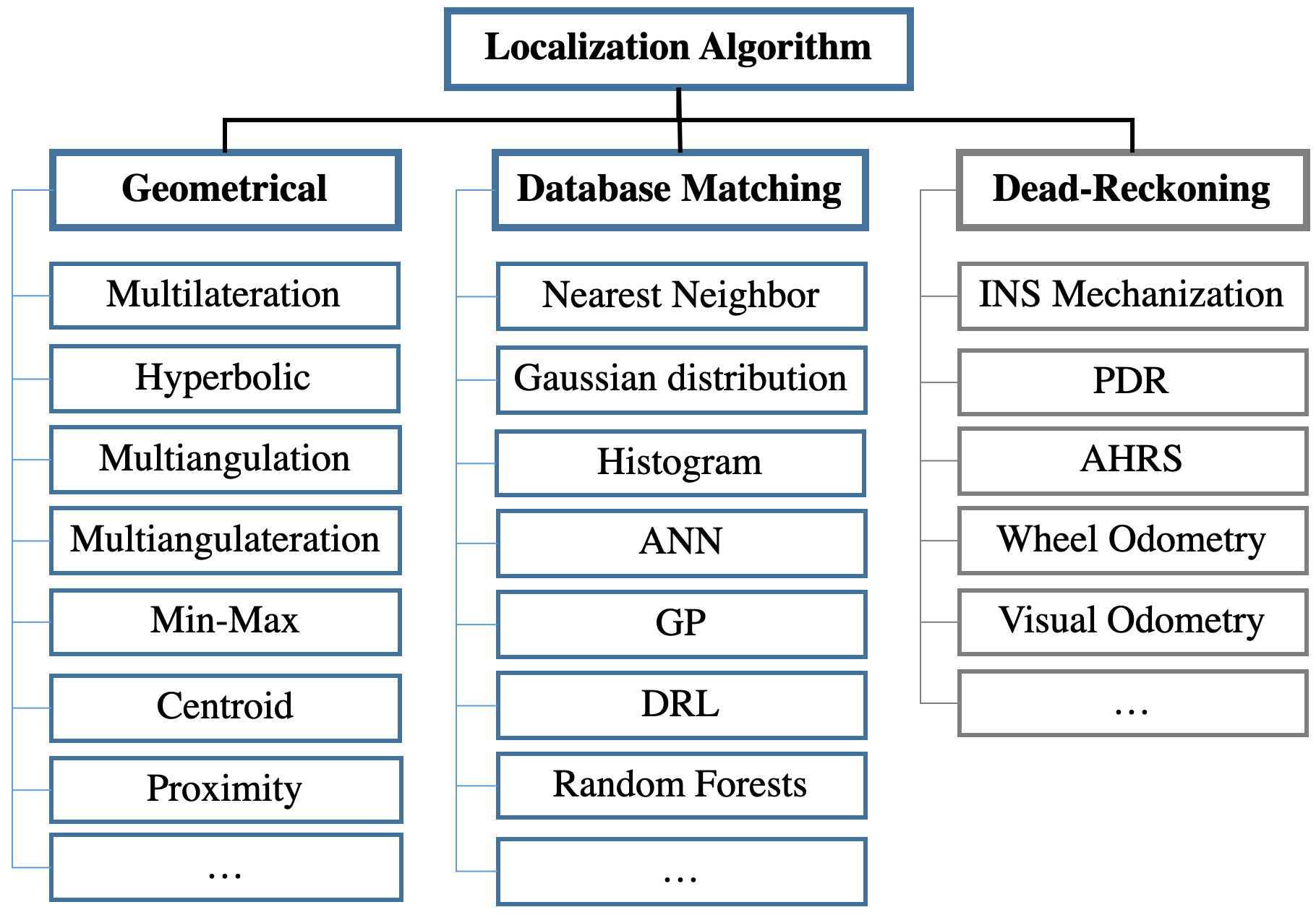}
           \caption{Localization algorithms}
           \label{fig:loc-algorithm}
         \end{figure} 

\section{IoT Localization Methods}
\label{sec-loc-method}
This section will answer the following questions: (1) what are the state-of-the-art localization approaches; and (2) what are the advantages and challenges for each type of localization method. The existing surveys (e.g., \cite{Zafari-2019}, \cite{Shit-2018}, and \cite{ZhuangY-VLP-2018}) already have detailed reviews on localization approaches. Most of these surveys classify the existing methods by sensor types or localization signal types. With the development of ML techniques and the diversification of modern localization scenarios, new localization methods have emerged. Specifically, over one decade ago, the majority of localization methods were geometrical ones, which are realized on geometric measurements such as distances and angles. By contrast, DB-M methods, which are data-driven, have been well developed during this decade. As a result, the IoT localization approaches can be divided into two categories: DB-M and geometrical localization. Meanwhile, there are DR methods which use sensors such as inertial, odometer, and vision ones. Figure \ref{fig:loc-algorithm} demonstrates part of the main localization algorithms.

\subsection{Database-Matching Localization Methods}
\label{sec-db-m}
\begin{figure}
           \centering
           \includegraphics[width=0.43 \textwidth]{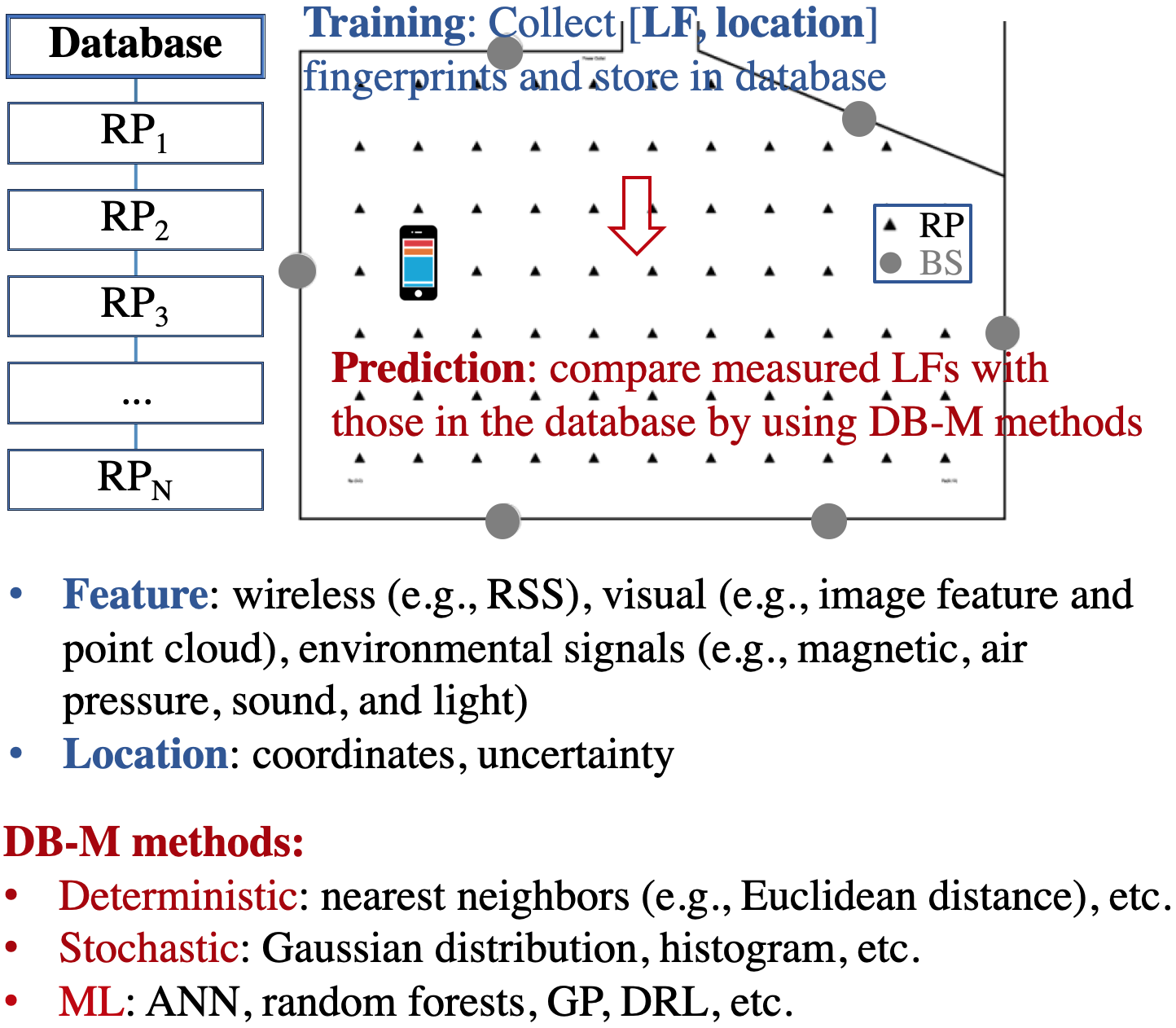}
           \caption{Principles of DB-M localization}
           \label{fig:db-m}
         \end{figure}

Although different localization technologies have various physical measurements and principles, they can generally be used for localization through DB-M. The basic principle for DB-M localization is to compute the difference between the measured fingerprints and the reference fingerprints in the database, and find the closest match. The DB-M process consists of three steps: (1) Localization feature (LF) extraction, (2) database training (or learning or mapping), and (3) prediction. Figure \ref{fig:db-m} demonstrates the principle of the training and prediction steps. The details of the steps are as follows.

\begin{itemize}
\item In LF extraction, valuable LFs are extracted from raw localization signals. A valuable LF should be stable over time and distinct over space. Examples of the LFs include RSS/CSI for wireless localization, magnetic intensity for magnetic matching, and visual features for vision localization. The extracted LFs are recorded and used for training and prediction. 
\item At the database-training step, [LF, location] fingerprints at multiple RPs are used to generate or update a database, which can also be regarded as a type of map. The database may be a data structure that stores the LFs at multiple RPs, or be the coefficients of parametric models. 
\item At the prediction step, the real-time measured LFs are compared with the database to locate the node. The likelihood (or weight) for each RP can be computed through (a) deterministic (e.g., nearest neighbors \cite{LimH-2006}), (b) stochastic (e.g., Gaussian distribution \cite{Haeberlen2004} and histogram \cite{Laoudias-C-2013}), and (c) ML (e.g., ANNs \cite{LiY-NN}, random forests \cite{GuoX-2018}, GP \cite{HeZ2018}, and DRL \cite{PengB-2019}) methods. These methods are described separately in the following subsections.
\end{itemize}

\subsubsection{Deterministic DB-M}
\label{sec-deterministic-db-m}
In these methods, only deterministic values (e.g., the mean value) of each LF at each RP are stored in the database. Thus, the LF values at each RP construct a vector, while the reference LF values at multiple RPs build a matrix. Each column vector in the matrix is the reference LF vector at one RP. At the prediction step, the similarity between the measured LF vector and each reference LF vector in the matrix is calculated by computing the vector distance. The RPs that have the highest similarity values are the nearest neighbors. To compute the similarity, the Euclidean distance \cite{LimH-2006} is widely used. Meanwhile, there are other types of vector distances, such as Manhattan distance \cite{NiuJ-2015}, Minkowski distance \cite{Torres-Sospedra-2015}, Spearman distance \cite{XieY-2015}, and information entropy \cite{Torres-Sospedra-2015}.

The deterministic DB-M methods have advantages of small database size and computational load. On the other hand, a main limitation is that the stochastic LFs, which may be caused by various factors in real-world practices, have not been involved.

\subsubsection{Stochastic DB-M}
\label{sec-stochastic-db-m}
Compared to the deterministic methods, stochastic DB-M approaches introduce stochastic LFs at each RP through various methods, such as the Gaussian-distribution \cite{Haeberlen2004} and histogram \cite{Laoudias-C-2013} methods. Meanwhile, likelihood \cite{LiY-JIOT-2019} between measured LFs and the reference ones in database is used to replace vector distance as the weight of similarity. 

It is notable that the majority of related works assume that various LFs (e.g., RSS values with different BSs) are independent with one another, so as to simplify the likelihood computation to the product (or summation) of the likelihood values for all LF components. Then, the likelihood-computation problem becomes how to estimate the likelihood value for each LF component. If the Gaussian-distribution method is used, the likelihood value for a LF can be estimated by applying Gaussian-distribution model \cite{Haeberlen2004}. 
 		
 	Although the Gaussian-distribution model is one of the most widely used models for stochastic errors in localization applications, there may be systematic LF measurement errors due to environmental factors. The existence of systematic errors theoretically breaks the Gaussian-distribution assumption. However, the Gaussian distribution is still widely used in engineering practices because real-world environmental factors is difficult to predict and model. One approach for reducing the degradation from systematic errors is to set relative larger variance values in localization filters to absorb the impact of such errors. 

	Moreover, WiFi \cite{Kaemarungsia-2012} and LoRa \cite{LiY-LoRa-2018} RSS values may have not only symmetric histograms but also asymmetric ones, such as left-skewed, bimodal, and other irregular histograms. Such asymmetric measurement distributions may be caused by environmental factors. An approach for mitigating the impact of asymmetric measurement distributions is to use histograms \cite{Laoudias-C-2013} or advanced stochastic models, such as those discussed in \cite{Radi-2019}.

The histogram method can obtain likelihood values without assumption on signal distributions. The histograms for all LF components at all RPs are calculated in the training step. In the prediction step, the measured fingerprint is compared with the corresponding histogram to find the likelihood. However, since it determines likelihood values through histogram matching, instead of using a parametric model, it may suffer from a large database size and overfitting.

\subsubsection{ML-based DB-M}
\label{sec-ml-db-m}
In recent years, ML has started to bring empowerment to numerous applications because of the increased data volume, the increased computing power, and the enhanced ML algorithms. This subsection reviews the typical ML algorithms that have been utilized for localization. 

\textit{- ANN}: ANN is a type of framework for using ML methods to process complex (e.g., nonlinear and non-Gaussian) data. ANN consists of one input layer, at least one hidden layer, and one output layer. Each layer contains at least one neuron. The neurons are connected via weights and biases, which are trained and stored. There are numerous publications (e.g., \cite{Gurney-1997}) on the principle of ANN. Also, there are various types of ANN, such as the recurrent neural network (RNN) \cite{ChoiJ2018}, convolutional neural network (CNN) \cite{Konings2018}, radial basis function neural network (RBF) \cite{LiZ2018}, and multi-layer perceptron (MLP) \cite{Abdallah2018}. 

As an example, MLP is used to illustrate the mechanism within ANN. MLP is a supervised-learning approach that is based on the error back-propagation algorithm, which optimizes the parameters (i.e., weights and biases) by minimizing the cost function (e.g., the sum of squared errors) of the neurons at the output layer. Specifically, the back-propagation algorithm can be divided into five steps \cite{Book-Goodfellow}: input, feed-forward, output error computation, error back-propagation, and output. Four standard back-propagation equations \cite{HaganM1994} can be used to calculate the errors at the output layer and then back-propagate these errors to update the weights and biases based on a learning rate. 

ANN techniques have been used for localization over a decade ago~\cite{Chiang-NN}; however, it has not been widely adopted until recent years. RSS (e.g., RSS from WiFi~\cite{Akram-2018}, BLE~\cite{LiJ2018}, ZigBee~\cite{GharghanS2018}, RFID~\cite{Berz2018}, cellular~\cite{Abdallah2018}, and photodiode~\cite{Konings2018}), RSS features (e.g., 2D RSS map~\cite{Jang2018}, differential RSS (DRSS)~\cite{LiZ2018}, and RSS statistics~\cite{Elbakly2018}), CSI~\cite{ChoiJ2018}, and AoA ~\cite{WangX2018} have been used. The majority of these works directly output node locations, while the others also generate the identification of floors~\cite{Elbakly2018}, rooms~\cite{Akram-2018}, and regions~\cite{ZhangX2018}, NLoS ~\cite{ChoiJ2018}, similarity of fingerprints~\cite{Dong2018}, localization success rate~\cite{Akail2018}, and localization accuracy prediction \cite{LiY-NN}. 

ANN has several advantages: (1) the algorithm has been well-developed and successfully in various fields (e.g., speech recognition \cite{Graves-2013} and image processing \cite{Gregor2015}). (2) The current ANN platforms and toolboxes are open and straightforward to use. On the other hand, the shortcomings of ANN include: (1) an ANN model is similar to a black box for most users. It is difficult to determine an explicit model representation of how the ANN works. (2) It is difficult to understand and adjust the internal algorithms. For example, although the majority of localization works above use one to three hidden layers, they set the numbers of hidden layers and neurons through brute-force data processing, instead of following a theoretical guide. Such specifically-tuned ANN parameters may be not suitable for varying localization environments.

\textit{- GP}: GP is a supervised ML method for regression and probabilistic classification \cite{Rasmussen-2006}. A GP is a set of random variables which have joint Gaussian distributions. Therefore, for localization applications, GP can involve the correlation among all RPs. This characteristic makes GP different from many other localization methods which treat each RP separately. Another characteristic for GP is that it can be uniquely determined by a mean function and a kernel function (i.e., covariance function). In the localization area, the mean function may be set by using geometrical LF models \cite{LiY-MMT-2019}; meanwhile, a zero mean function is used in some scenarios \cite{HeZ2018}. In contrast, there are various types of covariance functions, such as the constant, linear, squared exponential, Matérn, and periodic ones \cite{Rasmussen-2006}. The geometrical LF models have not been involved in covariance functions in the existing works. The research in \cite{Ferris-2006} has presented a hyperparameter estimation model for learning the GP model parameters.

The use of GP for RSS localization has been proposed in \cite{AntonT-2004} for cellular networks. The paper \cite{Ferris-2006} extends the work by introducing a Bayesian filter that builds on a graphed space representation. Afterwards, GP has been used in processing data from various localization sensors, such as magnetometers \cite{Solin-2018} and RFID \cite{SecoF-2010}. Furthermore, GP has become one of the main techniques for localization-database prediction (or interpolation), that is, to predict LFs at unvisited or out-of-date RPs based on training data at other RPs. Reference \cite{ChoiW-2018} compares the performance of DB prediction by using GP and geometrical (e.g., linear, cubic, thin plate, and quintic polynomial) interpolation methods. 

GP has advantages such as: (1) it has a physical meaning and an explicit model representation, compared to many other ML methods. (2) GP captures both the predicted solution and its uncertainty. The latter is not provided in ANN. (3) GP has a small number of parameters; thus, its engineering implementation is straightforward. The challenges for GP include: (1) it is based on a Gaussian-process assumption, which may be degraded in challenging localization scenarios. Integrating GP with geometrical localization models may be a possible method to mitigate this degradation \cite{LiY-MMT-2019}. (2) GP has a small number of parameters. Thus, in localization scenarios that have complex environments and massive data, GP may not be able to exploit the potential of complex databases as well as other ML methods (e.g., ANN).

\textit{- Random forests}: The random forests algorithm is an ensemble classifier that uses a set of decision trees (i.e., classification and regression trees) for supervised classification \cite{Scikit-2011}. The paper \cite{Rodriguez-2006} has a detailed description on its principle and theoretical formulae. Random forests can be implemented through three steps: subsampling, decision-tree training, and prediction. In the subsampling step, the algorithm randomly selects a subsample that contains a fixed number of randomly-selected features from the original dataset. The subsample is trained with a decision in the decision-tree training step. The training process creates the if-then rules of the tree. One typical method for this process is Gini impurity \cite{Menze-2009}, which is a measure of how often a randomly selected element will be incorrectly labeled if the element is randomly labeled according to the label distribution in the subset. 

When splitting a branch in the tree, all possible conditions are considered and the condition with the lowest Gini impurity is chosen as the new node of the decision tree. If the split is perfect, the Gini impurity of that branch would be zero. There are other criteria (e.g., information gain \cite{Quinlan-1986}). For prediction, each tree in the ensemble gives a prediction result. Based on the votes from all trees, a probabilistic result can be generated. 

In the localization area, the random-forest approach has been applied for RSS fingerprinting \cite{GuoX-2018}, CSI fingerprinting \cite{WangY-2018}, vision localization \cite{Massiceti-2017}, NLoS condition identification \cite{Ramadan-2018}, loop-closure detection \cite{KuaJ-2012}, and VLP \cite{GuoXS-2017}. The advantages of random forests include \cite{Horning-2019}: (1) compared to other decision-tree based methods, it is less sensitive to outliers in training data. (2) The random-forest parameters can be set easily. (3) Random forests can generate variable importance and accuracy together with prediction solutions. The shortcomings of random forests include: (1) it is not efficient in computational load. A large number of trees are needed for an accurate vote. This phenomenon leads to large databases and computational loads. (2) It may be over-fitted. Also, it is sensitive to noise \cite{Briem-2002}.

\textit{- DRL}: DRL, which is the core algorithm for AlphaGo, has attracted intensive attention. It combines deep learning and reinforcement learning. The former provides a learning mechanism, while the later provides goals for learning \cite{Hessel-2018}. In general, DRL allows the agent to observe states and act to collect long-term rewards. The states are mapped to an action through a policy \cite{PengB-2019}. 

The DRL algorithm has experienced stages such as Deep Q-Networks (DQN), Asynchronous Advantage Actor-Critic (A3C), and UNsupervised REinforcement and Auxiliary Learning (UNREAL). Specifically, DQN \cite{Mnih-2015} introduces value networks to represent the critic module, and constructs value networks according to specific applications by using ANNs (e.g., Long Short-Term Memory (LSTM) networks and CNN). Then, A3C \cite{Mnih-2016} applies the actor-critic framework and asynchronous learning. The basic idea of actor-critic is to evaluate the output action and tune the possibility of actions based on evaluation results. Compared to A3C, the UNREAL algorithm is closer to the human-learning mode. Specifically, UNREAL enhances the actor-critic mechanism through multiple auxiliary tasks. The research in \cite{HuX-COMM-2018} has pointed out three components for a DRL solution: basis/core (e.g., state definition, action definition, and reward definition), basic units (e.g., Q-network, action selection, replay memory, and target network), and state reformulation (i.e., the method for state-awareness data processing). 

DRL has been used for navigation in Atari games \cite{Mnih-2015}, mazes \cite{Dhiman-Banerjee-2019}, and the real world  \cite{Zhang-Springenberg-2019}. Meanwhile, DRL has been applied for navigation using data from monocular camera \cite{Zhang-Springenberg-2019}, 360-degree camera \cite{Bruce-Sunderhauf-2017}, LiDAR \cite{Tai-Paolo-2017}, magnetic sensor \cite{BejarE-2018}, wireless sensor \cite{Mohammadi-Fuqaha-2018}, and Google street view \cite{Mirowski-Grimes-2018}. Many of the recent DRL research works focus on navigation without a map \cite{Tai-Paolo-2017}, in new environments \cite{Zhelo-Zhang-2018}, and with varying targets \cite{Zhu-Mottaghi-2018}. However, most of these methods are designed for navigation, instead of localization. Navigation and localization both use data from wireless, environmental, and vision sensors as inputs, 
they have different principles. Navigation is the issue of finding the optimal motion path between the node and a target location; in contrast, a localization module outputs the node location. It is straightforward to model a navigation process as a Markov decision process; thus, it can be processed by DRL. By contrast, localization is closer to a deep learning problem. 

	The advantages of DRL include: (1) it can obtain not only the optimal solution at the current moment, but also the long-term reward. (2) DRL can reduce the computational complexity caused by re-optimization due to factors such as environment changes \cite{HuX-COMM-2018}. The challenges of DRL include: (1) the set of reward definition is key to the DRL performance. However, it is challenging to determine a theoretical model for reward definition. (2) The DRL algorithm itself has met several challenges \cite{Scholz-2019}, such as hyperparameter sensitivity, sample efficiency, off-policy learning, and imitation learning. (3) The future DRL algorithms may need supports from ML chips due to their large computational loads.

There are also other ML methods for enhancing localization. For example, the Hidden Markov Model (HMM) approach has been used for RSS fingerprinting \cite{SunS-2019}, trajectory modeling \cite{Jitta-2017}, and room recognition \cite{Carrera-2018}. Also, the Support Vector Machine (SVM) method has been applied for CSI localization \cite{Sanam-2018}, RSS localization \cite{AhmadiH-2017}, and wide-area localization \cite{Timoteo-2016}. Meanwhile, the fuzzy-logic method has been used for smartphone localization \cite{Orujov-2018} and VLP \cite{PauG-2019}. Various ML methods have different advantages and thus are suitable for different localization use cases.

\subsection{Geometrical Localization Methods}
\label{sec-geometric-loc}
\begin{figure}
           \centering
           \includegraphics[width=0.48 \textwidth]{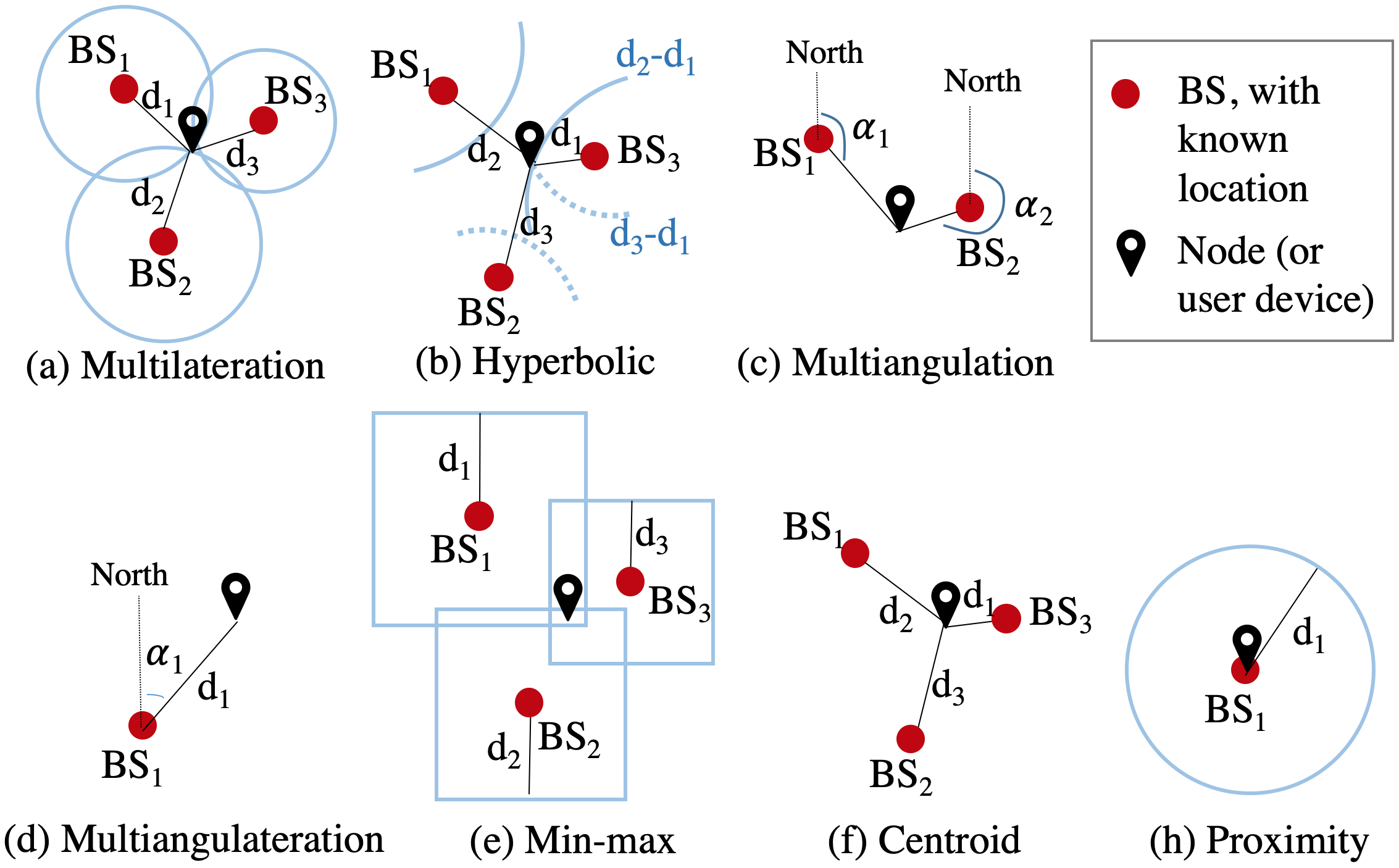}
           \caption{Principle of geometrical localization methods}
           \label{fig:geometrical-loc-method}
         \end{figure}

Geometrical localization methods have been researched for decades. To use such methods, BS locations are commonly known or can be estimated. The main measurements are BS-node distances and angles. The main localization methods include multilateration, hyperbolic localization, multiangulation, multiangulateration, and other simplified methods such as min-max, centroid, and proximity. Previous survey papers \cite{Zafari-2019} and \cite{ZhuangY-VLP-2018} already have detailed descriptions on these methods. Thus, this survey only summarizes their main characteristics. Figure \ref{fig:geometrical-loc-method} illustrates the principle of several geometrical localization methods.

\subsubsection{Multilateration}
\label{sec-multilateration}
Multilateration can be used to estimate node location by using locations of at least three BSs and their distances to the node. Its basic principle is to estimate the intersection between spheres (for 3D localization) and circles (for 2D localization). This method has been widely used in GNSS and wireless Ad Hoc network node \cite{Guvenc-2009} or BS \cite{ChengY-2005} location estimation. The common estimation techniques are least squares and Kalman filter (KF). The multilateration performance may be enhanced through improving ranging accuracy by mitigating the impact of environment-related and receiver errors \cite{Plets-2012}. There are also well-developed blunder-detection and accuracy-evaluation mechanisms \cite{Petovello-2003} as well as geometry indicators such as the Dilution of Precision (DOP) \cite{Langley1999}. 

\subsubsection{Hyperbolic Localization}
\label{sec-hyperbolic}
Hyperbolic localization, which was developed for Loran navigation \cite{Hefley-1972}, is the main method for TDoA localization. It is based on the distance differences between the node and various BSs. Because hyperbolic localization has eliminated the requirement for precise timing on nodes, it has strong potential for LPWAN localization. There is also accuracy analysis \cite{Kaune-2011} for this method.

\subsubsection{Multiangulation}
\label{sec-multiangulation}
Multiangularation can be implemented by measuring the angles between the node and at least two BSs \cite{BaiL2008}. Theoretically, two BS-node angles can determine a 2D point. When considering angle-measurement errors, a quadrangle can be determined. The multiangulation method has been used in wireless Ad Hoc networks to reduce the requirement on BS (or anchor) locations \cite{Niculescu-2003}. The performance analysis of multiangulation has also been provided \cite{WangY-2015}. 

\subsubsection{Multiangulateration}
\label{sec-multiangulateration}
Multiangulateration localizes a node by at least one BS-node angle and one BS-node distance. This method has been widely used in traverse networks in engineering surveying \cite{Schofield-2001}. For indoor localization, it is feasible to use one AoA (e.g., VLP \cite{ZhuangY-VLP-2018} and BLE \cite{Quuppa2018}) BS on the ceiling with known height for localization. 

\subsubsection{Min-Max}
\label{sec-minmax}
Min-max is a variant of multilateration. Its geometrical principle is to calculate the intersection between cubes (for 3D localization) and squares (2D localization), instead of spheres and circles. The benefit for using cubes and squares is that their intersections can be directly computed through deterministic equations \cite{Will-2012}, which have significantly lighter computational loads than least squares. The limitation of min-max is that it has not considered the stochastic measurements. Meanwhile, compared to spheres, cubes deviate from spatial distribution with a certain BS-node distance. The min-max method can be used to generate coarse localization solutions \cite{Savvidesm-2012}, which provide initial positions for fine-localization approaches such as multilateration. 

\subsubsection{Centroid}
\label{sec-centroid}
Centroid is a simplification of multilateration. It estimates the node location by weighted average of the locations of various BSs. The weights for BS locations are commonly determined by BS-node distances \cite{Pivato-2011}. Similar to min-max, the centroid has a low computational load but has not considered stochastic measurements. Furthermore, the centroid result will be limited within the region that has a boundary formed by BS locations. Similar to min-max, the centroid is commonly used for coarse localization.

\subsubsection{Proximity}
\label{sec-proximity}
Proximity can be regarded as a further simplification of the centroid approach. The possible location region is determined by using the location of one BS as the circle center and using the BS-node distance as the radius. Proximity is commonly used in RFID \cite{Bouet-2008} and cell-identification \cite{BsharaM-2011} localization. The method has the lowest computational load, the lowest number of BS, but the largest location uncertainty. Thus, it is commonly used for near-field localization, coarse localization, or for bridging the outages when the other localization methods do not have sufficient BSs.

Meanwhile, there are other geometrical localization methods, such as the zone-based \cite{WuY-2016}, compressed-sensing \cite{KirschF-2018}, law-of-cosines \cite{Lagias-2018} methods. 

\subsection{Summary and Insight on Database-Matching and Geometrical Localization Methods}
\label{sec-comm-loc-methods}
DB-M and geometrical localization methods have several similarities, such as

\begin{itemize}
\item Both methods have the training and prediction steps. In DB-M methods, a [LF, location] database is generated through training; in contrast, in geometrical methods, the coefficients for parametric models are estimated and stored at the training step. 
\item They have some common error sources, such as device diversity, orientation diversity, and human-body effects. 
\end{itemize}

On the other hand, these methods have differences (or complementary characteristics), such as
\begin{itemize}
\item Geometrical methods are more suitable for scenarios (e.g., outdoor and indoor open environments) that can be explicitly modeled and parameterized. In contrast, DB-M approaches are more suitable for complex scenarios (e.g., wide-area urban and indoor areas) that are difficult to be parameterized. 
\item For geometrical methods, environmental factors such as NLoS and multipath conditions are error sources that need to be modeled and mitigated. By contrast, DB-M methods may use the measurements of these factors as fingerprints to enhance localization. 
\item Geometrical methods are based on parametric models; thus, the databases contain only model parameter values and thus are relatively small. On the other hand, it is difficult to describe complex scenarios by using parametric models with limited numbers of parameters. In contrast, DB-M methods directly describe localization scenarios by using data at all points in the space; thus, both database and computational loads are large, especially for wide-area applications. However, DB-M approaches have more potential to provide higher resolution and more details on complex localization scenarios.
\item Progresses in ML algorithms have brought great potential to localization methods, especially for the applications that have complex scenarios that are difficult to model, parameters that are difficult to determine and tune, and have nonlinear, non-parametrical, correlated measurements that are caused by environmental and motion factors. Although ML methods are classified into the DB-M group in this survey according to the existing works, ML can also be used to enhance geometrical localization, such as training of parametric models and their parameters. 
\end{itemize}

		 \begin{figure*}
           \centering
           \includegraphics[width=0.80 \textwidth]{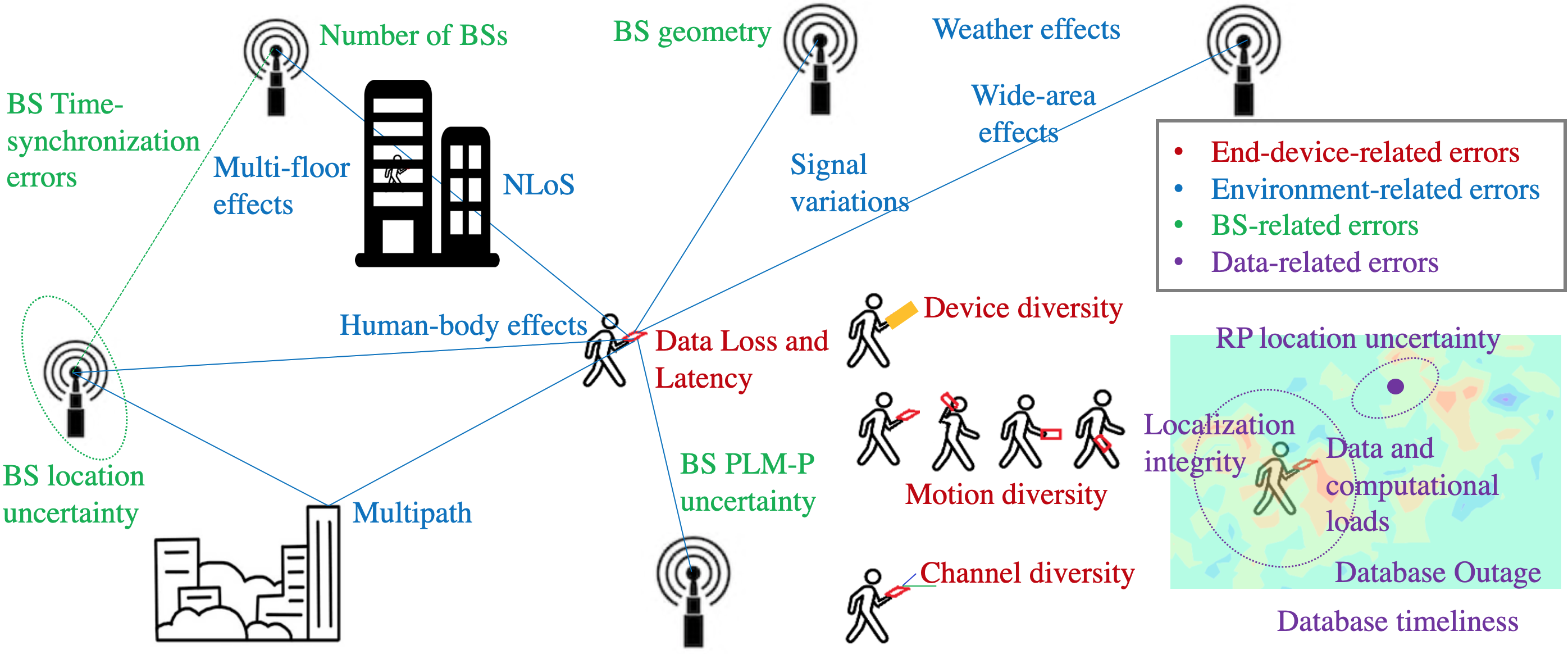}
           \caption{IoT localization error sources}
           \label{fig:localization-error-source}
         \end{figure*} 

Due to their complementary characteristics, geometrical and DB-M methods can be integrated. For example, geometrical methods can be used to reduce the computational load of DB-M \cite{ZhangJ-2017}, to aid database prediction \cite{LiY-MMT-2019}, and to provide localization uncertainty prediction \cite{LiY-JIOT-2019}. It is also feasible to directly integrate geometrical and DB-M methods for more robust localization solutions \cite{Kodippili-2010}. 

\section{IoT Localization Error Sources and Mitigation}
\label{sec-err-mitigation}
This section describes the main IoT-localization error sources and their mitigation. These two topics are key components in a LE-IoT system. This section answers two questions: (1) what are the error sources for IoT localization; and (2) how to mitigate or eliminate the impact of these error sources. The localization error sources are classified into four groups as

\begin{itemize}
\item \textit{End-device-related errors}: device diversity, motion/attitude diversity, data loss and latency, and channel diversity.
\item \textit{Environment-related errors}: multipath, NLoS, wide-area effects, multi-floor effects, human-body effects, weather effects, and signal variations. 
\item \textit{Base-station-related errors}: the number of BSs, BS geometry, BS location uncertainty, BS PLM-P uncertainty, and BS time synchronization errors.
\item \textit{Data-related errors}: database timeliness/training cost, RP location uncertainty, database outage, intensive data, data and computational loads, and localization integrity.
\end{itemize}

Figure \ref{fig:localization-error-source} shows the main IoT-localization error sources. The details for the listed error sources are provided in the following subsections.

\subsection{End-Device-Related Errors}
\label{sec-err-end-device}
This subsection introduces the error sources on the node side. The error sources include device diversity, motion/attitude diversity, data loss and latency, and channel diversity.

\subsubsection{Node (Device) Diversity}
\label{sec-node-diversity}
A large numbers of low-cost MEMS sensors are used in IoT nodes. Such low-cost sensors may have diversity when measuring the same physical variable. For example, the RSS diversity for the same-brand BLE nodes may reach 20 dBm \cite{LiY-SensJ-2019}. Such node diversity directly leads to localization errors \cite{LiY-SensJ-2019}. 

To alleviate the impact of node diversity, there are calibration-based methods that fit data from multiple nodes through the use of histograms \cite{Laoudias2012}, least squares \cite{Lee2015}, ANNs \cite{Tsui2009}, multi-dimensional scaling \cite{MaL-2019}, and motion states from DR \cite{HeS2018}. Reference \cite{Figuera-2011} provides details about several node-diversity calibration methods, such as nonlinear adjustment, sensitivity threshold correction, and interdependence across BSs. 

Meanwhile, there are calibration-free differential-measurement-based approaches. Examples of such methods include the differential approach that subtracts the data from a selected datum BS from those from the other BSs \cite{Hossain2013}, takes the differences between data from all BS couples \cite{Dong2009}, uses the average data from all BSs \cite{Laoudias2014} or selected BSs \cite{Yang2013} as the datum, and adopts the advanced datum-selection method \cite{Ye2016}. Additionally, differential measurements are used in multilateration \cite{Lin2013} and other geometrical \cite{LeeJ2009} methods. There are two challenges for using differential signals: (1) it is challenging to switch the datum for differential computation in a wide-area IoT application. (2) The differential computation increases the noise levels in measurements. 

\subsubsection{Motion/Attitude Diversity}
\label{sec-motion-diversity}
IoT nodes may experience various motion modes, such as being held horizontally, dangling in the hand, mounted on chest, and stored in pockets or bags. This phenomenon leads to changes in node attitude (or orientation). Meanwhile, movements of nodes also lead to changes in the relative attitude angles between node- and BS-antenna directions. Such attitude changes lead to localization errors \cite{LiY-SensJ-2019}.

To reduce the impact of motion/attitude diversity, the research in \cite{King2006} proposes a compass-aided method by using attitude-matched databases; furthermore, the paper \cite{Deng2018} extends the method to dynamic localization scenarios. Meanwhile, the papers \cite{Papapostolou2009} and \cite{Sanchez2012} present a heading-aided methods that add the node attitude into fingerprints and decision trees, respectively. Moreover, reference \cite{Fang2015} uses histogram equalization for orientation-effect compensation, while the research in \cite{LiY-SensJ-2019} presents an orientation-compensation model to compensate for orientation diversity. 

The orientation angles of BS antennae are commonly constant, while the node attitude angles are dynamic and can be estimated in real time through the Attitude and Heading Reference System (AHRS) algorithm \cite{DaiH-2018}. In such an algorithm, gyro measurements are used to construct the system model, while accelerometer and magnetometer data provide measurement updates. Meanwhile, autonomous gyro calibration is important to enhance attitude estimation \cite{LiY-sensj-calibration}.

\subsubsection{Data Loss and Latency}
\label{sec-data-loss}
Response rate is a practical factor in wireless localization. Data from some BSs may be wrongly missed in the scanning duration \cite{ZhuangY-2016}. Also, the response rate may vary across BSs and are related with the level of signals \cite{ZhuangY-2016}. BSs with lower RSS values tend to have lower response rates. The response rate can also be used as fingerprint information \cite{LedlieJ-2011}.

Meanwhile, to meet the requirement of low power consumption, IoT nodes commonly have low sampling rates. Furthermore, the existence of massive numbers of nodes may lead to data collision, loss \cite{Bor-2016}, and latency \cite{Oliveira-2018}. Both data loss and latency may directly lead to localization errors. In particular, the loss or delay of data from an important BS (e.g., a BS that is geographically close to the node) \cite{FangS-2010} may lead to significant degradation on localization performance.

A possible method to mitigate this issue is to predict localization signals by using approaches such as time-series analysis \cite{Hamilton-1994} and ML \cite{HeZ-2019}. Furthermore, data from other localization sensors or vehicle-motion constraints may be used to construct node-motion models, which can be used for localization-signal prediction.

\subsubsection{Channel Diversity}
\label{sec-channel-diverity}
In IoT applications, especially those with a high node density (e.g., in animal-husbandry applications), multi-channel mechanisms may be utilized to reduce data collision \cite{HeZ-2019}. The difference between training and prediction data channels may lead to localization errors. To alleviate such errors, one method is to calibrate channel diversity through parametric models \cite{ChoiT-2016} or ML \cite{HeZ-2019}. Meanwhile, data from multiple channels may be combined \cite{Zanella-2014} or treated as data from various BSs \cite{ZhuangY-sens-2016} to enhance localization.

\subsection{Environment-Related Errors}
\label{sec-environm-error}
This subsection describes the environment-related localization error sources, including multipath, NLoS, wide-area effects, multi-floor effects, human-body effects, weather effects, and signal variations.

\subsubsection{Multipath}
\label{sec-multipath}
Multipath is a common issue when using wireless signals, especially in indoor and urban areas. It is difficult to discriminate between multipath propagations and the direct path. There are papers that evaluate the impact of multipath in localization. For example, the paper \cite{De-La-Llana-Calvo-2016} compares the impact of multipath on AoA and PoA methods with a photodiode and found that AoA localization was affected by an order of magnitude lower than PoA. Meanwhile, the research \cite{YangZ-survey-2013} has investigated the impact of multipath on RSS and CSI localization and found that CSI localization suffered from less degradation. In the localization area, the research on multipath are mainly on two aspects: mitigating its degradation on localization solutions and using it to enhance localization. 

There are various methods for mitigating the degradation from multipath. Some research reduce multipath errors by using specific node design, such as adopting antenna arrays \cite{SamanK-2019}, beamforming \cite{TaoC-2017}, frequency-hopping \cite{TaoC-2017}, a modified delay locked loop \cite{Dierendonck-1992}, and multi-channel signals \cite{Zanella-2014}. Meanwhile, there are approaches for distinguishing the direct path from multipath reflections \cite{SenS-2013} and extracting individual multipath propagation delays \cite{Zayets2017}. Using 3D city models to assist multipath detection, which is a combination of localization and mobile mapping, has also been researched \cite{KumarRakesh-2017} in recent years. Furthermore, some papers focus on modeling the multipath components by using different Path-Loss Models (PLMs) for direct-path and multipath signals \cite{DeyleT-2008}, involving multipath components in PLMs \cite{ShenJ-2019}, and estimating multipath parameters in real time by using SLAM \cite{Vanderveen-1998} and Extended KF (EKF) \cite{ZhuM-2015} algorithms. 

Moreover, due to the new features such as MIMO, dense minimized BS, and mmWave systems in 5G, using multipath signals to enhance localization is attracting research interest. The research in \cite{Witrisal-2015} describes the principle and methodology for multipath-assisted localization. Meanwhile, reference \cite{Meissner-2014} has derived the statistical performance bounds and evaluation models, while the paper \cite{Leitinger-2015} derives the Cramér-Rao lower bound (CRLB) for multipath-assisted localization. Multipath-assisted localization has been utilized in outdoor \cite{Ljungzell-2018} and indoor \cite{GroB-2018} single-BS systems. Moreover, the research in \cite{Gentner-2016} treats multipath components as signals emitted from virtual transmitters and uses them to assist localization within a SLAM algorithm, while the paper \cite{PopleteevA-2019} improves the accuracy of RSS fingerprinting by introducing multipath-sensitive signal features. Furthermore, there are approaches that directly use multipath signals for DB-M \cite{KupershteinE-2013} and multilateration \cite{Soltanaghaei-2013} computation.

\subsubsection{NLoS}
\label{sec-nlos}
Similar to the multipath effect, the NLoS effect is an important error source for wireless localization. Actually, NLoS is one factor that will cause multipath propagation. This subsection focuses on other effects of NLoS. It has been revealed that NLoS environments may significantly reduce the coverage range of IoT signals \cite{Bor-M-2016} and change their PLM-P values \cite{Bor-2016}. The paper \cite{Shi-G-2019} investigates the effect of NLoS signals on RSS localization and attempts to quantify the relation between localization errors and NLoS, while the research in \cite{ZhouBP-TC-2019} derives the NLoS-based CRLB. In general, the research on NLoS can be classified into three groups: identification, modeling, and mitigation.

To identify NLoS signals, algorithms such as ANN \cite{Heidari-M-2019}, random forests \cite{Ramadan-2018}, SVM \cite{Marano-S-2019}, the Dempster-Shafer evidence theory \cite{Li-Sh-2016}, and the Neyman-Pearson test \cite{XiaoZ2015} have been applied. There is also a method that detects NLoS by comparing the difference between signals in multiple frequency bands \cite{Jo-H-2018}. 

Examples of NLoS-modeling approaches include that use advanced PLMs involving walls \cite{Lemic-2016} and building floors \cite{Seidel-1992}. Furthermore, there are research that have considered the thickness of obstacles and intersection angles between obstacles and the direct path \cite{Ngo-Q-2015} as well as the wall and interaction loss factors \cite{Plets-2012}.

To mitigate the NLoS effect, there are methods that use various estimation techniques, including Monte-Carlo Gaussian smoothing \cite{Vila-Valls-2017}, residual analysis \cite{Wang-Y-2018}, least trimmed squares \cite{Khodjaev-J-2016}, and improved filtering techniques, such as the cubature KF \cite{Li-W-2013}, skew-t variational Bayes filter \cite{Nurminen-H-2015}, Particle Filter (PF) \cite{Boccadoro-M-2012}, Unscented KF (UKF) \cite{Huerta-J-2009}, finite impulse response filter \cite{Pak-J-2017}, and biased KF \cite{Rohrig-C-2009}. Advanced models (e.g., the radial extreme value distribution model \cite{Okusa-K-2017}) have also been used. Furthermore, it is feasible to realize NLoS mitigation by introducing external localization sensors, such as vision \cite{Su-Z-2017} and inertial \cite{Yavari-M-2017} sensors.

\subsubsection{Wide-Area Effects}
\label{sec-wide-area}
Wide-area localization is more challenging than local localization due to factors such as lower RSS, SNR, and response rate values \cite{LiY-LoRa-2018}, and stronger multipath and fading effects \cite{Bshara-M-2010}. Moreover, it is challenging to obtain wide-area PLM-P values because they change significantly with factors such as environment type (e.g., highway, rural, and urban) \cite{Karedal-M-2011}, terrain category (e.g., hilly, flat, with light/moderate/heavy tree densities, and on water/ground) \cite{Petajajarvi-J-2015}, and even BS-node distances and BS antenna heights \cite{Erceg-MV-1999}.	

Advanced PLMs have been used to enhance wide-area localization. Example of these PLMs include the multi-slope PLM for BS-node distances from meters to hundreds of meters \cite{Hrovat-A-2010}, the higher-order PLM \cite{Sallouha-2017}, and the height-dependent PLM \cite{Sallouha-H-2010}. ML methods (e.g., ANN) have been used to determine the PLM-P values \cite{Hosseinzadeh-S-2017}. Meanwhile, there are other improved wide-area localization algorithms, such as the BS-identity \cite{Bshara-M-2010} and minimum-mean-square-error \cite{Xiao-L-2017} ones. Considering the popularization of small BSs, it may become a trend to use wide-area IoT signals for coarse positioning and use local IoT signals for fine localization.

\subsubsection{Multi-Floor Effects}
\label{sec-multi-floor}
Most of the existing works are focusing on 2D localization. However, in the IoT era, nodes may be used in 3D scenarios. A practical 3D-localization approach is to find the floor at which the node is located and then localize the node in the floor. Thus, robust floor detection is required. For this purpose, researchers have used data from various sensors, such as wireless sensors \cite{Elbakly2018}, a barometer \cite{Haque2018}, inertial sensors \cite{Ramana2016}, a floor plan \cite{LoC2018}, and a user-activity probability map \cite{Herrera2014}. There are also various estimation techniques, such as KF \cite{Haque2018}, PF \cite{LoC2018}, and ANNs \cite{Elbakly2018}. The impact of multi-floor effects on PLM-P values has also been researched \cite{Seidel-1992}.

\subsubsection{Human-Body Effects}
\label{sec-human-body}
It is necessary to consider the human-body effect in some IoT-localization applications, especially when the nodes are placed on the human body. There are four types of research on the human-body effect: evaluation, modeling, mitigation, and utilization.

There are papers that evaluate the impact of the human body on multilateration \cite{Kilic-Y-2012}, centroid \cite{Pathanawongthum-2009}, and several other localization algorithms \cite{Schmitt-S-2014}. It has been found that the human body may cause degradation on localization; also, the degradation is more significant when the node is carried on human body, compared to the use case when a human blocks the BS-node LoS \cite{Pathanawongthum-2009}. Meanwhile, the research \cite{Trogh-2016} has revealed that influence of the human-body effect varies with the node position and orientation. Also, the tests in \cite{PopleteevA-2019} show that the human body leads to short-term RSS fluctuations, instead of long-term signal shifts. 

There are approaches for modeling the human-body effect. For example, the literature \cite{Geng-Y-2013} models the impacts on both ToA ranging and localization. Also, the paper \cite{Trogh-2016} presents a compensation model for the human-body effect by introducing the user orientation toward fixed infrastructure. For modern localization applications, it may be feasible to use other sensors, such as vision, to detect the human body and compensate for its effect.

Moreover, to mitigate the human-body effect, the research in \cite{Cully-E-2012} treats it as a NLoS signal and mitigates its effect by using NLoS-mitigation methods. Also, the paper \cite{Trogh-2016} combines the data from multiple nodes at different places on a human body to control the human-body effect. 

Furthermore, human-body effects have been utilized for device-free localization \cite{Adib-F-2014}. Signals such as range \cite{Kilic-Y-2014}, RSS \cite{Saeed-A-2014}, CSI \cite{Gao-Q-2017}, and AoA \cite{Zhang-L-2018} are used. Moreover, there are other works that use the human-body effect for activity recognition \cite{Wang-J-2018}, fall detection \cite{Wang-Y-2017}, and people counting \cite{Xu--C-2013}. 

\subsubsection{Weather Effects}
\label{sec-weather}
Researchers have studied the weather effect on wireless signal propagation. Theoretically, large-wavelength wireless signals are not susceptible to external factors, such as precipitation and vegetation \cite{P-620}. Meanwhile, the research in \cite{PopleteevA-2019} indicates that FM RSS has a weak correlation with temperature, relative humidity, air pressure, wind speed, and aviation-specific runway visibility. The research has not considered weather-dependent environment changes such as movements of trees and power-line wires as well as changes of ground conductivity and the multipath effects. 

However, on-site experimental results in \cite{Fang-SH-2016} show that variations in rain conditions and wind speeds significantly degraded accuracy in distance estimation using Global System for Mobile communication (GSM) RSS. Meanwhile, reference \cite{Petajajarvi-J-2015} shows that LoRa RSS measurements suffer from significant noise and drifts when operating near lakes. Thus, whether the weather change has a significant impact on a certain type of IoT signal need specific evaluation. Meanwhile, integrating IoT signals with self-contained localization technologies, such as INS, may reduce the weather impact on positioning solutions.

\subsubsection{Signal Variations}
\label{sec-sig-var}
Fluctuations is an issue inherent to wireless signals, especially for those in indoor and urban areas. The LoRa RSS variation can reach several dBm and over 10 dBm under static and dynamic motions, respectively \cite{LiY-LoRa-2018}. Although signal variations cannot be eliminated, their effects can be mitigated through denoising methods such as averaging \cite{Small-J-2000}, autoregressive \cite{Nassar-S-2003}, wavelet \cite{Chang-S-2000}, KF \cite{Peesapati-S-2013}, and ANN \cite{Xie-J-2012}. 

Many denoising methods assume the noise follows a Gaussian distribution. However, both LoRa \cite{LiY-LoRa-2018} and WiFi \cite{Kaemarungsia-2012} RSS variations can follow either symmetric or asymmetric distributions. To model an irregular distribution, stochastic-signal analysis methods such as Allan variance \cite{LuoC-2018} and multisignal wavelet variance \cite{Radi-2019} can be used.

\subsection{Base-Station-Related Errors}
\label{sec-bs-error}
Examples of localization error sources on the BS side include insufficient number of BSs, poor geometry, BS location uncertainty, BS PLM-P uncertainty, and BS time synchronization errors.

\subsubsection{Number of BSs}
\label{sec-num-bs}
Public telecommunication and IoT BSs are mainly deployed for communication, instead of localization. Communication uses require signals from at least one BS, while localization needs signals from multiple BSs. Thus, the dependence on the number of BSs is a challenge for telecommunication- and IoT-signal based localization. To alleviate this issue, there are several approaches, such as using single-BS localization techniques, adding motion constraints, choosing localization algorithms that need fewer BSs, and integration with other sensors. 

Single-BS localization techniques can be conducted by using various measurements, such as ToA, AoA, and RSS. Similar to multi-BS localization, ToA \cite{Xie2018} and AoA \cite{Guerra2018} based localization may provide high-accuracy locations but require professional nodes for distance and angle measurements, respectively. RSS-based single-BS localization can also be implemented through either DB-M \cite{Rzymowski2016} or parametric-model-based methods \cite{LiY-TIM-2019}. Compared to multi-BS localization methods, single-BS localization has great potential to be used in existing telecommunication systems but face new challenges, such as the difficulty to detect outliers in its results \cite{LiY-TIM-2019}. 

To mitigate this issue, other approaches are needed. First, adding motion constraints can reduce the requirement for localization-signal measurements. For example, adding a constant-height constraint can reduce one state to be estimated \cite{Petovello-2003}. Meanwhile, coarse-localization algorithms (e.g., min-max, centroid, and proximity) require localization signals from less BSs. Furthermore, it is feasible to fuse wireless signal measurements with data from other sensors (e.g., inertial and vision sensors) for tightly-coupled localization \cite{ZhuangY2018-IoT}. 

\subsubsection{BS Geometry}
\label{sec-bs-geo}
The wireless-localization performance is directly correlated with BS geometry. The research in \cite{AbdulwahidM} and \cite{Plets-2012} investigate BS location optimization from the indoor communication and multi-floor signal coverage perspective, respectively. In the localization field, poor BS geometry may lead to the problem of location ambiguity \cite{ZhuangY-2016}. The relation between BS geometry and localization performance has been investigated through simulation \cite{Kihara-M-1994} and field testing \cite{Meng-X-2004}. Furthermore, there are indicators, such as DOP \cite{Langley1999}, to quantify the geometry. DOP can be used to predict the multilateration accuracy given BS locations. A small DOP value is a necessary condition for accurate multilateration.

On the other hand, a small DOP value is not a sufficient condition for accurate localization because DOP can reflect only the geometrical BS-node relation, instead of many other error sources, such as NLoS, multipath, and stochastic errors. Therefore, other approaches for BS location optimization have been presented. For example, the research in \cite{Rajagopal-Chayapathy-2016} combines DOP and a floor plan to address the problem. Meanwhile, there are methods for BS-geometry evaluation and optimization through CRLB analysis \cite{Laitinen-E-2016} and the Genetic algorithm \cite{Campos-R-2019}. The research in \cite{Baala-O-2019} analyzes the influence of BS geometry on indoor localization and has involved the impact of obstacles. According to the existing works, it is worthwhile to evaluate the BS geometry in advance and optimize it based on the actual localization environment.

In the future, more BSs will become available for localization due to the popularization of small IoT BSs. The existence of more BSs is generally beneficial for localization; however, too many BSs may increase the complexity of location estimation. For example, if several BSs are located in close proximity, a location ambiguity problem may occur around the region in which these BSs are located. Thus, it is also necessary to use BSs selectively based on their importance. The research in \cite{FangS-2010} quantifies the importance of each BS by the signal discrimination between distinct locations. Meanwhile, the paper \cite{Laitinen-E-2015} examines several BS-selection criteria, such as the Kullback-Leibler divergence, dissimilarity, maximum value, and entropy. The literature \cite{Laitinen-E-2012} adds more criteria, including the number of RPs, the average, and variance values. Moreover, there are BS-selection approaches through the use of theoretical-error analysis \cite{Jia-B-2019}, adaptive cluster splitting \cite{Liang-D-2015}, and strong-signal detection \cite{Jiang-P-2015}. Meanwhile, the research in \cite{Bel-A-2015} investigates optimal BS selection through a tradeoff btween localization accuracy and energy consumption.

\subsubsection{BS Location Uncertainty}
\label{sec-bs-loc-uncertainty}
Geometrical localization methods commonly need known BS locations, a requirement which cannot be met in many IoT applications. To solve this issue, BS-localization approaches have been researched. The literature \cite{ChengY-2005} reflects the idea of estimating the locations of BSs by using their distances to multiple RPs that has known locations. The research in \cite{Han-D-2009} improves the accuracy of BS localization by using RSS gradient. Meanwhile, there are improved BS-localization methods by using particle-swarm optimization \cite{Awad-F-2017}, Fresnel-zone identification \cite{Chen--Y--2018}, and recursive partition \cite{Liu-W--2018}. The BS-localization methods have been applied in 5G mmWave systems \cite{Landstrom-A--2016}, Ad-Hoc networks \cite{Niculescu-2003}, and cooperative-localization systems \cite{Awad-F--2018}. Meanwhile, the literature \cite{Yu-J--2012} estimates BS PLM-P values before BS-location determination. 

Another challenge is that the locations of IoT BSs may change, which cause localization errors. Thus, SLAM (e.g., Foot-SLAM \cite{Bruno-L--2011} and Fast-SLAM \cite{Seco-F--2017}) and crowdsourcing-based methods (e.g., \cite{ZhuangY-2016}) have been applied to estimate node and BS locations simultaneously. Meanwhile, there is research that involves the changes of BS locations in node localization \cite{He-S--2017}. 

\subsubsection{BS Path-Loss Model Parameter Uncertainty}
\label{sec-bs-plmp-uncertainty}
To achieve accurate RSS-based ranging and geometrical localization, BS PLM-P values are commonly estimated. The SLAM \cite{Seco-F--2017} and crowdsourcing \cite{ZhuangY-2016} methods estimate PLM-P values simultaneously with BS locations. Meanwhile, there are other PLM-P estimation methods, such as those based on geometrical probability \cite{Hassan-Ali-M--2002}, the finite-difference time-domain technique \cite{Wu-Y--2008}, and the Cayley-Menger determinant \cite{Mao-G--2007}. Also, there are improved estimation techniques such as the closed-form weighted total least squares \cite{Hu-Y--2007} and PF \cite{Rodas2010DynamicPE}. The papers \cite{Li-XR-2006} and \cite{Salman-N-2012} have derived the CRLB and hybrid CRLB models for PLM-P estimation errors, respectively.

Meanwhile, to enhance the ranging and localization performance, researchers have applied advanced PLMs, such as the third-order polynomial long-distance PLM \cite{Tian-Y-2012}, its combination with Gaussian models \cite{Vallet-YJ-2012}, and the models that involve signal degradation by walls, floors, and topological features \cite{Ahmadi-H-2016}. Furthermore, specific localization signals are introduced to improve PLM-P estimation. For example, the research in \cite{Alam-N-2010} introduces the measurements of the Doppler effect, while \cite{Liu-W--2018} adds extra directional detection hardware. Moreover, the research in \cite{Liang-C-2010} implements dynamic PLM-P estimation in cooperative localization, while the literature \cite{Liang-C-2011} performs BS selection simultaneously with PLM-P estimation. 

\subsubsection{BS Time-Synchronization Errors}
\label{sec-bs-time-sync}
The BS timing-synchronization error is a practical and important error source for time (e.g., ToA, TDoA, and RTT) based localization methods. The research in \cite{Leugner-2016} investigates multiple wired and wireless time-synchronization approaches and the relation between BS time-synchronization errors and localization errors, while the research in \cite{McElroy-C-2014} evaluates several wireless time-synchronization approaches for indoor devices. Also, the literature \cite{WangY-HoC-2013} examines the degradation in localization accuracy caused by time-synchronization errors through the CRLB analysis under Gaussian noise, while the paper \cite{ChenX-WangD-2018} derives the CRLB for TDoA localization when multiple BSs subject to the same time-synchronization error.

\subsection{Data-Related Errors}
\label{sec-data-error}
IoT localization systems also meet challenges at the data level. Examples of such challenges include those for database timeliness and training cost, RP location uncertainty, database outage, intensive data, data and computational loads, and the degradation of localization integrity in challenging environments. 

\subsubsection{Database Timeliness/Training Cost}
\label{sec-db-train-cost}
To maintain localization accuracy in public areas, periodical database update is required. The most widely used database-training method is to collect data at every RP. Such a method can improve database reliability by averaging LFs at each RP \cite{ChengJ2014}. However, the static data-collection process becomes extremely time-consuming and labor-intensive when dense RPs are needed to cover a large area \cite{Bolliger2008}. To reduce time and manpower costs, dynamic-survey methods have been applied through the use of landmarks (e.g., corners and intersections with known positions and corridors with known orientations) on floor plans and a constant-speed assumption \cite{Nguyen-2013} or DR solutions \cite{LiY-WiFi-Mag-2015}. To further reduce user intervention, there are other types of database-update methods based on crowdsourcing \cite{ZhouB-ITS-2015} or SLAM \cite{Kok2018}. The research in \cite{ZhangP2018} divides crowdsourcing approaches to active and passive ones. The former allows users to participate in the database-updating process, while the latter is an unsupervised method which completes data uploading and processing automatically without user participation. A challenge for crowdsourcing is to obtain robust RP positions, which is discussed in the next subsection.

\subsubsection{RP Location Uncertainty}
\label{sec-rp-loc-uncertainty}
The uncertainty in RP locations will lead to drifts in localization databases. In particular, for dynamic-survey or crowdsourcing based localization database training, it is difficult to assure the reliability of RP locations. DR can provide autonomous localization solutions \cite{YuN2018}; however, it is challenging to obtain long-term accurate DR solutions with low-cost sensors because of the existence of sensor errors \cite{LiY-sensj-calibration}, the requirement for position and heading initialization, and the misalignment angles between the vehicle (e.g., the human body) and the node \cite{PeiL-access}. Thus, constraints are needed to correct for DR errors. Vehicle-motion constraints, such as Zero velocity UPdaTes (ZUPT), Zero Angular Rate Updates (ZARU), and Non-Holonomic Constraints (NHC) \cite{LiY-thesis}, are typically adopted. However, these motion constraints are relative constraints, which can only mitigate the drifts of DR errors, instead of eliminating them.
Absolute constraints, such as GNSS position \cite{LiY-JIOT-2019} and user activity constraint \cite{ZhouB-ITS-2015} updates, can be used to ensure the quality of DR solutions in the long term. However, such position updates are not always available in indoor environments. 

From the perspective of geo-spatial big data, only a small part of crowdsourced data is robust enough for updating databases. Thus, the challenge becomes how to select the crowdsourced data that have the most reliable DR solutions. Reference \cite{ZhangP2018} presents a general framework for assessing sensor-data quality. This framework contains the impact of multiple factors, such as indoor localization time, user motion, and sensor bias. The research in \cite{LiY-JIOT-2019} enhances this framework and introduces stricter quality-assessment criteria. 

\subsubsection{Database Outage}
\label{sec-db-outage}
To obtain a reliable database, sufficient training data are required for all accessible areas. However, such a requirement is difficult to meet in wide-area applications \cite{ShinH2015}. If there is an outage of databases in certain areas, it will be impossible to locate the user correctly with traditional DB-M methods. To mitigate the database-outage issue, geometrical interpolation (e.g., mean interpolation \cite{mi}, inversed distance weighted interpolation \cite{idw}, bilinear interpolation \cite{bi}, and Kriging interpolation \cite{Kriging}) and ML methods (e.g., GP \cite{Ferris-2006} and support vector regression \cite{svr-interp}) have been presented for DB prediction, that is, predicting LFs at arbitrary locations based on training data at other locations. The research in \cite{ChoiW-2018} compares the performance of DB prediction by using GP and geometrical interpolation methods such as linear, cubic, and thin plate interpolation. Furthermore, as demonstrated in \cite{LiY-MMT-2019}, the combination of geometrical and ML-based methods may provide database-prediction solutions with higher accuracy and resolution. 

\subsubsection{Data and Computational Loads}
\label{sec-data-load}
The localization database becomes large in wide-area loT applications. This phenomenon brings three challenges: large data load, large computational load, and potentially large mismatch (i.e., the node is localized to a place that is far from its actual position) rate. Therefore, for wide-area IoT localization, a coarse-to-fine localization strategy can be used. The coarse-localization solutions from algorithms such as min-max, centroid, and proximity may be used to limit the search region for fine localization. A practical strategy is to use LPWAN or cellular signals for coarse localization and use local wireless (e.g., WiFi, BLE, and RFID) signals for fine localization. Meanwhile, there are other methods that use wireless signals for coarse localization and use magnetic measurements for fine localization \cite{LiY-WiFi-Mag-2015}. 

Furthermore, region-based algorithms can be applied to reduce the search space within the database. Examples of these algorithms include region division \cite{Shang-F-2016}, correlation-database filtering \cite{Zekavat-R-2011}, genetic algorithms \cite{Zekavat-R-2011}, and timing advanced-based algorithms \cite{Klozar-L-2012}. Region-based algorithms can effectively reduce the data and computational loads. On the other hand, correct region detection and handover between adjacent regions are key for region-based methods.

\subsubsection{Localization Integrity}
\label{sec-loc-integrity}
The integrity of localization solutions, which describes the consistency between actual localization errors and the estimated localization uncertainty, is even more important than the localization accuracy. To be specific, it may be difficult to provide accurate localization solutions due to the limitation of physical environment. At this time, the localization system should be able to output an accurate indicator of the location uncertainty, which makes it possible to reduce the weight of an inaccurate localization result. 

However, most of the widely-used DB-M approaches do not have an indicator for the uncertainty of location outputs. Also, DB-M methods such as random forests and GP can only provide an indicator for relative probability (i.e., the probability of the selected RPs compared to other RPs), instead of the absolute location uncertainty. In contrast, geometrical methods have theoretical location-accuracy prediction indicators such as DOP \cite{Langley1999}, CRLB \cite{ZhouBP-IT-2017}, and observability \cite{Lourenco2013}. However, these indicators have not involved many real-world error sources, such as the environment-related and motion-related errors. This phenomenon leads to poor localization integrity. For example, smartphones may provide over-estimated GNSS location accuracy when a user stands at indoor window areas. The reason for this phenomenon is that the device can actually receive an enough number of wireless signal measurements but has not detected the presence of environment-related errors such as multipath. 

Thus, assuring location integrity for IoT systems is challenging but important. In particular, the future IoT systems will have a higher requirement on scalable localization, which is less dependent on the human perception and intervention. Thus, the existing localization solutions, which do not have an accuracy metric, will limit the promotion of localization techniques in IoT. 

To predict the uncertainty of localization solutions, both field-test \cite{Yiu2016} and simulation \cite{Nguyen2017} methods are utilized to localize the node and compare with its reference locations. Meanwhile, since IoT signals are susceptible to environmental factors, data from other localization sensors (e.g., inertial sensors) can be used to detect unreliable wireless signal measurements through adaptive KFs based on variables such as residuals \cite{YuM2012} and innovations \cite{Aghili2016}. Furthermore, to enhance localization-uncertainty prediction for wireless localization, mathematical model \cite{LiY-JIOT-2019} and ML \cite{LiY-NN} based methods can be used. However, it is still a challenge to assure localization integrity in indoor and urban areas due to the complexity and unpredictability of localization environments.

\begin{table*}
           \centering
           \begin{tabular}{p{1.5cm} p{16.5cm} }
             \hline
               &  \textbf{End-device-related errors}	\\
            Node (Device) Diversity 	& 	\begin{itemize}
            \item Calibration-based methods: least squares \cite{Lee2015}, histograms \cite{Laoudias2012}, ANN \cite{Tsui2009}, multi-dimensional scaling \cite{MaL-2019}, integration with DR \cite{HeS2018}, nonlinear adjustment, and sensitivity-threshold correction \cite{Figuera-2011}. 
            \item Calibration-free methods \#1: real-time estimation through crowdsourcing \cite{Zhao2018} and SLAM \cite{Zhang2018}.
            \item Calibration-free methods \#2: use differential measurements such as pairwise difference \cite{Dong2009}, signal strength difference \cite{Hossain2013}, mean difference \cite{Laoudias2014}; use advanced datum BS selection \cite{Ye2016}, and averaged measurements from selected BSs \cite{Yang2013}; use differential measurements in multilateration \cite{Lin2013} and other geometrical \cite{LeeJ2009} methods.
			\end{itemize} \\ 
			
			   Motion/Attitude Diversity  	& 	\begin{itemize}
            \item DB-M based approaches: attitude-appended fingerprints \cite{Papapostolou2009} and decision trees \cite{Sanchez2012}, magnetometer-aided databases\cite{King2006}, and dynamic AHRS-aided databases \cite{Deng2018}. 
            \item Combination of DB-M and parametric models: attitude compensation through histogram equalization \cite{Fang2015}.
            \item Parametric model based methods: orientation-compensation model \cite{LiY-SensJ-2019}.  
			\end{itemize} \\ 
			
			 Data Loss and Latency  	& 	\begin{itemize}
            \item Predict localization signals: time-series analysis \cite{Hamilton-1994}, multi-channel mechanisms \cite{HeZ-2019}, and ML \cite{HeZ-2019}.  
            \item Evaluate and control impact of data loss ﻿\cite{Bor-2016}, lagency ﻿\cite{Oliveira-2018}, ﻿response rate ﻿\cite{ZhuangY-2016}\cite{LedlieJ-2011}, especially those for important BSs ﻿\cite{FangS-2010}.
			\end{itemize} \\ 
			   
			 Channel Diversity  	& 	\begin{itemize}
            \item Calibrate channel diversity through parametric models \cite{ChoiT-2016} and ML \cite{HeZ-2019}. 
            \item Data from multiple channels may be combined \cite{Zanella-2014} or treated as data from various BSs \cite{ZhuangY-sens-2016}.
			\end{itemize} \\ \hline 
			
			   & \textbf{Environment-related errors} \\
			
			 Multipath  	& 	\begin{itemize}
            \item Multipath detection: direct-path and multipath separation \cite{SenS-2013}, multipath-delay extraction \cite{Zayets2017}, and 3D city model assisting \cite{KumarRakesh-2017}.
            \item Multipath modeling: different PLMs for direct-path and multipath signals \cite{DeyleT-2008}, multipath components in PLMs \cite{ShenJ-2019}, and real-time multipath-parameter estimation by using SLAM \cite{Vanderveen-1998} and EKF \cite{ZhuM-2015}. 
            \item Multipath-effect mitigation: antenna arrays \cite{SamanK-2019}, beamforming \cite{TaoC-2017}, frequency-hopping \cite{TaoC-2017}, a modified delay locked loop \cite{Dierendonck-1992}, and multi-channel signals \cite{Zanella-2014}.
            \item Multipath-assisted localization: principle and methodology \cite{Witrisal-2015},  statistical performance bounds and evaluation models \cite{Meissner-2014}, CRLB \cite{Leitinger-2015}, and uses in outdoor \cite{Ljungzell-2018} and indoor \cite{GroB-2018} signal-BS systems; SLAM that treats multipath components as signals emitted from virtual BSs \cite{Gentner-2016}; the use of multipath-sensitive signal features \cite{PopleteevA-2019}; methods that use multipath signals for localization computation, such as DB-M \cite{KupershteinE-2013} and multilateration \cite{Soltanaghaei-2013}.
			\end{itemize} \\ 
			
			    NLoS  	& 	\begin{itemize}
            \item NLoS identifiction: algorithms such as ANN \cite{Heidari-M-2019}, random forests \cite{Ramadan-2018}, SVM \cite{Marano-S-2019}, the Dempster-Shafer evidence theory \cite{Li-Sh-2016}, and Neyman-Pearson test \cite{XiaoZ2015}; use multi-channel data \cite{Jo-H-2018}; relation between NLoS and location errors \cite{Shi-G-2019}; NLoS-based CRLB \cite{ZhouBP-TC-2019}.               
            \item NLoS modeling: advanced PLMs that involve walls \cite{Durantini-2005}\cite{Lemic-2016}, building floors \cite{Seidel-1992}, factors such as thickness of obstacles and intersection angles between obstacles and direct path \cite{Ngo-Q-2015}, and losses of walls and interactions \cite{Plets-2012}.
            \item NLoS-effect mitigation: estimation techniques such as Monte-Carlo Gaussian smoothing \cite{Vila-Valls-2017}, residual analysis \cite{Wang-Y-2018}, Least Trimmed Squares \cite{Khodjaev-J-2016}, and improved filtering techniques, such as the cubature KF \cite{Li-W-2013}, skew-t variational Bayes filter \cite{Nurminen-H-2015}, PF\cite{Boccadoro-M-2012}, UKF \cite{Huerta-J-2009}, finite impulse response filter \cite{Pak-J-2017}, and biased KF \cite{Rohrig-C-2009}; advanced models such as radial extreme value distribution \cite{Okusa-K-2017}; integration with external sensors (e.g., vision \cite{Su-Z-2017} and inertial \cite{Yavari-M-2017} sensors).
			\end{itemize} \\ 

			  Wide-Area Effects  	& 	\begin{itemize}
            \item Evaluate wide-area PLM changes with factors such as environment type (e.g., highway, rural, and urban) \cite{Karedal-M-2011}, terrain category (e.g., hilly, flat, with light/moderate/heavy tree densities, and on water/ground) \cite{Erceg-MV-1999}\cite{Petajajarvi-J-2015}, and BS antenna heights \cite{Erceg-MV-1999}.
            \item Use advanced wide-area PLMs: the multi-slope PLM \cite{Hrovat-A-2010}, the higher-order PLM \cite{Sallouha-2017}, and the height-dependent PLM \cite{Sallouha-H-2010}; ML methods for determining the PLM-P values \cite{Hosseinzadeh-S-2017}. 
            \item Improved wide-area localization algorithms such as BS identity \cite{Bshara-M-2010} and minimum-mean-square-error \cite{Xiao-L-2017}; use wide-area and local IoT signals for coarse and fine localization, respectively.
			\end{itemize} \\ 

			Multi-Floor Effects  	& 	\begin{itemize}
            \item Floor detection using data from wireless sensors \cite{Elbakly2018}, a barometer \cite{Haque2018}, inertial sensors \cite{Ramana2016}, a floor plan \cite{LoC2018}, and a user-position probability map \cite{Herrera2014}; use estimation techniques such as KF \cite{Haque2018}, PF \cite{LoC2018}, and ANN \cite{Elbakly2018}. 
            \item Evaluate the impact of multi-floor effects on PLMs \cite{Seidel-1992}.
			\end{itemize} \\ 

			 Human-Body Effects  	& 	\begin{itemize}
            \item Evaluate the human-body impact on multilateration \cite{Kilic-Y-2012}, centroid \cite{Pathanawongthum-2009}, and other localization algorithms \cite{Schmitt-S-2014}; evaluate the human-body effect when the node is carried on the human body and when a human is close to a BS \cite{Pathanawongthum-2009}; evaluate the influence when the node is located at various places and with different orientations on human body \cite{Trogh-2016}, and short-term and long-term human-body effects \cite{PopleteevA-2019}.
           \item Human-body effect modeling: model it on ToA ranging and localization \cite{Geng-Y-2013}; model its relation with node orientation toward fixed BSs \cite{Trogh-2016}; may use other sensors (e.g., vision) to detect human bodies and aid the modeling of their effects. 
            \item Human-body effect mitigation: treat it as a NLoS signal and use NLoS-mitigation methods \cite{Cully-E-2012}; combine data from multiple nodes on different human body locations \cite{Trogh-2016}.
            \item Use human-body effects for device-free localization \cite{Adib-F-2014} by using signals such as ranges \cite{Kilic-Y-2014}, RSS \cite{Saeed-A-2014}, CSI \cite{Gao-Q-2017}, and AoA \cite{Zhang-L-2018}; use human-body effects for activity recognition \cite{Wang-J-2018}, fall detection \cite{Wang-Y-2017}, and people counting \cite{Xu--C-2013}. 
			\end{itemize} \\ 

			  Weather effects 	& 	\begin{itemize}
            \item Evaluate the relation between ranging/localization performance and factors such as temperature, relative humidity, air pressure, aviation-specific runway visibility \cite{PopleteevA-2019}, rain condition and wind speed \cite{Fang-SH-2016}, and water \cite{Petajajarvi-J-2015}.
            \item Integrate IoT signals with self-contained localization technologies (e.g., INS) to reduce weather effects.
			\end{itemize} \\ 

			    Signal Variations  	& 	\begin{itemize}
            	 \item Model stochastic signals by using methods such as Allan variance \cite{LuoC-2018} and multisignal wavelet variance \cite{Radi-2019}.
				\item Mitigation through denoising methods such as averaging \cite{Small-J-2000}, autoregressive \cite{Nassar-S-2003}, wavelet \cite{Chang-S-2000}, KF \cite{Peesapati-S-2013}, and ANN \cite{Xie-J-2012}.
			\end{itemize} \\ 
			   &  \\ \hline
           \end{tabular}
           \caption{ Localization error sources and mitigation, part \#1: end-device- and environment-related errors   }
           \label{tab-loc-err-1}
         \end{table*}

\begin{table*}
           \centering
           \begin{tabular}{p{1.5cm} p{16.5cm} }
             \hline
               &  \textbf{Base-station-related errors}	\\
           			
			      Number of BSs  	& 	\begin{itemize}
            \item Single-BS localization techniques: ToA \cite{Vasisht2016}\cite{Xie2018}, AoA \cite{Quuppa2018}\cite{Guerra2018}, and RSS DB-M \cite{Rzymowski2016} and parameter-model \cite{LiY-TIM-2019} based ones.
            \item Motion constraints, such as the uniform-motion assumption, height constraint \cite{Petovello-2003}, ZUPT, ZARU, and NHC \cite{LiY-thesis} constraints.
            \item Use coarse-localization algorithms, such as min-max \cite{Will-2012}, centroid \cite{Pivato-2011}, and proximity \cite{BsharaM-2011}.
            \item Fuse wireless signal measurements with data from other sensors (e.g., inertial and vision sensors) for tightly-coupled localization \cite{ZhuangY2018-IoT}.
			\end{itemize} \\ 

			     BS geometry 	& 	\begin{itemize}
            \item BS-geometry evaluation and optimization: DOP \cite{Langley1999}, CRLB analysis \cite{Laitinen-E-2016},  location ambiguity analysis \cite{ZhuangY-2016}, and Genetic algorithms \cite{Campos-R-2019}; involve the ﻿impact of the obstacles ﻿\cite{Baala-O-2019}; investigate the relation between BS geometry and localization performance through simulation \cite{Kihara-M-1994} and field testing \cite{Meng-X-2004}. 
            \item BS selection: BS importance evaluation\cite{FangS-2010}; BS-selection criteria (e.g., Kullback-Leibler divergence, dissimilarity, maximum value, entropy \cite{Laitinen-E-2015}, mean value, variance, and the number of RPs \cite{Laitinen-E-2012}); theoretical-error analysis \cite{Jia-B-2019}, adaptive cluster splitting \cite{Liang-D-2015}, strong-signal detection \cite{Jiang-P-2015}, and tradeoff between localization accuracy and energy consumption \cite{Bel-A-2015}.
			\end{itemize} \\ 
			
			     BS Location Uncertainty  	& 	\begin{itemize}
            \item BS localization: wide-area war-driving \cite{ChengY-2005}, RSS gradient \cite{Han-D-2009}, particle-swarm optimization \cite{Awad-F-2017}, Fresnel-zone identification \cite{Chen--Y--2018}, and recursive partition \cite{Liu-W--2018}; applied in 5G mmWave systems \cite{Landstrom-A--2016}, Ad-Hoc networks \cite{Niculescu-2003}, and cooperative-localization systems \cite{Awad-F--2018}. 
            \item SLAM (e.g., Foot-SLAM \cite{Bruno-L--2011} and Fast-SLAM \cite{Seco-F--2017}) and crowdsourcing \cite{ZhuangY-2016} methods that estimate node and BS locations simultaneously; consider changes of BS locations \cite{He-S--2017}. 
			\end{itemize} \\ 
			
			 BS PLM-P Uncertainty  	& 	\begin{itemize}
            \item PLM-P estimation: geometrical probability \cite{Hassan-Ali-M--2002}, finite-difference time-domain \cite{Wu-Y--2008}, Cayley-Menger determinant \cite{Mao-G--2007}, and estimation techniques such as closed-form weighted total least squares \cite{Hu-Y--2007} and PF \cite{Rodas2010DynamicPE}; derive CRLB \cite{Li-XR-2006} and hybrid CRLB \cite{Salman-N-2012} models.
            \item Advanced PLMs: third-order polynomial long-distance PLM \cite{Tian-Y-2012}, its combination with Gaussian models \cite{Vallet-YJ-2012}, and models that involve walls, floors, and topological features \cite{Ahmadi-H-2016}; introduce specific localization signals such as Doppler \cite{Alam-N-2010} and directional detection \cite{Liu-W--2018}; dynamic PLM-P estimation in cooperative localization \cite{Liang-C-2010} or simultaneously with BS selection \cite{Liang-C-2011}. 
			\end{itemize} \\ 
			
			BS Time-Synchronization Error  	& 	\begin{itemize}
            \item Evaluate relation between BS time-synchronization errors and localization errors \cite{Leugner-2016}, relation between localization accuracy and time-synchronization errors through CRLB analysis under Gaussian noise \cite{WangY-HoC-2013}, and CRLB for TDoA localization when multiple BSs subject to the same time-synchronization error \cite{McElroy-C-2014}.
            \item Investigate multiple wired \cite{Leugner-2016} and wireless \cite{Leugner-2016}\cite{McElroy-C-2014} BS time-synchronization approaches.
			\end{itemize} \\ 

			\hline

			&  \textbf{Data-related errors}	\\
           						
			     Database Timeliness/ Training Cost  	& 	\begin{itemize}
            \item Static survey: \cite{ChengJ2014}, which is time-consuming and labor-intensive \cite{Bolliger2008}. 
            \item Dynamic survey: use of landmarks, floor plans, and the constant-speed assumption \cite{Nguyen-2013} or short-term DR trajectories ﻿\cite{Nguyen-2013}\cite{LiY-WiFi-Mag-2015}.
            \item Crowdsourcing \cite{ZhouB-ITS-2015} and SLAM \cite{Kok2018} based database-updating; active and passive crowdsourcing approaches \cite{ZhangP2018}.
			\end{itemize} \\ 

			   RP Location Uncertainty  	& 	\begin{itemize}
            \item Enhanced DR: autonomous sensor calibration \cite{LiY-sensj-calibration}, misalignment estimation \cite{PeiL-access}, motion constraints (e.g., ZUPT, ZARU, and NHC) \cite{LiY-thesis}, position-fixing constraints \cite{LiY-JIOT-2019}, and user-activity constraints \cite{ZhouB-ITS-2015}.  
            \item Sensor data quality control: crowdsourced data quality-assessment framework \cite{ZhangP2018} and improved quality-assessment criteria \cite{LiY-JIOT-2019}. 
			\end{itemize} \\ 
			
			  Database Outage  	& 	\begin{itemize}
            \item Geometrical database prediction: mean interpolation \cite{mi}, inversed distance weighted interpolation \cite{idw}, bilinear interpolation \cite{bi}, and Kriging interpolation \cite{Kriging}.  
            \item ML-based database prediction: GP \cite{Ferris-2006}, SVM \cite{svr-interp}, and their comparison \cite{ChoiW-2018}; combination of ML and geometrical methods \cite{LiY-MMT-2019}.
			\end{itemize} \\ 
			
			 Data and Computational Load  	& 	\begin{itemize}
            \item Coarse-to-fine localization: use coarse locations from algorithms (e.g., min-max, centroid, and proximity) and sensors (e.g., LPWAN and cellular) to limit search regions for fine localization; use wireless signals for coarse localization and use magnetic measurements for fine localization \cite{LiY-WiFi-Mag-2015}.  
            \item Region-based algorithms: region division \cite{Shang-F-2016}, correlation-database filtering \cite{Zekavat-R-2011}, genetic algorithms \cite{Zekavat-R-2011}, and timing advanced-based algorithms \cite{Klozar-L-2012}.
			\end{itemize} \\ 
			
			Localization Integrity  	& 	\begin{itemize}
            \item Theoretical location-performance prediction indicators such as DOP \cite{Langley1999}, CRLB \cite{Abu-Shaban-Z-2018}, and observability \cite{Lourenco2013}. 
            \item Localization uncertainty prediction: field testing \cite{Yiu2016}, simulation \cite{Nguyen2017}, use KF residuals \cite{YuM2012} and innovations \cite{Aghili2016}, mathematical models \cite{LiY-JIOT-2019}, and ML \cite{LiY-NN}.
			\end{itemize} \\ 

			   &  \\ \hline
           \end{tabular}
           \caption{ Localization error sources and mitigation, part \#2: base-station- and data-related errors   }
           \label{tab-loc-err-2}
         \end{table*}

\subsection{Summary and Insight on IoT Error Sources and Mitigation}
\label{sec-comm-loc-err-mit}
\begin{itemize}
\item Different IoT applications may suffer from various degrees of localization errors. For example, professional IoT use cases commonly have limited application areas, which make it more straightforward to model and mitigate environment-related errors; also, the use of high-end sensors and in-the-lab sensor calibration can effectively reduce many end-device-related errors; meanwhile, the availability of specifically designed and deployed BSs can control BS-related errors; finally, the availability of powerful communication and computation hardware can alleviate the data-related errors. Therefore, it is more straightforward to promote localization techniques in professional IoT applications. By contrast, it is commonly not affordable for mass-market IoT applications to add specific node or BS hardware or implement in-the-lab sensor calibration; meanwhile, the mass-market localization environment varies significantly; finally, only low-cost localization, communication, and processing sensors can be used. Thus, it is necessary to involve more error sources when designing localization algorithms for mass-market IoT applications.

\item In real-world localization scenarios, especially those for dynamic applications, the actual localization error is a combination of multiple error sources. Each error source may change in real time. The complexity and diversification of actual IoT application scenarios have greatly increased the challenge to mitigate some errors (e.g., many environment-related errors). Thus, although the reviewed approaches can effectively reduce or eliminate the influence of some errors in some scenarios, it is challenging to mitigate other errors due to the physical environment limitations. To mitigate this issue, integrating IoT signals with data from other sensors such as inertial sensors is a feasible approach. IoT signals may provide long-term and wide-area absolute location solutions, while DR solutions from inertial sensors can provide short-term reliable and smooth relative location solutions. Also, DR solutions can bridge short-term outages and resist outliers in IoT signals.

\item Some error sources occur in both DB-M and geometrical localization methods. These error sources include the majority of end-device-related errors (e.g., device diversity, motion/attitude diversity, data loss/latency, and channel diversity), the minority of environment-related errors (e.g., wide-area effects, weather effects, and signal variations), the minority of BS-related errors (e.g., the number of BSs and geometry), and the minority of data-related errors (e.g., RP location uncertainty and localization integrity issues). By contrast, there are error sources that mainly exist in geometrical methods. These error sources include several main environment-related errors (e.g., multipath, NLoS, multi-floor effects, and human-body effects) and the majority of BS-related errors (e.g., BS location uncertainty, BS PLM-P uncertainty, and BS time-synchronization errors). The error sources that mainly exist in DB-M approaches include the majority of data-related errors (e.g., database timeliness/training cost, database outage, intensive data, and data and computational loads). Thus, in-depth knowledge of localization error sources is vital for selecting localization sensors and algorithms. 

\item Although some factors (e.g., multipath, NLoS, and human-body effects) have been listed as localization error sources, they can also be used as valuable measurements to enhance localization. Furthermore, these factors can even be used as the key component for new localization approaches, such as device-free localization. This phenomenon can also partly explain why DB-M methods are commonly more suitable for complex environments (e.g., indoor environments). A main reason is that the existence of these factors is beneficial to DB-M.
\end{itemize}

\section{IoT Localization-Performance Evaluation}
\label{sec-loc-per-eval}
This section describes the existing localization-performance evaluation approaches, including theoretical analysis, simulation analysis, in-the-lab testing, field testing, and signal grafting. These methods are important for evaluating and predicting the localization performance of LE-IoT systems. Meanwhile, these approaches reflect the tradeoff between performance and cost. In general, this section answers the following questions: (1) what are the existing localization-performance evaluation methods; and (2) what are the advantages and limitations of these methods. 

\subsection{Theoretical analysis}
\label{sec-thero-analysis}
There are various theoretical-analysis methods, including DOP, CRLB, and observability analysis. The existing research on DOP have been described in Subsection \ref{sec-bs-geo}. The CRLB analysis methods have been applied on various localization applications, such as those for multilateration \cite{Shen-2012}, hyperbolic positioning \cite{Kaune-2011}, multiangulation \cite{WangY-2015}, DB-M \cite{Nikitin-A-2017}, generic 3D localization \cite{Abu-Shaban-Z-2018}, and the combination of multiple localization methods such as multilateration and multiangulation \cite{Abu-Shaban-Z-2018}. The influence of other factors, such as SNR \cite{Peral-Rosado-J-2016}, has also been investigated. 

Besides DOP and CRLB, observability describes the ability in correctly estimating the states in a system \cite{Ham-Brown-1983}. With better observability, it is more straightforward to estimating a state in the localization filter. By contrast, an unobservable state cannot be estimated correctly even when the system does not have measurement noise \cite{Goshen-Meskin-Bar-Itzhack-1992}. Observability-analysis methods are commonly applied in inertial sensor-based localization \cite{LiY-NHC-OBS} and have also been used in wireless positioning \cite{Lourenco2013}.

Theoretical-analysis methods have advantages such as: (1) they have a rigorous theoretical derivation. (2) They can be used to analyze not only a certain error source, but also the relation between multiple error sources. (3) They have a low hardware complexity and algorithm computational load. (4) They can be directly transposed to other localization methods. On the other hand, the limitations of theoretical-analysis methods include: (1) it is difficult to involve actual localization error sources in theoretical analysis. Thus, theoretical-analysis methods can only provide the necessary conditions and a lower bound on performance. (2) Real-world localization scenarios such as the environmental and motion factors are difficult to model and derive.

\subsection{Simulation Analysis}
\label{sec-simu-analysis}
Simulation analysis is also a widely used localization-performance evaluation approach, especially for frontier applications (e.g., LPWAN and 5G) in which field-test data is difficult to obtain. Simulation analysis can be implemented in advance and its outputs can guide the design of subsequent tests such as in-the-lab and field testing. The simulation may be conducted based on localization software \cite{LiY-CSNC-OBS-2012} or specific simulation platforms \cite{Baik-Lee-2018}. 

There are several advantages for simulation analysis: (1) it has a low hardware cost. (2) It can analyze the impact of a certain error source (e.g., BS geometry, BS-node range, motion mode, and noise level) \cite{Langendoen-Reijers-2003} as well as the relation between multiple error sources \cite{LiY-CSNC-OBS-2012}. (3) It is straightforward to set and tune parameter values and assess performance trends. On the other hand, the limitations of simulation analysis include: (1) most of the existing simulation-analysis models are relying on simplified models and have not reflected complex environmental and motion factors. (2) It is difficult to reflect actual localization situation through simulation. 

\subsection{In-The-Lab Testing}
\label{sec-in-the-lab}
In-the-lab testing is a localization-performance evaluation method between simulation analysis and field testing. Compared to simulation, in-the-lab testing uses real hardware and sensors for data collection. Compared to real-field testing, in-the-lab testing is implemented in a controlled environment and is commonly affordable to use specific calibration and testing equipment such as a turntable \cite{Coca-Popa-2013} or shake table \cite{Xu-Shi-2013}.
 
The advantages of in-the-lab testing include: (1) the localization sensor data are real. (2) It is feasible to test a certain type of error source (e.g., antenna radiation pattern \cite{Coca-Popa-2013} and obstructions \cite{Durantini-2005}) in lab environment. However, in-the-lab testing methods have limitations such as (1) it may be difficult to reflect some error sources in lab environments due to the limitation of available equipment. Thus, it is still challenging to reflect real localization scenarios by in-the-lab testing. (2) In-the-lab tests may require specific equipment, which is not affordable for many low-cost IoT applications. 

\subsection{Signal Grafting}
\label{sec-sig-grafting}
The signal-grafting method is a tradeoff between field testing and in-the-lab testing \cite{Niu-Wang-2016}. The idea of signal grafting is to add data from low-cost sensors to the data obtained from higher-end sensors, so as to mimic low-cost sensor data. Specifically, real localization data are collected by taking high-end sensors and moving under real scenarios. Thus, both the localization environmental and motion factors are real; also, the data from aiding sensors (e.g., GNSS, magnetometers, and odometers) are real. Meanwhile, the data of low-cost sensors are collected in the lab and grafted into the field-tested high-end sensor data. 

The advantages of signal grafting include: (1) it is significantly more cost-effective than field testing. In particular, it is not necessary to conduct specific field test for each low-cost node. Only typical high-end sensor data in real localization scenarios and low-cost sensor data in the lab are needed. (2) It has real localization environment, movement, and aiding information. Thus, it is closer to real localization situation than in-the-lab testing. On the other hand, its challenges include: (1) it is difficult to reflect the sensor errors that are correlated with real localization scenarios. (2) This method has not yet been applied in published IoT-signal-based localization studies. 

\subsection{Field Testing}
\label{sec-field-test}
\begin{figure}
           \centering
           \includegraphics[width=0.48 \textwidth]{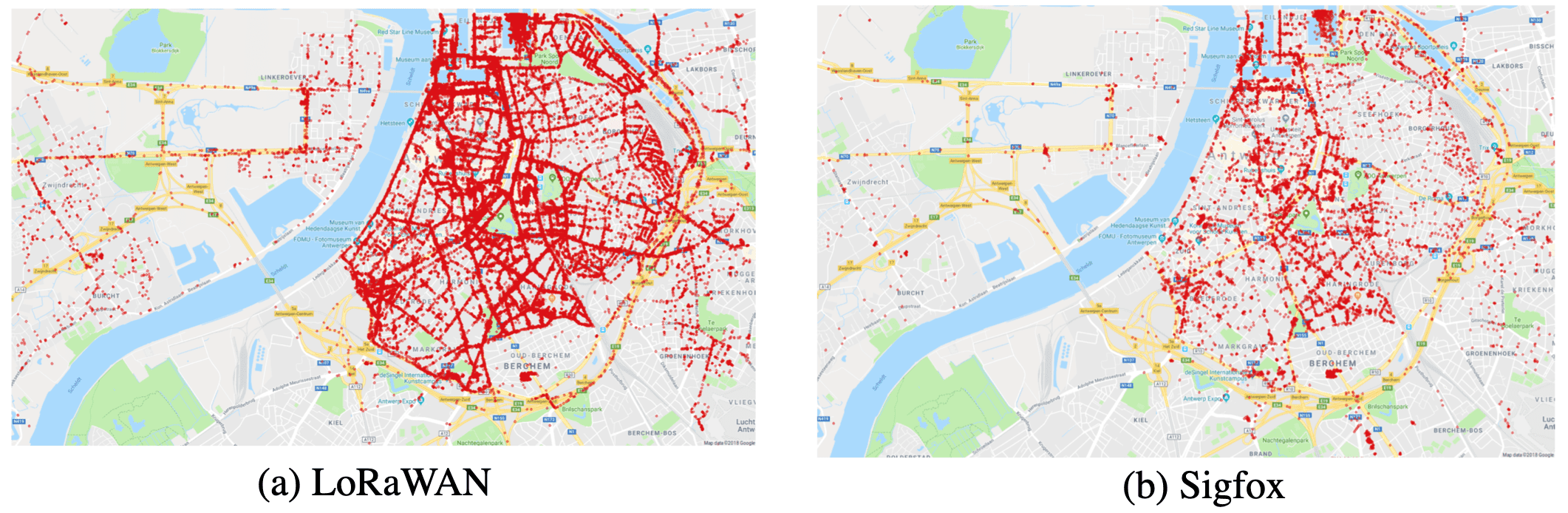}
           \caption{Example wide-area LoRaWAN and Sigfox data \cite{Aernouts-Berkvens-2018}. Red dots represent data-collection points}
           \label{fig:lora-sigfox-data}
         \end{figure}

The field-testing method has been widely used in IoT positioning applications, such as Sigfox and LoRaWAN DB-M \cite{Aernouts-Berkvens-2018}, TDoA \cite{Plets-Podevijn-2018}, and RSS localization \cite{LiY-LoRa-2018}. Figure \ref{fig:lora-sigfox-data} demonstrates an example wide-area LoRaWAN and Sigfox data. In field testing, the whole process from sensor selection to data processing is real. However, a main challenge for field testing is the high cost, especially for wide-area applications. Meanwhile, for low-cost wide-area IoT applications, it is not affordable to carry out a field test that may cost much more than the actual IoT nodes. 

\subsection{Summary and Insight on Localization-Performance Evaluation Methods}
\label{sec-comm-per-evaluation}
The evaluation of localization performance is a tradeoff between performance and cost. For example, from the perspective of reflecting real localization scenarios, the order of the methods from the best to the worst is field testing, signal grafting, in-the-lab testing, simulation analysis, and theoretical analysis. In contrast, from the cost-effectivity perspective, the opposite order prevails. In LE-IoT applications, especially the low-cost ones, it is preferred to implement localization-performance evaluation through the methods in the order of theoretical analysis, simulation analysis, in-the-lab testing, signal grafting, and finally field testing.

An important factor for in-the-lab, signal-grafting, and field-testing methods is the acquirement of location references. When higher-accuracy external position-fixing techniques (e.g., GNSS Real-time kinematic, optical tracking, vision localization, UWB, ultrasonic, and RFID) are available, their positioning results can be used as location references. Otherwise, manually selected landmark points may be used for evaluating the localization solutions when the node has passed these points. Meanwhile, the constant-speed assumption \cite{Nguyen-2013} or short-term DR solutions \cite{LiY-WiFi-Mag-2015} may be used to bridge the gap between landmark points. In this case, it is important to assure the accuracy of landmark positions and motion-assumption/DR solutions. Compared with the handheld and wearable modes, foot-mounted DR \cite{Niu-Li-RF-DR} provide significantly more robust location solutions. Thus, it is feasible to combine foot-mounted DR and external landmark and position-fixing data to generate location references. 

\section{Localization Opportunities From LPWAN and 5G}
\label{sec-loc-opport}
	The newly-emerged LPWAN and 5G signals are bringing changes into the localization field. For example, they are expected to bring new technologies \cite{Andrews-2014}\cite{Zhang-Lu-2017} such as smaller BSs, smarter devices, mmWave, MIMO, the support for D2D communications, and device-centric architectures. These new technologies may bring opportunities and changes to IoT localization. 
	
\subsection{Cooperative Localization}
\label{sec-coop-loc}
	The research of cooperative localization mainly covers two topics. First, there are cooperative-localization methods that use the connection between multiple nodes and localization sensors on each node \cite{Li-Hedley-2015}. Meanwhile, there may be master nodes, which are equipped with higher-end sensors, and slave nodes, which have lower-end sensors \cite{Lee-Grejner-Brzezinska-2012}. The theoretical model and accuracy bounds of non-cooperative \cite{Ouyang-Wong-2010} and cooperative \cite{Ouyang-Wong-2010}\cite{ZhouBP-TC-2016} localization systems have been researched. Meanwhile, the research in \cite{Kim-Lee-2018} provides theoretical and simulation analysis of vehicle cooperative localization through vehicle-to-vehicle and vehicle-to-infrastructure communications. 
	
	Meanwhile, there are cooperative-localization approaches that use multiple sets of nodes (e.g., smartphone, smart watch, and smart glasses) at various places on the human body \cite{Omr-Georgy-2014}. The research in \cite{Lan-Thesis} has a detailed description on the soft and hard constraints as well as the state-constrained KF for localization using multiple devices on the same human body. 
	
	In the coming years, the characteristics of dense BSs and D2D communication capability may make it possible to provide accurate cooperative localization. The research in \cite{Zhang-Lu-2017} has reviewed 5G cooperative localization and pointed out that cooperative localization can be an important feature for 5G networks. 
	
\subsection{Machine Learning / Artificial Intelligence }
\label{sec-ml-ai}
Subsection \ref{sec-ml-db-m} has demonstrated the use of ML methods in IoT localization. Furthermore, the papers \cite{Alsheikh-Lin-2014} and \cite{Li-Xu-2019} have illustrated some challenges for using ML in sensor networks and location-based services. Examples of these challenges include how to improve ML effectiveness under localization scenario changes and how to collect, transfer, and store massive localization data. It is expected that ML will be more widely and deeply used in IoT localization applications due to factors such as the popularization of IoT BSs and nodes, the emergence of geo-spatial big data, and the further development of ML platforms and algorithms. How to combine the existing geometrical and DB-M localization methods with the state-of-the-art ML techniques will be a significant direction for IoT localization.

\subsection{Multi-Sensor Integration}
\label{sec-ms-integration}
Multi-sensor integration is becoming a mainstream technique for enhanced IoT localization. Technologies such as LPWAN, 5G, GNSS, WiFi, and BLE can provide long-term location updates \cite{LiY-JIOT-2019}. Meanwhile, inertial sensors and magnetometers can be used to provide attitude updates \cite{Wang-Wu-2015}, which can be used for compensating for orientation diversity in IoT signals \cite{LiY-SensJ-2019}. Moreover, it is possible to obtain a DR solution by fusing data from inertial sensors, odometers \cite{WuY-WuM-2015}, air-flow sensors \cite{Li-Zahran-2019}, and vision sensors \cite{Nister-Naroditsky-2004}. Furthermore, to enhance localization performance, barometers \cite{LiY-Access-2018} and Digital Terrain Models (DTMs) \cite{Nowak-A-2017} can provide height or floor constraints, while road networks \cite{Quddus-Ochieng-2003} and floor plans \cite{Link-Smith-2011} can provide map constraints. Observability analysis \cite{LiY-NHC-OBS} can be applied to indicate the unobservable or weakly-observable states in the localization system; then, it becomes possible to determine which types of sensors can be added to enhance localization. 

\subsection{Motion Constraints}
\label{sec-motion-constrain}
For many low-cost IoT applications, it is not affordable to enhance localization through adding extra sensor hardware. Therefore, the use of motion constraints within the algorithm has a great potential. The typical constraints including vehicle kinematic constraints (e.g., NHC, ZUPT, and ZARU) \cite{Shin-Thesis-2005}, vehicle dynamic models (e.g., steering constraints \cite{Niu-Zhang-2010}, accelerating/braking constraints \cite{Ahmed-Tahir-2017}, multi-device constraints \cite{Lan-Thesis}, aerodynamic forces \cite{Keshmiri-Mirmirani-2004}, air-flow constraints \cite{Li-Zahran-2019}, and path information \cite{Nowak-A-2017}), and control inputs (e.g., \cite{Blosch-Weiss-2010}). It is also possible to evaluate the contribution of algorithm constraints through methods such as observability \cite{LiY-NHC-OBS} and CRLB \cite{Abu-Shaban-Z-2018} analyses.

\subsection{Airborne-Land Integrated Localization}
\label{sec-airborne-land}
With the development of small satellite and Low Earth Orbits (LEO) communication technologies, it has become possible to extend IoT-signal coverage by using LEOs \cite{Cluzel-Franck-2018}. The research in \cite{Liu-Hu-2018} introduces the optimization of LEO signals, while the paper \cite{Wang-Chen-2018} characterizes the performance of localization signals from LEOs.

Besides LEOs, enhancing localization signals from airborne platforms may also become a trend. For example, the research in \cite{Sallouha-H-2010} uses UAVs as BSs for IoT localization. There is great academic and industrial potential to integrate airborne- and land-based signals for enhanced IoT localization.  

\subsection{Multipath-Assisted Localization}
\label{sec-multipath-assist}
Due to new features such as MIMO, dense miniaturized BS, and mmWave systems in 5G, using multipath signals to enhance localization, instead of reducing the multipath effect is attracting research interest. The research in \cite{Witrisal-2015} describes the principle and methodology for multipath-assisted localization, which has shown its potential to provide high-accuracy localization solutions.

\subsection{Fog/Edge Computing}
\label{sec-fog-edge}
Fog and edge computing are being intensively researched in the IoT field. The papers \cite{Oteafy-Hassanein-2018} and \cite{Shi-Cao-2016} have described in detail the relation between IoT and fog/edge computing. However, the influence of fog/edge computing on localization has not been investigated. The research in \cite{Tong-Sun-2019} points out that there is a trend to use fog computing technology to achieve low-latency localization and location awareness solutions. On the other hand, an accurate IoT localization solution may contribute to the use of fog/edge computing. Therefore, the integration of IoT localization and fog/edge computing needs further investigation.

\subsection{Blockchain}
\label{sec-blockchain}
Blockchain is another newly-emerged technology that has gained wide attentions in numerous fields. The paper \cite{Novo-O-2018} has reviewed the relation between blockchain and IoT but has not involved IoT localization. For localization, the research in \cite{Victor-Zickau-2018} presents a blockchain-based geofencing and localization strategy. In general, the combination of blockchain and IoT localization is still at an early stage and thus requires further investigation. 

\section{Conclusion}
 \label{sec-conclusion}
This paper reviews the IoT localization system through the sequence of IoT localization system review, localization data sources, localization algorithms, localization error sources and mitigation, localization performance evaluation, and new localization opportunities. Specifically, 

Section \ref{sec-overview} overviews the existing IoT technologies, followed by IoT localization applications, system architecture, and signal measurements. 

Section \ref{sec-loc-method} demonstrates the state-of-the-art IoT localization methods, including DB-M (e.g., deterministic, stochastic, and ML based DB-M) and geometrical (e.g., multilateration, hyperbolic positioning, multiangulation, and multiangulateration) localization. 

Afterwards, Section \ref{sec-err-mitigation} systematically reviews IoT localization error sources and mitigation. The localization errors are divided into four parts: (1) end-device-based errors (e.g., device diversity, motion/attitude diversity, data loss/latency, and channel diversity), (2) propagation errors (e.g., multipath, NLoS, wide-area effects, multi-floor effects, human-body effects, weather effects, and signal variations), (3) base-station-based errors (e.g., number of BSs, BS geometry, BS location uncertainty, BS PLM-P uncertainty, and BS time synchronization errors), and (4) data-based errors (e.g., database timeliness/training cost, RP location uncertainty, database outage, data and computational loads, and localization integrity). 

Then, Section \ref{sec-loc-per-eval} illustrates IoT localization performance evaluation methods, including theoretical analysis, simulation analysis, in-the-lab testing, signal grafting, and field testing. 

Finally, Section \ref{sec-loc-opport} shows the possible localization opportunities, such as cooperative localization, AI, multi-sensor integration, motion constraints, fog/edge computing, blockchain, airborne-land integration, and multipath-assisted localization. 

In general, the emergence of LPWAN and 5G technologies have brought not only great advantages but also new challenges to localization applications. These technologies have attractive features such as long-range, low-power, and low-cost IoT signals, massive node connections, small and high-density BSs, the communication capacity. Therefore, it is definitely worthwhile to conduct further research on exploring localization functionality for future LE-IoT systems.


\clearpage

\begin{IEEEbiography}[{\includegraphics[width=1in,height=1.25in,clip,keepaspectratio]{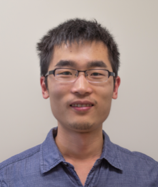}}]{You Li} (M'16) received Ph.D. degrees with the University of Calgary and Wuhan University in 2016. He is a Senior Researcher at the University of Calgary, and has been the R\&D Lead at Appropolis Inc., the Lead Scientist of EZRoad Ltd, the Algorithm Designer at Trusted Positioning Inc. (acquired by InvenSense Inc.). His research interests include location, motion, and related problems. He has co-authored 70 academic papers and has over 20 patents filed or pending. Also, he has been the winner of the IEEE EvAAL indoor localization competition and four best paper awards from IEEE, ISPRS or ION conferences. He serves as an Associate Editor of the IEEE Sensors Journal and the TPC of the International Conference on Mobile Mapping Technology (MMT).
\end{IEEEbiography}

\begin{IEEEbiography}[{\includegraphics[width=1in,height=1.25in,clip,keepaspectratio]{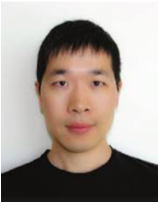}}]{Yuan Zhuang} (M'16) received the bachelor degree in information engineering from Southeast University, Nanjing, China, in 2008, the master degree in microelectronics and solid-state electronics from Southeast University, Nanjing, China, in 2011, and the Ph.D. degree in geomatics engineering from the University of Calgary, Canada, in 2015. Since 2015, He is a lead scientist in Bluvision Inc. (acquired by HID Global), Fort Lauderdale, FL, USA. His current research interests include real-time location system, personal navigation system, wireless positioning, multi-sensors integration, Internet of Things (IoT), and machine learning for navigation applications. To date, he has co-authored over 50 academic papers and 11 patents and has received over 10 academic awards. He is an associate editor of IEEE Access, the guest editor of the IEEE Internet of Things Journal and IEEE Access, and a reviewer of over 10 IEEE journals.
\end{IEEEbiography}

\begin{IEEEbiography}[{\includegraphics[width=1in,height=1.25in,clip,keepaspectratio]{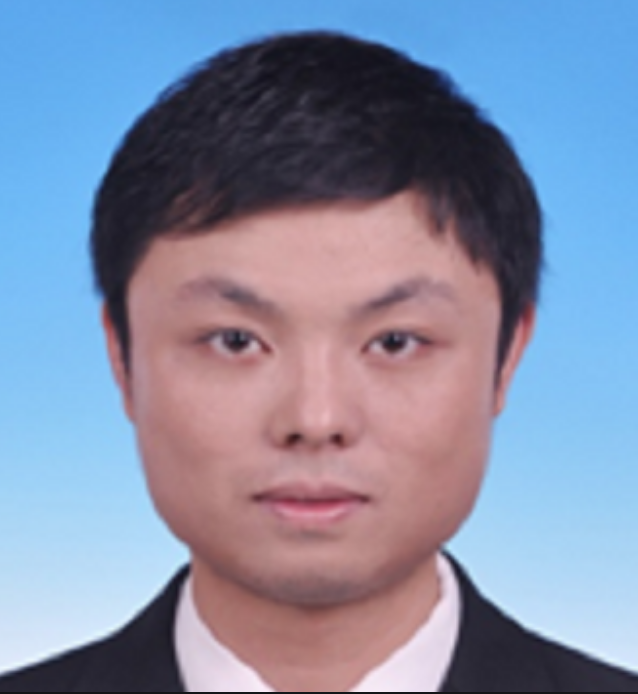}}]{Xin Hu} (M'16-SM’19) received the Ph.D. degree from the Institute of Electrics, Chinese Academy of Sciences, in 2012. He is currently an Associate Professor with the Information and Electronics Technology Lab, Beijing University of Posts and Telecommunications. He has been a visit scholar at the Department of Electrical and Computer Engineering, University of Calgary. His research interests include smart signal processing, space and ground information integration, aerospace electronic information synthesis, and communication/navigation integration.
\end{IEEEbiography}

\begin{IEEEbiography}[{\includegraphics[width=1in,height=1.25in,clip,keepaspectratio]{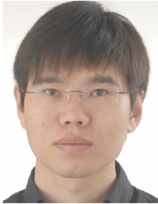}}]{Zhouzheng Gao} received his bachelor degree and master degree from China University of Geosciences Beijing, China in 2008 and 2012, and he received the PhD degree in school of Geodesy and Geomatics at Wuhan University in 2016. During 2014 and 2017, He worked in German Research Center for Geosciences (GFZ) in Potsdam, Germany as a visiting scholar (2014-2016) and post-Doctor (2016-2017), respectively. Currently he is a researcher in School of Land Science and Technology at China University of Geosciences Beijing. To date, he has one authorized software copyright and publishes 30 papers. Currently, his research interest focusses on GNSS precise positioning algorithms (PPP and RTK), GNSS/INS integration, multi-sensor integration, multi-constellation GNSS real-time positioning, and the application of multi-GNSS/multi-sensor integration.
\end{IEEEbiography}

\begin{IEEEbiography}[{\includegraphics[width=1in,height=1.25in,clip,keepaspectratio]{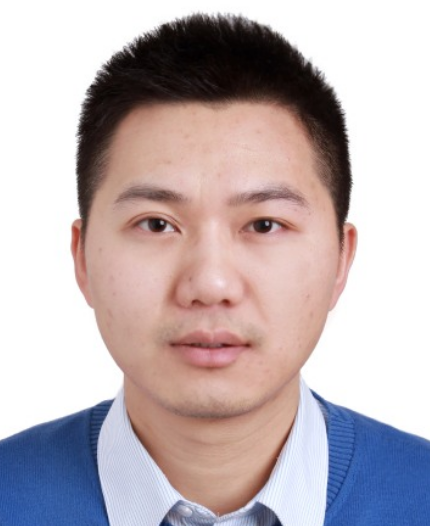}}]{Jia Hu} is  a Lecturer of computer science with the University of Exeter, Exeter, U.K.  His research has been supported by   the U.K. EPSRC, EU, China NSFC, and industry such as Huawei. His current research interests include performance evaluation, next generation networking, resource allocation and optimization, and network security. He has authored or co-authored over 50 research papers in the above areas in prestigious international journals and at reputable international conferences. Dr. Hu was a recipient of the Best Paper Award of IEEE SOSE’16 and IUCC’14. He serves on Editorial Boards and has guest-edited  many special issues in major international journals. He has served as the chair/co-chair of many international conferences.
\end{IEEEbiography}

\begin{IEEEbiography}[{\includegraphics[width=1in,height=1.25in,clip,keepaspectratio]{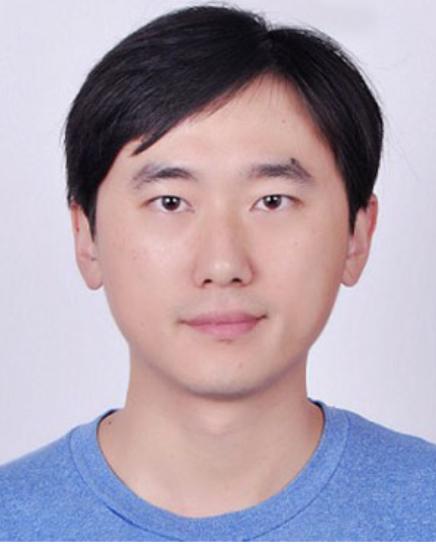}}]{Long Chen} (M'14-SM'19) received the B.Sc. degree in communication engineering and the Ph.D. degree in signal and information processing from Wuhan University, Wuhan, China, in 2007 and 2013, respectively. From 2010 to 2012, he was a co-trained Ph.D. Student with the National University of Singapore. He is currently an Associate Professor with the School of Data and Computer Science, Sun Yat-sen University, Guangzhou, China. His areas of interest include autonomous driving, robotics, artificial intelligence, where he has contributed more than 70 publications. He received the IEEE Vehicular Technology Society 2018 Best Land Transportation Paper Award, the IEEE Intelligent Vehicle Symposium 2018 Best Student Paper Award, and the Best Workshop Paper Award. He serves as an Associate Editor for the IEEE Transactions on Intelligent Transportation Systems and the IEEE Technical Committee on Cyber-Physical Systems newsletter. He serves as the Guest Editor for the IEEE Transactions on Intelligent Vehicles  and the IEEE Internet of Things Journal.
\end{IEEEbiography}

\begin{IEEEbiography}[{\includegraphics[width=1in,height=1.25in,clip,keepaspectratio]{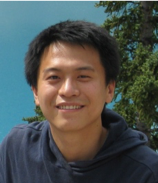}}]{Zhe He} (M'16) received his B.S. and M.S. degrees in Shanghai Jiao Tong University, majoring in Navigation Guidance and Control, and Ph.D. degree in Geomatics Engineering in University of Calgary in Canada. He worked as a post-doctoral fellow in the Position, Location And Navigation (PLAN) Group in the Department of Geomatics Engineering at the University of Calgary and is with a IoT company in Calgary. His research interests are statistical estimation theory applied to GNSS, inertial and integrated navigation systems, low power wide area networks LBS systems.
\end{IEEEbiography}

\begin{IEEEbiography}[{\includegraphics[width=1in,height=1.25in,clip,keepaspectratio]{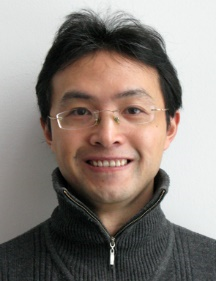}}]
{Ling Pei} (M'13) received the Ph.D. degree from South- east University, Nanjing, China, in 2007. From 2007 to 2013, he was a Specialist Research Scientist with the Finnish Geospatial Research Institute. He is currently an Associate Professor with the School of Electronic Information and Electrical Engineering, Shanghai Jiao Tong University. He has authored or co-authored over 90 scientific papers. He holds 24 patents and pending patents. His main research is in the areas of indoor/outdoor seamless positioning, ubiquitous computing, wireless positioning, bio-inspired navigation, context-aware applications, location-based services, and navigation of unmanned systems. He was a recipient of the Shanghai Pujiang Talent in 2014.
\end{IEEEbiography}

\begin{IEEEbiography}[{\includegraphics[width=1in,height=1.25in,clip,keepaspectratio]{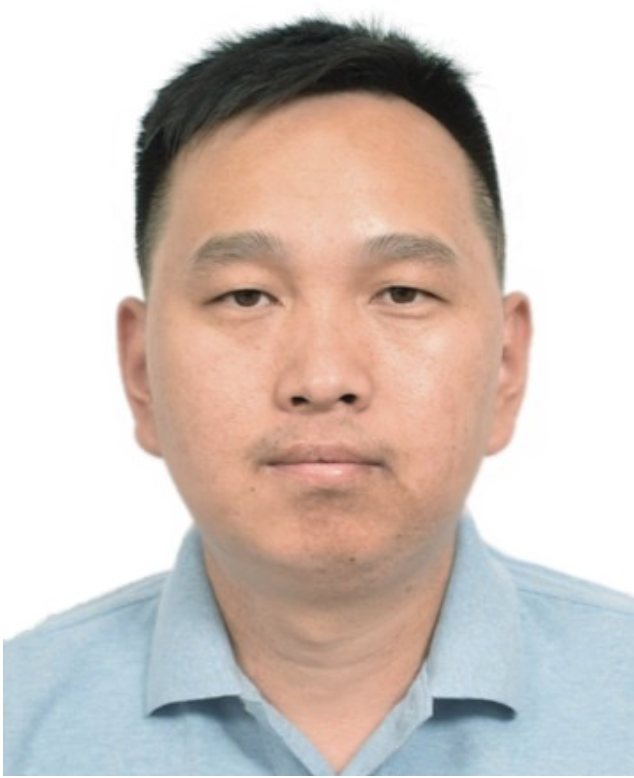}}]
{Kejie Chen} is now an assistant professor of the Department of Earth and Space Sciences, Southern University of Science and Technology, China, and the focus of his current mainly lies in precise GNSS data processing and its geophysical application. Dr. Chen received his PhD degree from University of Potsdam, Germany in 2016. From May 2016 to May 2018, he worked as a postdoc at National Aeronautics and Space Administration’s Jet Propulsion Laboratory, where he had employed real-time GNSS data to build an operational tsunami early warning system.  From May 2018 to September 2019, he was an assistant research scientist in Seismological Laboratory, California Institute of Technology, where he extended his knowledge from geodetic to geophysical field.
\end{IEEEbiography}

\begin{IEEEbiography}[{\includegraphics[width=1in,height=1.25in,clip,keepaspectratio]{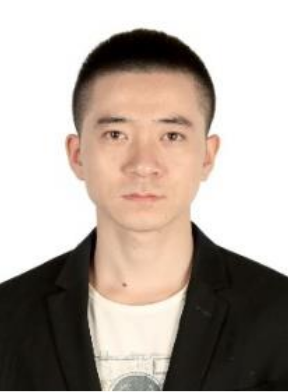}}]
{Maosong Wang}  received  the  B.S. degree from   Harbin   Engineering   University   in 2012,  and the  M.S. and Ph.D. degrees  from National University of Defense Technology in 2014 and 2018, respectively. From  September  2016  to  March 2018,  he was  a  Visiting  Student  Researcher  at  the University  of  Calgary,  Canada.  Currently, he   is   a   lecturer   at   the   National   University   of   Defense Technology.  His  research  interests  include  inertial  navigation algorithm,  and  multi-sensor  integrated  navigation  theory  and application.
\end{IEEEbiography}

\begin{IEEEbiography}[{\includegraphics[width=1in,height=1.25in,clip,keepaspectratio]{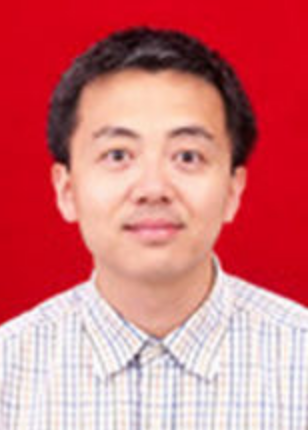}}]
{Xiaoji Niu} is currently a Professor with the GNSS Research Center, Wuhan University, China. He received the Bachelor’s and Ph.D. degrees from the Department of Precision Instruments, Tsinghua University, in 1997 and 2002, respectively. He did his Postdoctoral Research with the University of Calgary. He was a Senior Scientist with SiRF Technology Inc. He has published more than 100 academic papers and own 28 patents. He leads the Multi-Sensor Navigation Group, which focuses on GNSS/INS integrations, low-cost navigation sensor fusion, and its new applications.
\end{IEEEbiography}

\begin{IEEEbiography}[{\includegraphics[width=1in,height=1.25in,clip,keepaspectratio]{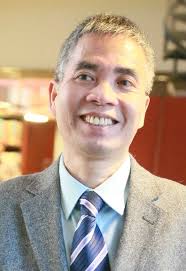}}]
{Ruizhi Chen}  is the Director of the State Key Laboratory of Information Engineering in Surveying, Mapping and Remote Sensing, Wuhan University. He has been an Endowed Chair and Professor in Texas A \& M University Corpus Christ, U.S. and Head \& Professor of the Department of Navigation and Positioning at the Finnish Geodetic Institute, Finland. He has co-authored two books, 5 book chapters, and over 200 scientific papers. Dr. Chen is the general chair of the IEEE conferences ``Ubiquitous Positioning, Indoor Navigation and Location-based Services", Editor-in-Chief the J Global Positioning Systems and associate editor of the J Navigation. He was the President of the International Association of Chinese Professionals in Global Positioning Systems (2008) and board member of the Nordic Institute of Navigation (2009-2012). His research interests include smartphone positioning indoors/outdoors, context awareness and satellite navigation.
\end{IEEEbiography}

\begin{IEEEbiography}[{\includegraphics[width=1in,height=1.25in,clip,keepaspectratio]{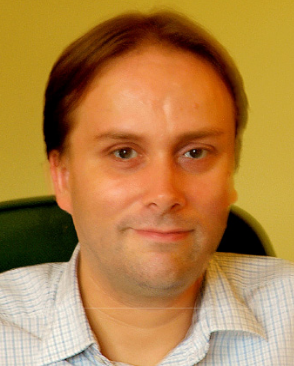}}]
{John Thompson} (M'94-SM'13-F'16)  is  currently the   Personal   Chair   in   Signal   Processing   and Communications  with  the  University  of  Edinburgh,U.K. His main research interests are in wireless communications,  sensor  signal  processing  and  energy efficient communications networks, and smart grids. He has published around 300 papers in these topics. He  was  a  recipient  of  the  Highly  Cited  Researcher Award  from  Thomson  Reuters  in  2015  and  2016. He  also  currently  leads  the  European  Marie  Curie Training   Network   ADVANTAGE, which   trains 13 Ph.D. students in smart grids. He was a Distinguished Lecturer on green topics  for  ComSoc  in  2014  and  2015.  He  is  also  currently  an  Editor  of  the Green Series of IEEE Communications Magazine and an Associate Editor for the IEEE TRANSACTIONS ON GREEN COMMUNICATIONS ANDNETWORKS.
\end{IEEEbiography}

\begin{IEEEbiography}[{\includegraphics[width=1in,height=1.25in,clip,keepaspectratio]{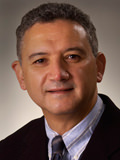}}]
{Fadhel M. Ghannouchi} (S'84-M'88-SM'93-F'07) is a Professor, a Fellow of several institutions, including the Institution of Electrical and Electronics Engineers (IEEE), the Royal Society of Canada (RSC), the Engineering Institute of Canada (EIC), the Canadian Academy of Engineering (CAE), and the Institution of Engineering and Technology (IET), a Canada Research Chair of Green Radio Systems, and the Director of the iRadio Laboratory with the Department of Electrical and Computer Engineering, University of Calgary, Canada. He has experience in wireless communication, positioning, and navigation for over thirty years and have authored or co-authored over 700 refereed publications and 25 U.S. patents (5 pending), 6 books, and 3 spun-off companies in these fields. 
\end{IEEEbiography}

\begin{IEEEbiography}[{\includegraphics[width=1in,height=1.25in,clip,keepaspectratio]{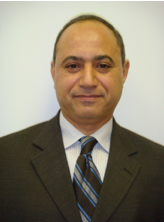}}]{Naser El-Sheimy} is a Professor at the Department of Geomatics Engineering, the University of Calgary. He is a Fellow of the Canadian Academy of Engineering and the US Institute of Navigation, and a Tier-I Canada Research Chair in Geomatics Multi-sensor Systems. His research expertise includes Geomatics multi-sensor systems, GPS/INS integration, and mobile mapping systems. He is also the founder and CEO of Profound Positioning Inc. He published two books, 6 book chapters and over 450 papers in academic journals, conference and workshop proceedings, in which he has received over 30 paper awards. He supervised and graduated over 60 Masters and PhD students. He is the recipient of many national and international awards including the ASTech ``Leadership in Alberta Technology” Award, and the Association of Professional Engineers, Geologists, and Geophysicists of Alberta (APEGGA) Educational Excellence Award.
\end{IEEEbiography}




\end{document}